\documentclass[runningheads]{svmult}

\usepackage[dvips]{graphicx}
\usepackage{epsfig}
\usepackage{amsmath}

\def\d{\partial}
\def\vxi{\mbox{\boldmath $\xi$}}
\def\vecxi{\mbox{\boldmath $\xi$}}

\def\vecB{\mbox{\boldmath $B$}}
\def\vecU{\mbox{\boldmath $U$}}
\def\vecn{\mbox{\boldmath $n$}}
\def\vxi{\mbox{\boldmath $\xi$}}
\def\Rreal{{\rm I\kern-.2em R}}

\font\tenbi=cmmib10
\newfam\bifam \def\bi{\fam\bifam\tenbi} \textfont\bifam=\tenbi
\def\veccA{\mbox{\boldmath $c_A$}}

\def\vecU{{\textstyle{\bi U}}}
\def\veck{{\textstyle{\bi k}}}

\def\vece{{\textstyle{\bi e}}}
\def\vecr{{\textstyle{\bi r}}}
\def\vecu{{\textstyle{\bi u}}}

\newcommand{\aap}{    {\rm Astron. Astrophys.}\ }

\newcommand{\apj}{    {\rm Astrophys. J.}\ }
\newcommand{\apjl}{   {\rm Astrophys. J. Lett.}\ }

\newcommand{\mnras}{  {\rm Mon. Not. Roy. Astron. Soc.}\ }
\newcommand{\nat}{    {\rm Nature}\ }

\newcommand{\pasj}{   {\rm Pub. Astron. Soc. Japan}\ }

\newcommand{\solphys}{{\rm Solar Phys.}\ }

\usepackage{amsfonts}   
\begin{document}
\title*{Advances in Global and Local Helioseismology: an Introductory Review}
\toctitle{Advances in Global and Local Helioseismology: an Introductory Review}
%
%
\titlerunning{Advances in Global and Local Helioseismology}
%

\author{Alexander G. Kosovichev}
\authorrunning{Alexander G. Kosovichev}

\institute{W.W. Hansen Experimental Physics Laboratory, Stanford University \\
Stanford, CA 94305, USA \\
E-mail: \texttt{sasha@sun.stanford.edu}}


\maketitle              

\begin{abstract}
\noindent Helioseismology studies the structure and dynamics of the
Sun's interior by observing oscillations on the surface. These
studies provide information about the physical processes that control
the evolution and magnetic activity of the Sun. In recent years,
helioseismology has made substantial progress towards the
understanding of  the physics of solar oscillations and the physical
processes inside the Sun, thanks to observational, theoretical and
modeling efforts. In addition to the global seismology of the Sun
based on measurements of global oscillation modes, a new field of
local helioseismology, which studies oscillation travel times and
local frequency shifts, has been developed. It is capable of
providing 3D images of the subsurface structures and flows. The
basic principles, recent advances and perspectives of global and local helioseismology are reviewed
in this article.
\end{abstract}

\section{Introduction}

\begin{figure}[t]
\begin{center}
  \includegraphics[width=\linewidth]{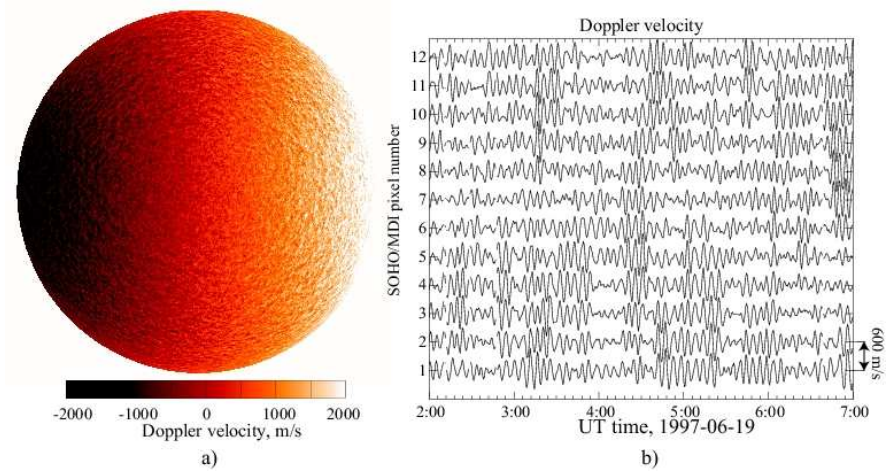}
  \caption{ a) Image of the line-of-sight (Doppler) velocity of the solar surface
  obtained by the Michelson Doppler Imager (MDI) instrument on board SOHO spacecraft on
  1997-06-19, 02:00 UT; b) Oscillations of the Doppler velocity, measured by MDI at the solar disk center in 12 CCD pixels separated by $\sim 1.4$~Mm on the Sun.
   }\label{fig1}
\end{center}
\end{figure}

In 1926 in his book {\it The Internal Constitution of the Stars} Sir
Arthur Stanley Eddington \cite{Eddington1926} wrote:
\begin{quotation}At first sight it would
seem that the deep interior of the sun and stars is less accessible
to scientific investigation than any other region of the universe.
Our telescopes may probe farther and farther into the depths of
space; but how can we ever obtain certain knowledge of that which is
hidden behind substantial barriers?  What appliance can pierce
through the outer layers of a star and test the conditions within?
\end{quotation}
The answer to this question was provided a half a century later by
{\it helioseismology}. Helioseismology studies the conditions inside the
Sun by observing and analyzing oscillations and waves on the
surface. The solar interior is not transparent to light but it is
transparent to acoustic waves. Acoustic (sound) waves on the Sun are
excited by turbulent convection below the visible surface
(photosphere) and travel through the interior with the speed of
sound. Some of these waves are trapped inside the Sun and form
{\it resonant oscillation modes}. The travel times of acoustic waves and
frequencies of the oscillation modes depend on physical conditions of the
internal layers (temperature, density, velocity of mass flows, etc). By measuring the travel times and frequencies one
can obtain information these condition. This is the basic principle
of helioseismology. Conceptually it is very similar to the Earth's
seismology. The main difference is that the Earth's seismology
studies mostly individual events, earthquakes, while helioseismology
is based on the analysis of acoustic noise produced by solar
convection. However, recently the local helioseismic techniques have been
applied for ambient noise tomography of Earth's structures. The
solar oscillations are observed in variations of intensity of solar
images or, more commonly, in line-of-sight velocity of the surface
elements, which is measured from the Doppler shift of spectral
lines (Fig.~\ref{fig1}). Variations caused by these oscillations are very small, much
smaller than the noise produced by turbulent convection. Thus, their
observation and analysis requires special procedures.

Helioseismology is a relatively new discipline of solar physics and
astrophysics. It has been developed over the past few decades by a
large group of remarkable observers and theorists, and is continued
being actively developed. The history of helioseismology has been
very fascinating, from the initial discovery of the solar 5-min
oscillations and the initial attempts to understand the physical
nature and mechanism of these oscillations to detailed diagnostics
of the deep interior and subsurface magnetic structures associated
with solar activity. This development was not straightforward. As
this always happens in science controversial results and ideas
provided inspiration for further more detailed studies.

In a brief historical introduction, I describe some key contributions.
It is very interesting to follow
the line of discoveries that led to our current understanding of the oscillations
and helioseismology techniques. Then, I overview the basic concepts and results of helioseismology. The launch of the Solar Dynamics Observatory in 2010 opens a new era in helioseismology. The Helioseismic and Magnetic Imager (HMI) instrument will provide uninterrupted high-resolution Doppler-shift and vector magnetogram data over the whole disk. These data will provide a complete information about the solar oscillations and their interaction with solar magnetic fields.

\section{Brief history of helioseismology}
\label{history}

Solar oscillations were discovered in 1960 by Robert Leighton,
Robert Noyers and George Simon \cite{Leighton1962} by analyzing
series of Dopplergrams obtained at the Mt. Wilson Observatory. Instead
of the expected turbulent behavior of the velocity field they found
two distinct classes: large-scale horizontal cellular motions,
which they called {\it supergranulation}, and vertical {\it
quasi-periodic oscillations} with a period of about 300 seconds (5
min) and a velocity amplitude of about 0.4 km$\,$s$^{-1}$. It turned out that these
oscillations are the dominant vertical motion in the lower
atmosphere (chromosphere) of the Sun. It is remarkable that they
realized the diagnostic potential  noting that
these oscillations "offer a new means of determining certain local properties of
the solar atmosphere, such as the temperature, the vertical
temperature gradient, or the mean molecular weight". They also
pointed out that the oscillations might be excited in the Sun's
granulation layer, and account for a part of the energy transfer
from the convection zone into the chromosphere.

This discovery was confirmed by other observers, and for several
years it was believed that the oscillations represent transient atmospheric
waves excited by {\it granules}, small convective cells on the solar
surface, $1-2\times 10^3$ km in size and $8-10$ min lifetime.
The physical nature of the oscillation at that time was
unclear. In particular, the questions whether these
oscillations are acoustic or gravity waves, and if they represent
traveling or standing waves remained unanswered for almost a decade after the discovery.


Pierre Mein \cite{Mein1966} applied a two-dimensional Fourier
analysis (in time and space) to observational data obtained by John
Evans and his colleagues at the Sacramento Peak Observatory in
1962-65. His idea was to decompose the oscillation velocity field
into {\it normal modes}. He calculated the {\it oscillation power
spectrum} and investigated the relationship between the period and
horizontal wavelength (or {\it frequency-wavenumber diagram}). From
this analysis he concluded that the oscillations are acoustic waves
that are {\it stationary (evanescent)} in the solar atmosphere. He
also made a suggestion that the horizontal structure of the
oscillations may be imposed by the convection zone below the
surface.

Mein's results were confirmed by Edward Frazier \cite{Frazier1968}
who analyzed high-resolution spectrograms taken at the Kitt Peak
National Observatory in 1965. In the wavenumber-frequency diagram he
noticed that in addition to the primary 5-min peak peak there is a
secondary lower frequency peak, which was a new puzzle.

This puzzle was solved by Roger Ulrich \cite{Ulrich1970} who
following the ideas of Mein and Frazier, calculated the spectrum of
standing acoustic waves trapped in a layer below the photosphere. He
found that these waves may exist only along discrete line in the
wavenumber-frequency ($k-\omega$) diagram, and that the two peaks
observed by Frazier correspond to the first two harmonics ({\it normal
modes}). He formulated the conditions for observing the discrete
acoustic modes: observing runs must be longer longer one hour, must
cover a sufficiently large region of, at least, 60,000 km in size;
the Doppler velocity images must have a spatial resolution of 3,000
km, and be taken at least every 1 minute.

At that time the observing runs were very short, typically, 30-40
min. Only in 1974-75 Franz-Ludwig Deubner \cite{Deubner1975} was
able to obtain three 3-hour sets of observations using a
magnetograph of the Fraunhofer Institute in Anacapri. He measured
Doppler velocities along a $\sim 220,000$ km line on the solar disk
by scanning it periodically at 110 sec intervals with the scanning
steps of about 700 km. The Fourier analysis of these data provided
the frequency-wavenumber diagram with three or four {\it mode
ridges} in the oscillation {\it power spectrum} that represents the
squared amplitude of the Fourier components as a function of
wavenumber and frequency. Deubner's results provided unambiguous
confirmation of the idea that the 5-min oscillations observed on the
solar surface represent the standing waves or resonant acoustic
modes trapped below the surface. The lowest ridge in the diagram is
easily identified as the {\it surface gravity wave} because its
frequencies depend only on the wavenumber and surface gravity. The
ridge above is the first acoustic mode, a standing acoustic waves
that have one node along the radius. The ridge above
this corresponds to the second acoustic modes with two nodes, and so
on.

While these observations showed a remarkable qualitative agreement
with Ulrich's theoretical prediction, the observed power ridges in
the $k-\omega$ diagram were systematically lower than the
theoretical mode lines. Soon after, in 1975, Edward Rhodes, Ulrich
and Simon \cite{Rhodes1977} made independent observations at the
vacuum solar telescope at the Sacramento Peak Observatory and
confirmed the observational results. They also calculated the
theoretical mode frequencies for various solar models, and by
comparing these with the observations determined the limits on the
depth of the solar convection zone. This, probably, was the first
helioseismic inference.

However, it was believed that the acoustic (p) modes do not provide
much information about the solar interior because detailed
theoretical calculations  of their properties by Hiroyashi Ando and
Yoji Osaki \cite{Ando1977} showed that while these mode are
determined by interior resonances their amplitude ({\it
eigenfunctions}) is predominantly concentrated close to the surface.
Therefore, the main focus was shifted to observations and analysis
of global oscillations of the Sun with periods much longer than 5
min. This task was particularly important for explaining the
observed deficit of high-energy {\it solar neutrinos} \cite{Bahcall1969},
which could be either due to a low temperature (or heavy element
abundance - low metallicity) in the energy-generating core or
neutrino oscillations.

In 1975, Henry Hill, Tuck Stebbins and Tim Brown \cite{Hill1975}
reported on the detection of oscillations in their measurements of
solar oblateness.  The periods of these oscillations were between 10
and 40 min. They suggested that the oscillation signals might
correspond to global modes of the Sun. Independently, in 1976, two
groups, led by Andrei Severny at the Crimean Observatory
\cite{Severny1976} and George Isaak at the University of Birmingham
\cite{Brookes1976} found long-period oscillations in global-Sun
Doppler velocity signals. The oscillation with a period of 160 min
was particularly prominent and stable. The amplitude of this
oscillation was estimated close to 2 m/s. Later this oscillation was
found in observations at the Wilcox Solar Observatory
\cite{Scherrer1979} and at the geographical South Pole
\cite{Grec1980}. Despite significant efforts to identify this
oscillation among the solar resonant modes or to find a physical
explanation these results remain a mystery. This oscillations lost
the amplitude and coherence in the subsequent ground-based
measurements and was not found in later observations from SOHO
spacecraft \cite{Pall'e1998}. The period of this oscillation was
extremely close to 1/9 of a day, and likely was related to
terrestrial observing conditions.

Nevertheless, these studies played a very important role in
development of helioseismology and emphasized the need for long-term
stable and high-accuracy observations from the ground and space.
Attempts to detect long-period oscillations ({\it g-modes}) still
continue. However, the focus of helioseismology was shifted to
accurate measurements and analysis of the acoustic {\it p-modes}
discovered by Leighton.

The next important step was made in 1979 by the Birmingham group
\cite{Claverie1979}. They observed the Doppler velocity variations
integrated over the whole Sun for about 300 hours (but typically 8
hours a day) at two observatories, Izana, on Tenerife, and Pic du Midi
in the Pyrenees. In the power spectrum of 5-min oscillations they
detected several equally space lines corresponding to global ({\it
low-degree}) acoustic modes, radial, dipole and quadrupole. (In
terms of the angular degree these are labeled as $\ell=0, 1,$ and
2). Unlike, the previously observed local short horizontal
wavelength acoustic modes these oscillations propagate into the deep
interior and provide information about the structure of the solar
core. The estimated frequency spacing between the modes was 67.8
$\mu$Hz. This uniform spacing predicted theoretically by Yuri
Vandakurov \cite{Vandakurov1968} in the framework of a general
stellar oscillation theory corresponds to the inverse time that
takes for acoustic waves to travel from the surface of the Sun
through the center to the opposite side and come back. Thus, the frequency spacing
immediately gives an important constraint on the internal structure
of the Sun. A comparison with the solar models
\cite{Iben1976,Christensen-Dalsgaard1979} showed that the observed
spectrum is consistent with the spectrum of solar models with low
metallicity. This result was very exciting because if correct it
would provide a solution to the solar neutrino problem. Thus, the
determination of solar metallicity (or {\it heavy element
abundance}) became a central problem of helioseismology.

In the same year, 1974, Gerald Grec, Eric Fossat, and Martin Pomerantz
\cite{Grec1980} made 5-day continuous measurements at the
Amundsen-Scott Station at the South Pole of the global oscillations
and confirmed the Birmingham result. Also, they were able to resolve the fine
structure of the oscillation spectrum and in addition to the main
67.8 $\mu$Hz spacing ({\it large frequency separation}) between the
strongest peaks of $\ell=1$ and 2, observe a small 10-16 $\mu$Hz
splitting ({\it small separation}) between the $\ell=0$ and 2, and
$\ell=1$ and 3 modes. The small separation is mostly sensitive to
the central part of the Sun and provides additional diagnostic
power.

The comparison of the observed oscillation peaks in the frequency
power spectra with the p-mode frequencies calculated for solar
models showed that below the surface these oscillations correspond
to the standing waves with a large number of nodes along the radius
(or {\it high radial order}). The number of nodes is between 10 and
35, and it was difficult to determine the precise numbers for the
observed modes. This created an uncertainty in the helioseismic
determination of the heavy element abundance. Joergen
Christensen-Dalsgaard and Douglas Gough
\cite{Christensen-Dalsgaard1981} pointed out that while the South
Pole and new Birmingham data favor solar models with normal
metallicity the low metallicity models cannot be ruled out.

The uncertainty was resolved three years later in 1983 when Tom
Duvall and Jack Harvey \cite{Duvall1983} analyzed the Doppler velocity
data measured with a photo-diode array in 200 positions along the
North-South direction on the disk, and obtained the diagnostic
$k-\omega$ diagram for acoustic modes of degree $\ell$, from 1 to
110. This allowed them to connect in the diagnostic diagram the global
low-$\ell$ modes with the high-$\ell$ observed by Deubner. Since the
correspondence of the ridges on Deubner's diagram to solar
oscillation modes have been determined it was easy to identify the
low-$\ell$ modes by simply counting the ridges corresponding to the
low-$\ell$ frequencies. It turned out that the these modes are
indeed in the best agreement with the normal metallicity solar
model. This result had important implications for the solar neutrino
problem because it strongly indicated that the observed deficit of
solar neutrinos was not due to a low abundance of heavy elements on
the Sun but because of changes in neutrino properties (neutrino
oscillations) on their way from the energy-generating core to the
Earth. This was later confirmed by direct measurements of solar
neutrino properties \cite{Ahmad2002a}.

It was also important that the definite identification of the
observed solar oscillations in terms of normal oscillation modes
provided a solid foundation for developing diagnostic methods of
helioseismology based on the well-developed mathematical theory of
non-radial oscillations of stars
\cite{Pekeris1938,Cowling1941,Ledoux1958}. This theory provided
means for calculating eigenfrequencies and eigenfunctions of normal
modes for spherically symmetric stellar models. Mathematically, the problem is
reduced to solving a non-linear eigenvalue problem for a
fourth-order system of differential equations. This system has two
sequences of eigenvalues corresponding to p- and g-modes, and also a
degenerate solution, corresponding to f-modes (surface gravity
waves). The effects of rotation, asphericity and magnetic fields are
usually small and considered by a perturbation theory \cite{Gough1990,Dziembowski1984,Dziembowski1989,Dziembowski2005}.

An important prediction of the oscillation theory is that rotation
causes splitting of normal mode frequencies. Without rotation, the
normal mode frequencies are degenerate with respect to the azimuthal
wavenumber, $m$, that is the modes of the angular degree, $l$, and
radial order, $n$, have the same frequencies irrespective of the
azimuthal (longitudinal) wavelength. The stellar rotation removes
this degeneracy. Obviously, it does not affect the axisymmetrical
($m$=0) modes, but the frequencies of non-axisymmetrical modes are
split. Generally, these modes can be represented as a superposition
of two waves running around a star in two opposite directions
(prograde and retrograde waves). Without rotation, these modes have
the same frequencies and, thus, the same phase speed. In this case,
they form a standing wave. However, rotation increases the speed of
the prograde wave and decreases the speed of retrograde wave. This
results in an increase of the eigenfrequency of the prograde mode,
and a frequency decrease of the retrograde mode. This phenomenon is
similar to frequency shifts due to the Doppler effect. It is called
rotational frequency splitting.

The rotational frequency splitting was first observed by Ed Rhodes, Roger
Ulrich and Franz Deubner \cite{Rhodes1979,Ulrich1979,Deubner1979}.
These measurements provided first evidence that the rotation rate of
the Sun is not uniform but increases with depth. The rotational
splitting was initially measured for high-degree modes, but then the measurements were
extended to medium- and low-degree range by Tom Duvall and Jack
Harvey \cite{Duvall1984,Duvall1984a}, who made a long continuous
series of helioseismology observations at the South Pole. The
internal differential rotation law was determined from the data of
Tim Brown and Cherilynn Morrow \cite{Brown1987}.
 It was found that the differential
latitudinal rotation is confined in the convection zone, and that
the radiative interior rotates  almost uniformly, and also slower in
the equatorial region than the convective envelope \cite{Kosovichev1988,Brown1989}. Such rotation law was not
expected from theories of stellar rotation, which predicted that the
stellar cores rotate faster than the envelopes \cite{Rosner1985}.
The knowledge of the Sun's internal rotation law is of particular importance for understanding  the dynamo mechanism of magnetic field generation \cite{Parker1993}.

 It became clear that for long uninterrupted observations are
essential for accurate inferences of the internal structure and
rotation of the Sun.  Therefore, the observational programs focused
on development of global helioseismology networks, GONG
\cite{Harvey1988} and BiSON \cite{Brookes1978,Isaak1989}, and also
the Solar and Heliospheric Observatory (SOHO) space mission \cite{Domingo1995}. These
projects provided almost continuous coverage for helioseismic
observations and also stimulated development of new sophisticated
data analysis and inversion techniques.

In addition, the Michelson Doppler Imager (MDI) instrument on SOHO \cite{Scherrer1995}
and the GONG+ network upgraded to higher spatial resolution \cite{Leibacher2003}
provided
excellent opportunities for developing {\it local helioseismology},
which provides tools for three-dimensional imaging of the solar interior.
The local helioseismology methods are based on
measurements of local oscillation properties, such as frequency
shift in local areas or variations of travel times.

The idea of
using the local frequency shifts for inferring the subsurface flows
was suggested by Douglas Gough and Juri Toomre in 1983
\cite{Gough1983}. The method is now called {\it ring-diagram analysis} \cite{Hill1988},
because the dispersion relation of solar oscillations forms rings in
horizontal wavenumber plane at a given frequency. It measures shifts
of these rings, which are then converted into frequency shifts. Ten
years later, Tom Duvall and his colleagues \cite{Duvall1993}
introduced {\it time-distance helioseismology} method. In this method,
they suggested to measure travel times of acoustic waves from a
cross-covariance function of solar oscillations. This function is
obtained by cross-correlating oscillation signals observed at two
different points on the solar surface for various time lags. When
the time lag in the calculations coincides with the travel time of
acoustic waves between these points the cross-covariance function
shows a maximum. This method provided means for developing {\it acoustic
tomography} techniques \cite{Kosovichev1996,Kosovichev1997} for imaging 3D
structures and flows with the high-resolution comparable to the
oscillation wavelength. These and other methods of local area
helioseismology \cite{Chang1997,Lindsey2000}
have provided important results on the convective
and large-scale flows, and also on the structure and evolution of
sunspots and active regions. Their development continues.

The SOHO mission and the GONG network were primarily designed for observing solar oscillation modes of low- and medium-degree, needed for global helioseismology. Local helioseismology requires high-resolution
observations of high-degree modes. Because of the telemetry constraints such
data are available uninterruptedly from the MDI instrument on SOHO only for 2 months every year. These data provided only snapshots of the subsurface structures and dynamics associated with the solar activity. In order to fully investigate the evolving magnetic activity of the Sun, a new space mission Solar Dynamics Observatory (SDO) was launched on February 11, 2010. It carries Helioseismic and Magnetic Imager (HMI) instrument, which will provide continuous 4096x4096-pixel full-disk images of solar oscillations. These data will open new opportunities for investigation the solar interior by local helioseismology \cite{Kosovichev2007a}.

In the modern helioseismology, a very important role is played by
{\it numerical simulations}. Both, global and local helioseismology
analysis employ relatively simple for fitting the observational data
and performing inversions of the fitted frequencies and travel
times. For instance, the global helioseismology methods assume that
the structures and flows on the Sun are axisymmetrical and infer
only the axisymmetrical components of the sound speed and velocity
field. The local helioseismology methods are based on a simplified
physics of wave propagation on the Sun. The ring-diagram analysis
makes an assumption that that the perturbations and flows are
horizontally uniform within the area used for calculating the wave
dispersion relation, 5-15 heliographic degrees, while a typical size
of sunspots is about 1-2 degrees. Most of the time-distance
helioseismology inversions are based on a ray-path approximation and
ignore the finite wavelength effects that become important at small
scales, comparable with the wavelength. Also, all the methods,
global and local, do not take into account many effects of solar
magnetic fields. Properties of solar oscillations dramatically
change in regions of strong magnetic field. In particular, the
excitation of oscillations is suppressed in sunspots because the
strong magnetic field inhibits convection that drives the
oscillations. The magnetic stresses may cause anisotropy of wave
speed and lead to transformation of acoustic waves into various MHD
type waves. These and other effects have to be investigated and
taken into account in the data analysis and inversion procedures.
Because of the complexity, these processes can be fully investigated
only numerically. The numerical simulations of subsurface solar
convection and oscillations were pioneered by Robert Stein and
{\AA}ke Nordlund \cite{Stein1989}. These 3D radiative MHD
simulations include all essential physics and provide important
insights into the physical processes below the visible surface and
also artificial data for helioseismology testing. This type of
so-called ''realistic" simulations has been used for testing
time-distance helioseismology inferences \cite{Zhao2007}, and
continues being developed using modern turbulence models
\cite{Jacoutot2008}. In addition, for testing various aspects of
wave propagation and interaction with magnetic fields are studied by
solving numerically linearized MHD equations (e.g.
\cite{Hanasoge2007,Parchevsky2008,Hartlep2008}). The numerical
simulations become an important tool for verification and testing of
the helioseismology methods and inferences.

\section{Basic properties of solar oscillations}
\label{basic}

\subsection{Oscillation power spectrum}
\begin{figure}[t]
\begin{center}
  \includegraphics[width=\linewidth]{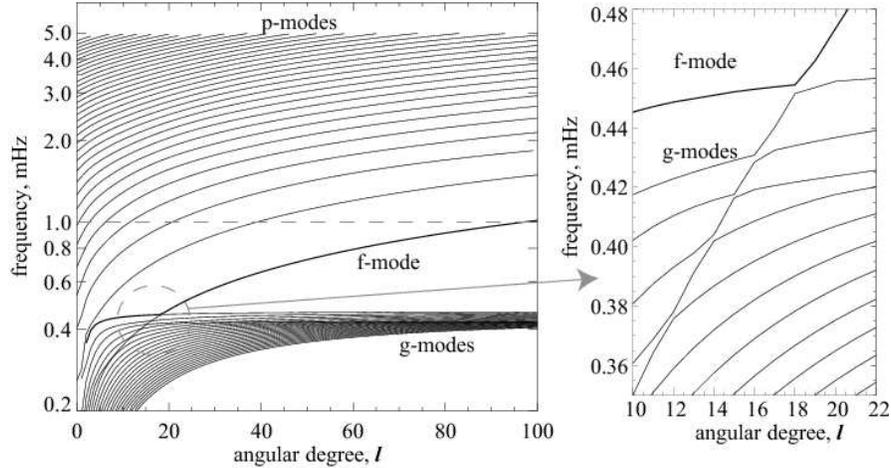}
  \caption{Theoretical frequencies of solar oscillation modes calculated for a standard
  solar model for a range of angular degree $l$ from 0 to 100, and for the frequency range
  from 0.2~mHz to 5~mHz. The solid curves connect modes corresponding to the different oscillation
  overtones (radial orders) The dashed grey horizontal line indicate the low-frequency observational limit: only the modes above this line have been reliably observed. The right panel shows an area of the avoided crossing of f- and g-modes (indicated by the gray dashed circle in left panel).
   }\label{fig2}
\end{center}
\end{figure}
The theoretical spectrum of solar oscillation modes shown in Fig.~\ref{fig2}
covers a wide range of frequencies and angular degrees. It includes
oscillations of three types: {\it acoustic (p) modes}, {\it surface gravity (f)
modes} and {\it internal gravity (g) modes}. In this spectrum, the modes are organized a series of curves corresponding to different overtones
of non-radial modes, which are characterized by the number of nodes along the radius (or by the radial
order, $n$). The angular degree, $l$, of the corresponding spherical
harmonics describes the horizontal wave number (or inverse
horizontal wavelength). The p-modes cover the frequency range from
0.3 to 5 mHz (or from 3 to 55 min in oscillation periods). The low
frequency limit corresponds to the first radial harmonic, and the
upper limit is set by the acoustic cut-off frequency of the solar
atmosphere. The g-modes frequencies have an upper limit corresponding to the
maximum Br\"unt-V\"ais\"al\"a frequency ($\sim 0.45$ mHz) in the
radiative zone and occupy the low-frequency part of the spectrum.
The intermediate frequency range of 0.3-0.4 mHz at low angular
degrees is a region of mixed modes. These modes behave like g-modes
in the deep interior and like p-modes in the outer region. The
apparent crossings in this diagram are not the actual crossings: the
mode branches become close in frequencies but do not cross each other. At these points the mode exchange their properties, and the mode branches are diverted. For instance, the f-mode ridge stays above the g-mode lines. A
similar phenomenon is known in quantum mechanics as {\it avoided
crossing}.

\begin{figure}[t]
\begin{center}
  \includegraphics[width=.9\linewidth]{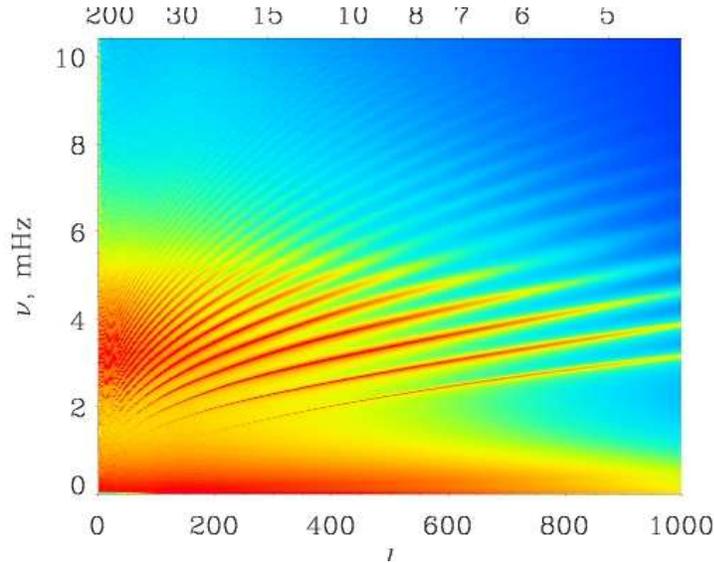}
  \caption{Power spectrum obtained from a 6-day long time series
  of solar oscillation data from the MDI instrument on SOHO in 1996 ($\nu$ is the cyclic frequency of the oscillations, $l$ is the angular degree, $\lambda_h$ is the horizontal wavelength in megameters).
  }\label{fig3}
\end{center}
\end{figure}

So far, only the upper part of the solar oscillation spectrum is
observed. The lowest frequencies of detected p- and f-modes are of
about 1 mHz. At lower frequencies the mode amplitudes decrease below
the noise level, and become unobservable. There have been several
attempts to identify low-frequency p-modes or even g-modes in the
noisy spectrum, but so far these results are not convincing.

\begin{figure}[t]
\begin{center}
\includegraphics[width=0.9\linewidth]{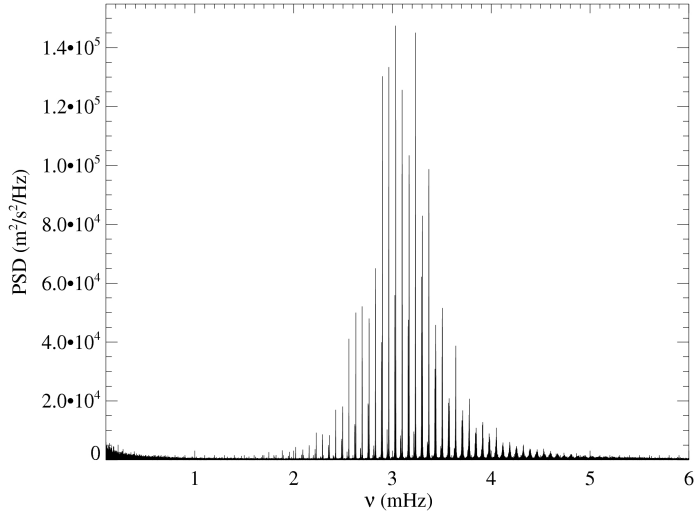}
\caption{Power spectral density (PSD) of low-degree solar oscillations, obtained from the integrated light
observations (Sun-as-a-star) by the GOLF instrument on SOHO, from 11/04/1996 to 08/07/2008.
}\label{fig4}
\end{center}
\end{figure}

The observed power spectrum is shown in Fig.~\ref{fig3}. The lowest ridge is
the f-mode, and the other ridges are p-modes of the radial order,
$n$, starting from $n=1$. The ridges of the oscillation modes
disappear in the convective noise at frequencies below 1 mHz.
The power spectrum is obtained from the SOHO/MDI data, representing 1024x1024-pixel images of the line-of-sight (Doppler) velocity of the solar surface taken every minute without interruption.
When the oscillations are observed in the integrated solar light ("Sun-as-a-star") then only the modes of low angular degree are detected
 in the power spectrum (Fig.~\ref{fig4}). These modes have a mean period
  of about 5 min, and represent p-modes of high radial order $n$ modes.
The
$n$-values of these modes can be determined by tracing in Fig.~\ref{fig3} the
the high-$n$ ridges of the high-degree modes into the low-degree
region. This provides unambiguous identification of the low-degree
solar modes. Obviously, the mode identification is much more
difficult for spatially unresolved oscillations of other stars.

\subsection{Excitation by turbulent convection}
Observations and numerical simulations have shown that
solar oscillations are driven by turbulent convection in a shallow
subsurface layer with a superadiabatic stratification,
where convective velocities are the
highest. However, details of the stochastic excitation mechanism are
not fully established. Solar convection in the superadiabatic layer
forms small-scale granulation cells. Analysis of the observations and
numerical simulations has shown that sources of solar oscillations
are associated with strong downdrafts in dark intergranular lanes
\cite{Rimmele1995}. These downdrafts are driven by radiative
cooling and may reach near-sonic velocity of several km/s. This
process has features of convective collapse \cite{Skartlien2000}.

\begin{figure}[t]
\begin{center}
  \includegraphics[width=0.9\linewidth]{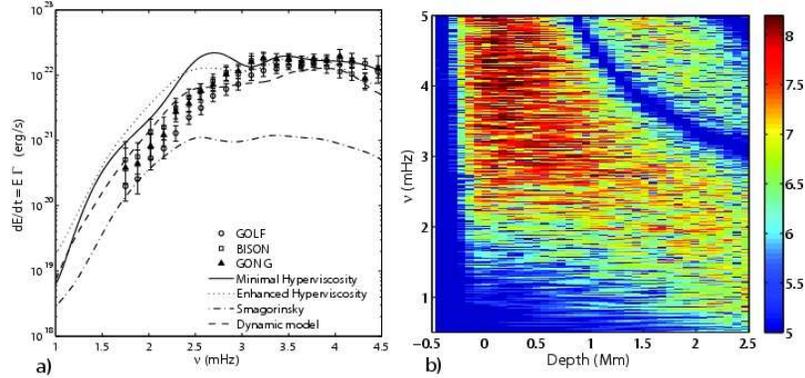}
  \caption{a) Comparison of observed and calculated rate of stochastic
energy input to modes for the entire solar surface ($erg\;s^{-1}$).
Different curves show the numerical simulation results obtained for 4 turbulence models: hyperviscosity (solid), enhanced hyperviscosity (dots), Smagorinsky (dash-dots), and dynamic model (dashes).
Observed distributions: $circles$ SoHo-GOLF, $squares$ BISON, and
$triangles$ GONG for $l=1$ \cite{Baudin2005}. b) Logarithm of the
work integrand in units of $erg\;cm^{-2}\;s^{-1}$), as a function of
depth and frequency for numerical simulations with the dynamic
turbulence model \cite{Jacoutot2008a}.}\label{fig5}
\end{center}
\end{figure}

Calculations of the work integral for acoustic modes using the
realistic numerical simulations of Stein and Nordlund
\cite{Stein2001} have shown that the principal contribution to the
mode excitation is provided by turbulent Reynolds stresses and that
a smaller contribution comes from non-adiabatic pressure
fluctuations. Because of the very high Reynolds number of the solar
dynamics the numerical modeling  requires an accurate
description of turbulent dissipation and transport on the numerical subgrid
scale. The recent radiative hydrodynamics modeling using the Large-Eddy
Simulations (LES) approach and various subgrid scale (SGS)
formulations \cite{Jacoutot2008} showed that among these
formulations the most accurate description in terms of the
reproducing the total amount of the stochastic energy input to the
acoustic oscillations is provided by a dynamic Smagorinsky model
\cite{Germano1991,Moin1991} (Fig.~\ref{fig5}a).

As we have pointed out, the observations show that the modal lines in
the oscillation power spectrum are not Lorentzian but display a
strong asymmetry \cite{Duvall1993a,Toutain1998}. Curiously, the
asymmetry has the opposite sense in the power spectra calculated
from Doppler velocity and intensity oscillations. The asymmetry
itself can be easily explained by interference of waves emanated by
a localized source \cite{Gabriel1992}, but the asymmetry reversal is
surprising and indicates complicated radiative dynamics of the
excitation process. The reversal has been attributed to a {\it correlated noise} contribution to the observed intensity oscillations \cite{Nigam1998}, but the physics of this effect is still not fully understood. However, it
is clear that the line shape of the oscillation modes and the
phase-amplitude relations of the velocity and intensity oscillations
carry substantial information about the excitation mechanism and,
thus, require careful data analysis and modeling.

\subsection{Line asymmetry and pseudo-modes}

\begin{figure}
\begin{center}
  \includegraphics[width=.6\linewidth]{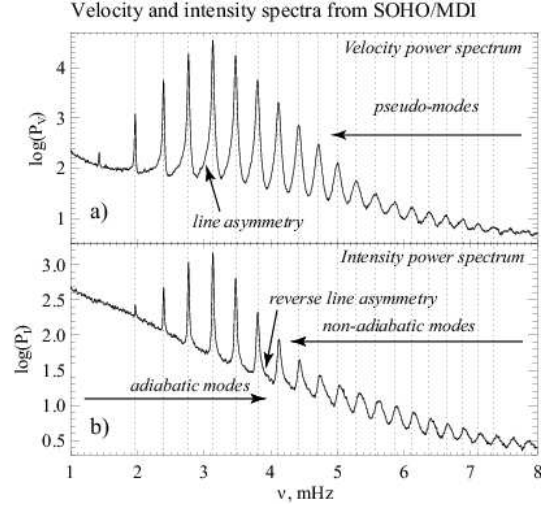}
  \caption{Power spectra of $l=200$ modes obtained from SOHO/MDI
observations of a) Doppler velocity, b) continuum intensity
\cite{Nigam1998}. }\label{fig6}
\end{center}
\end{figure}

Figure~\ref{fig6} shows the power spectrum for oscillations of the angular degree, $l=200$,
obtained from the SOHO/MDI Doppler velocity and intensity data \cite{Nigam1998}. The
line asymmetry is apparent, particularly, at low frequencies. In the
velocity spectrum, there is more power in the low-frequency wings
than in the high-frequency wings of the spectral lines. In the
intensity spectrum, the distribution of power is reversed. The data
also show that the asymmetry varies with frequency. It is the
strongest for the f-mode and low-frequency p-mode peaks. At higher
frequencies the peaks become more symmetrical, and extend well above
{\it the acoustic cut-off frequency} (Eq.~\ref{eq-cutoff}), which is $\sim 5-5.5$ mHz.

Acoustic waves with frequencies below the cut-off frequency are
completely reflected by the surface layers because of the steep
density gradient. These waves are trapped in the interior, and their
frequencies are determined by the resonant conditions, which depend
on the solar structure. But the waves with frequencies above the
cut-off frequency escape into the solar atmosphere. Above this
frequency the power spectrum peaks correspond to so-called
"pseudo-modes". These are caused by constructive interference of
acoustic waves excited by the sources located in the granulation
layer and traveling upward, and by the waves traveling downward, reflected
in the deep interior and arriving back to the surface. Frequencies
of these modes are no longer determined by the resonant conditions
of the solar structure. They depend on the location and properties
of the excitation source ("source resonance"). The pseudo-mode peaks
in the velocity and intensity power spectra are shifted relative to
each other by almost a half-width. They are also slightly shifted
relative to the normal mode peaks although they look like a
continuation of the normal-mode ridges in Figs~1b and 4a. This
happens because the excitation sources are located in a shallow
subsurface layer, which is very close to the reflection layers of
the normal modes. Changes in the frequency distributions below and
above the acoustic cut-off frequency can be easily noticed by
plotting the frequency differences along the modal ridges.
\begin{figure}
\begin{center}
  \includegraphics[width=\linewidth]{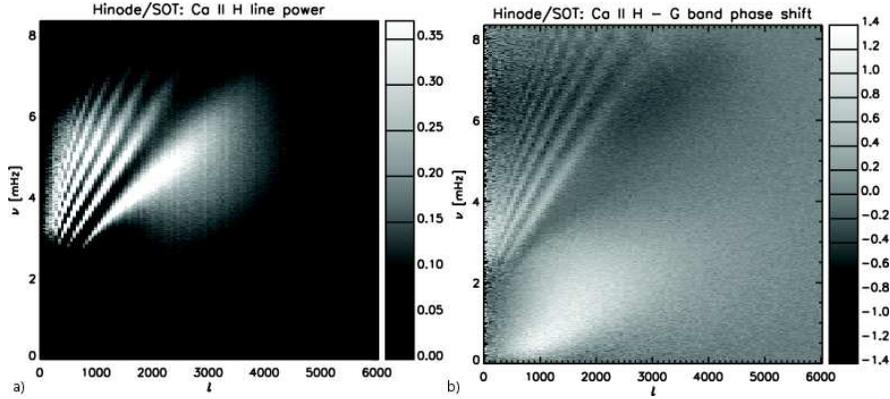}
  \caption{a) The oscillation power spectrum from Hinode CaII H line
  observations. b) The phase shift between CaII H and G-band
(units are in radians) \cite{Mitra-Kraev2008}.}\label{fig7}
\end{center}
\end{figure}

The asymmetrical profiles of normal-mode peaks are also caused by
the localized excitation sources. The interference signal between
acoustic waves traveling from the source upwards and  the waves
traveling from the source downward and coming back to the surface
after the internal reflection depends on the wave frequency.
Depending on the multipole type of the source  the interference signal can
be stronger at frequencies lower or higher than the resonant normal
frequencies, thus resulting in asymmetry in the power distribution
around the resonant peak. Calculations of Nigam et al.
\cite{Nigam1998} showed that the asymmetry observed in the velocity
spectra and the distribution of the pseudo-mode peaks can be
explained by a composite source consisting of a monopole term (mass
term) and a dipole term (force due to Reynolds stress) located in
the zone of superadiabatic convection at a depth of $\simeq 100$ km
below the photosphere. In this model, the reversed asymmetry in the
intensity power spectra is explained by effects of a correlated
noise added to the oscillation signal through fluctuations of solar
radiation during the excitation process. Indeed, if the excitation
mechanism is associated with the high-speed turbulent downdrafts in
dark lanes of granulation the local darkening contributes to the
intensity fluctuations caused by excited waves. The model also
explains the shifts of pseudo-mode frequency peaks and their higher
amplitude in the intensity spectra. The difference between the correlated and uncorrelated noise is that the correlated noise has some phase coherence with the oscillation signal, while the uncorrelated noise has no coherence.

While this scenario looks plausible and qualitatively explains the
main properties of the power spectra details of the physical
processes are still uncertain. In particular, it is unclear whether
the correlated noise affects only the intensity signal or both the
intensity and velocity. It has been suggested that the velocity
signal may have a correlated contribution due to convective
overshoot \cite{Roxburgh1997}. Attempts to estimate the correlated
noise components from the observed spectra have not provided
conclusive results \cite{Severino2001,Wachter2005}. Realistic
numerical simulations \cite{Georgobiani2003} have reproduced the
observed asymmetries and provided an indication that radiation
transfer plays a critical role in the asymmetry reversal.

Recent high-resolution observations of solar oscillation
simultaneously in two intensity filters, in molecular G-band and
CaII H line, from the Hinode space mission
\cite{Kosugi2007,Tsuneta2008} revealed significant  shifts in
frequencies of pseudo-modes observed in the CaII H and G-band
intensity oscillations \cite{Mitra-Kraev2008}. The phase of the
cross-spectrum of these oscillations shows peaks associated with the
p-mode lines but no phase shift for the f-mode (Fig.~\ref{fig7}b). The p-mode
properties can be qualitatively reproduced in a simple model with a
correlated background if the correlated noise level in the Ca II H
data is higher than in the G-band data \cite{Mitra-Kraev2008}.
Perhaps, the same effect can explain also the frequency shift of
pseudo-modes. The CaII H line is formed in the lower chromosphere
while the G-band signal comes from the photosphere. But how this may
lead to different levels of the correlated noise is unclear.

The Hinode results suggest that multi-wavelength observations of
solar oscillations, in combination with the traditional
intensity-velocity observations, may help to measure the level of
the correlated background noise and to determine the type of wave
excitation sources on the Sun. This is important for understanding the physical mechanism of the line asymmetry and for developing more accurate models and fitting formulae for determining the mode frequencies \cite{Nigam1998a}.

In addition, Hinode provided observations of non-radial acoustic and
surface gravity modes of very high angular degree. These
observations show that the oscillation ridges are extended up to $l
\simeq 4000$ (Fig.~\ref{fig7}a). In the high-degree range, $l \geq 2500$
frequencies of all oscillations exceed the acoustic cut-off
frequency. The line width of these oscillations dramatically
increases, probably due to strong scattering on turbulence
\cite{Duvall1998,Murawski1998}. Nevertheless, the ridge structure
extending up to 8 mHz (Nyquist frequency of these observations) is
quite clear. Although the ridge slope clearly changes at the
transition from the normal modes to the pseudo-modes.

\subsection{Magnetic effects: sunspot oscillations and acoustic halos}

In general, the main factors causing variations in oscillation
properties in magnetic regions, can be divided in two types: direct
and indirect. The direct effects are due to additional magnetic
restoring forces that can change the wave speed and may transform
acoustic waves into different types of MHD waves. The indirect
effects are caused by changes in convective and thermodynamic
properties in magnetic regions. These include depth-dependent
variations of temperature and density, large-scale flows, and
changes in wave source distribution and strength. Both direct and
indirect effects may be present in observed properties such as
oscillation frequencies and travel times, and often cannot be easily
disentangled by data analyses, causing confusions and
misinterpretations. Also, one should keep in mind that simple models
of MHD waves derived for various uniform magnetic configurations and
without stratification or with a polytropic stratification may not provide correct explanations to solar
phenomena. In this situation, numerical simulations play an
important role in investigations of magnetic effects.

Observed changes of oscillation amplitude and frequencies in
magnetic regions are often explained as a result of wave
scattering and conversion into various MHD modes. However, recent
numerical simulations helped us to understand that magnetic fields
not only affect the wave dispersion properties but also the
excitation mechanism. In fact, changes in excitation properties of
turbulent convection in magnetic regions may play a dominant role in
observed phenomena.

\subsubsection{Sunspot oscillations}

\begin{figure}
\begin{center}
  \includegraphics[width=\linewidth]{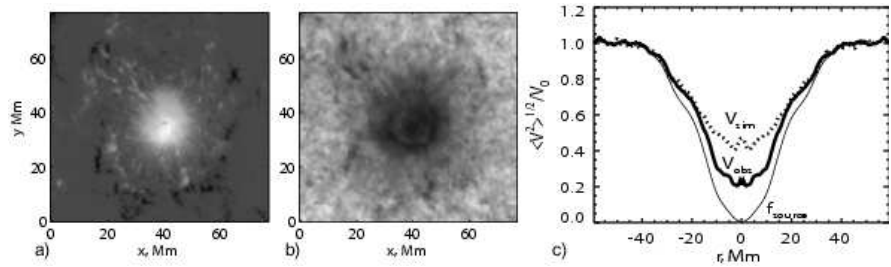}
  \caption{a) Line-of-sight magnetic field map of a sunspot (AR8243); b) oscillation amplitude map; c)
profiles of rms oscillation velocities at frequency 3.65 mHz for
observations (thick solid curves) and simulations (dashed curves);
the thin solid curve shows the distribution of the simulated source
strength \cite{Parchevsky2007a}.}\label{fig8}
\end{center}
\end{figure}

For instance, it is well-known that the amplitude of 5-min
oscillations is substantially reduced in sunspots. Observations show
that more waves are coming into the sunspot than going out of the
sunspot area (e.g. \cite{Braun1987}). This is often attributed to
absorption of acoustic waves in magnetic field due to conversion
into slow MHD modes traveling along the field lines (e.g.
\cite{Cally2009}). However, since convective motions are inhibited
by the strong magnetic field of sunspots, the excitation mechanism
is also suppressed. Three-dimensional numerical simulations of this
effect have shown that the reduction of acoustic emissivity can
explain at least 50\% of the observed power deficit in sunspots
(Fig.~\ref{fig8}) \cite{Parchevsky2007a}.

Another significant contribution comes from the amplitude changes
caused by variations in the background conditions. Inhomogeneities
in the sound speed may increase or decrease the amplitude of
acoustic wave traveling through these inhomogeneities. Numerical
simulations of MHD waves using magnetostatic sunspot models show
that the amplitude of acoustic waves traveling through sunspot
decreases when the wave is inside sunspot and then increases when
the wave comes out of sunspot \cite{Parchevsky2010}. Simulations
with multiple random sources show that these changes in the wave
amplitude together with the suppression of acoustic sources can
explain the whole observed deficit of the power of 5-min
oscillations. Thus, the role of the MHD mode conversion may be
insignificant for explaining the power deficit of 5-min photospheric
oscillations in sunspots. However, the mode conversion is expected
to be significant higher in the solar atmosphere where magnetic
forces become dominant.

We should note that while the 5-min oscillations in sunspots come
mostly from outside sources there are also 3-min oscillations, which
are probably intrinsic oscillations of sunspots. The origin of these
oscillations is not yet understood. They are probably excited by a
different mechanism operating in strong magnetic field.
\begin{figure}
\begin{center}
  \includegraphics[width=\linewidth]{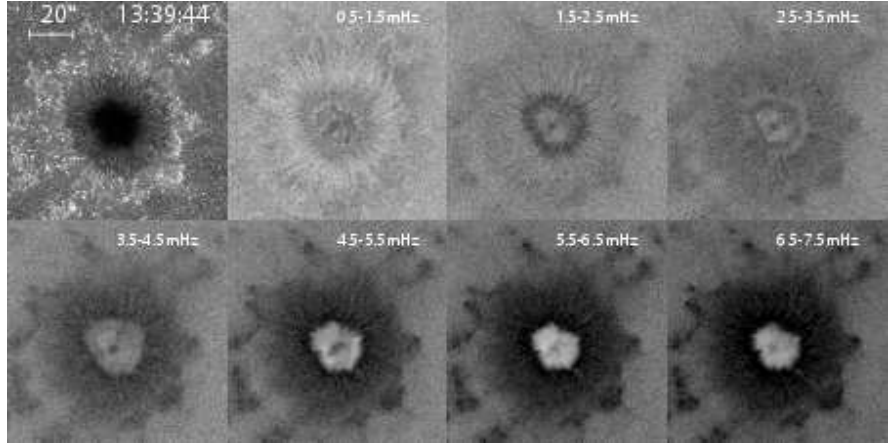}
  \caption{CaII H intensity image from Hinode observations
(top-left) and the corresponding power maps from CaII H intensity
data in five frequency intervals of active region NOAA 10935. The
field of view is 100 arcsec square in all the panels. The power is
displayed in logarithmic greyscaling \cite{Nagashima2007}.}
\end{center}\label{fig9}
\end{figure}

Hinode observations added new puzzles to sunspot oscillations.
Figure~\ref{fig9} shows a sample Ca\,II\,H intensity and the relative
intensity power maps averaged over 1 mHz intervals in the range from
1 mHz to 7 mHz with logarithmic greyscaling \cite{Nagashima2007}.
 In the Ca\,II\,H power maps, in all the frequency ranges,
there is a small area ($\sim$ 6 arcsec in diameter) near the center
of the umbra where the power was suppressed.  This type of `node'
has not been reported before. Possibly, the stable high-resolution
observation made by Hinode/SOT was required to find such a tiny
node, although analysis of other sunspots indicates that probably
only a particular type of sunspots, e.g., round ones with
axisymmetric geometry, exhibit such node-like structure.
 Above 4 mHz in the Ca\,II\,H power maps, power in the
umbra is remarkably high. In the power maps averaged over narrower
frequency range (0.05 mHz wide, not shown), the region with high
power in the umbra seems to be more patchy. This may correspond to
elements of umbral flashes, probably caused by overshooting
convective elements \cite{Schussler2006}. The Ca\,II\,H power maps
show a bright ring in the penumbra at lower frequencies. It probably
corresponds to the running penumbral waves.
 The power spectrum in the umbra has two peaks: one around 3
mHz and the other around 5.5 mHz. The high-frequency peak is caused by the oscillations that excited only in the strong magnetic field of sunspots.
The origin of these oscillations is not known yet.

\subsubsection{Acoustic halos}
\begin{figure}
\begin{center}
  \includegraphics[width=\linewidth]{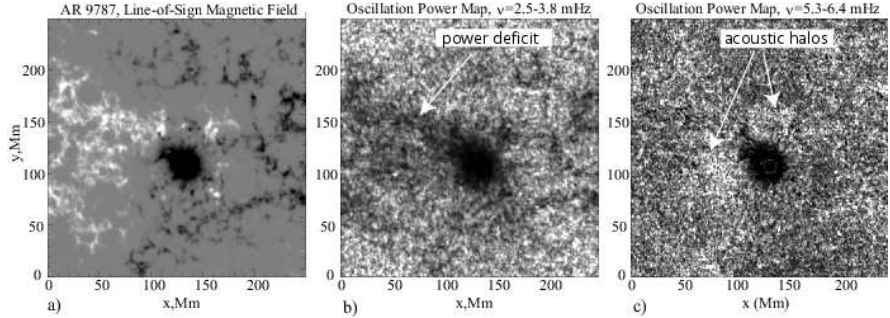}
  \caption{a) Line-of-sight magnetic field map of active region NOAA 9787 observed from SOHO/MDI
  on Jan. 24, 2002 and averaged over a 3-hour
  period; b) oscillation power map from Doppler velocity measurements for the same period in the frequency 2.5--3.8 mHz; c) power map for 5.3--6.4 mHz.}\label{fig10}
\end{center}
\end{figure}

\begin{figure}
\begin{center}
  \includegraphics[width=0.8\linewidth]{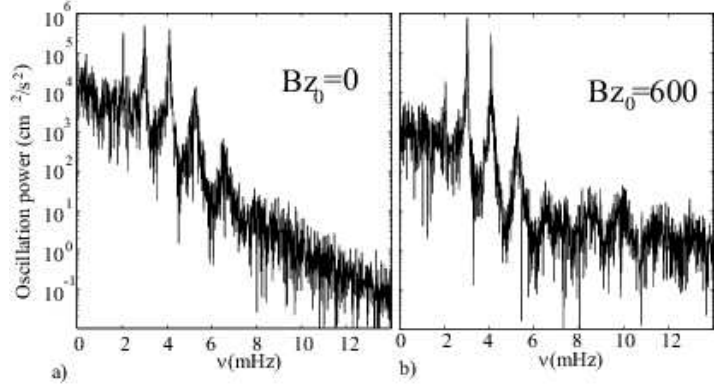}
  \caption{Power spectra of the horizontally averaged vertical velocity at the visible surface for
different initial vertical magnetic fields. The peaks on the top of
the smooth background spectrum of turbulent convection represent
oscillation modes: the sharp asymmetric peaks below 6 mHz are
resonant normal modes, while the broader peaks above 6 mHz, which
become stronger in magnetic regions, correspond to
pseudo-modes.\cite{Jacoutot2008a}}\label{fig11}
\end{center}
\end{figure}

In moderate field regions, such as plages around sunspot regions,
observations reveal enhanced emission at high frequencies, 5-7 mHz,
(with period $\sim 3$ min) \cite{Braun1992}. Sometimes this emission
is called the "acoustic halo" (Fig.~\ref{fig10}c).
There have been several attempts to explain this effect as a result of  wave transformation or scattering in magnetic structures (e.g. \cite{Hanasoge2008,Khomenko2009}). However,  numerical simulations show that magnetic field can change the excitation properties of solar granulation resulting in an enhanced high-frequency emission.
In particular, the radiative MHD simulations of solar
convection \cite{Jacoutot2008a} in the presence of vertical magnetic
field have shown that the magnetic field significantly changes the
structure and dynamics of granulations, and thus the conditions of
wave excitation. In magnetic field the granules become smaller, and
the turbulence spectrum is shifted towards higher frequencies. This
is illustrated in Figure~\ref{fig11}, which shows the frequency
spectrum of the horizontally averaged vertical velocity. Without a
magnetic field the turbulence spectrum declines sharply at
frequencies above 5 mHz, but in the presence of magnetic field it
develops a plateau. In the plateau region characteristic peaks
(corresponding to the "pseudo-modes") appear in the spectrum for
moderate magnetic field strength of about 300-600 G. These peaks may
explain the effect of the "acoustic halo". Of course, more detailed
theoretical and observational studies are required to confirm this
mechanism. In particular, multi-wavelength observations of solar
oscillations at several different heights would be important.
Investigations of the excitation mechanism in magnetic regions is
also important for interpretation of the variations of the frequency
spectrum of low-degree modes on the Sun, and for asteroseismic
diagnostics of stellar activity.

\subsection{Impulsive excitation: sunquakes}
\begin{figure}
\begin{center}
  \includegraphics[width=0.95\linewidth]{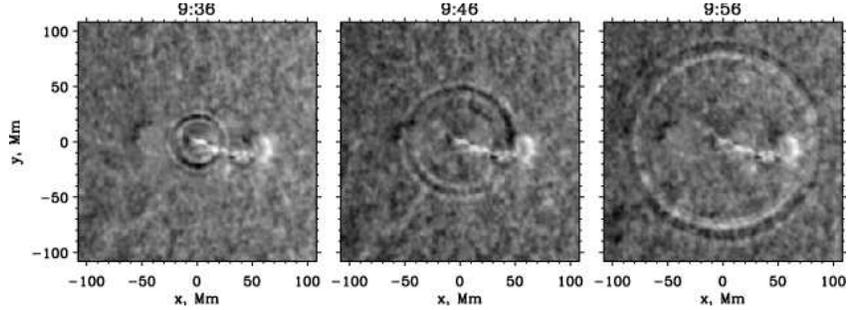}
  \caption{Observations of the seismic response  (``sunquakes'')
of the solar flare of 9 July, 1996, showing a sequence of Doppler-velocity images, taken by
the SOHO/MDI instrument. The signal of expanding ripples is enhanced by a factor 4 in
the these images.}\label{fig12}
\end{center}
\end{figure}

 ``Sunquakes", the helioseismic response to solar flares, are
caused by strong localized hydrodynamic impacts in the photosphere
during the flare impulsive phase. The helioseismic waves have been
observed directly as expanding circular-shaped ripples in SOHO/MDI
Dopplergrams \cite{Kosovichev1998} (Fig.~\ref{fig12}).

\begin{figure}
\begin{center}
  \includegraphics[width=0.8\linewidth]{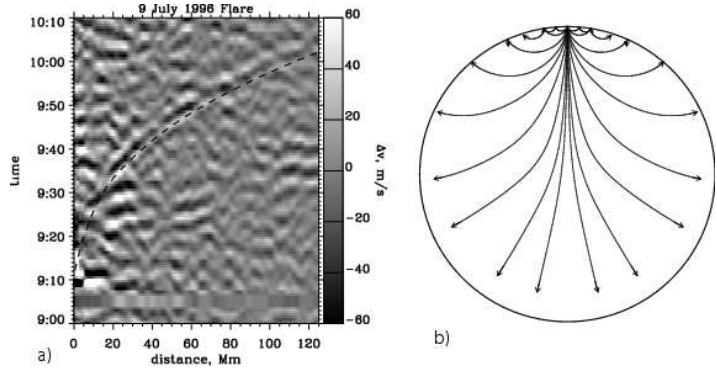}
  \caption{a) The time-distance diagram of the seismic response to the
  solar flare of 9 July, 1996. b) Illustration of acoustic ray paths of the
  flare-excited waves traveling through the Sun.}\label{fig13}
\end{center}
\end{figure}

These waves can be detected in Dopplergram movies and as a
characteristic ridge in time-distance diagrams (Fig.~\ref{fig13}a),
\cite{Kosovichev1998,Kosovichev2006a,Kosovichev2006b,Kosovichev2006c}, or indirectly by
calculating integrated acoustic emission \cite{Donea1999,Donea2005,Donea2006}. Solar flares are sources of high-temperature plasma and
strong hydrodynamic motions in the solar atmosphere. Perhaps, in all
flares such perturbations generate acoustic waves traveling through
the interior. However, only in some flares is the impact
sufficiently localized and strong to produce the seismic waves with
the amplitude above the convection noise level. It has been
established in the initial July 9, 1996, flare observations
\cite{Kosovichev1998} that the hydrodynamic impact follows the hard
X-ray flux impulse, and hence, the impact of high-energy electrons.
\begin{figure}
\begin{center}
  \includegraphics[width=\linewidth]{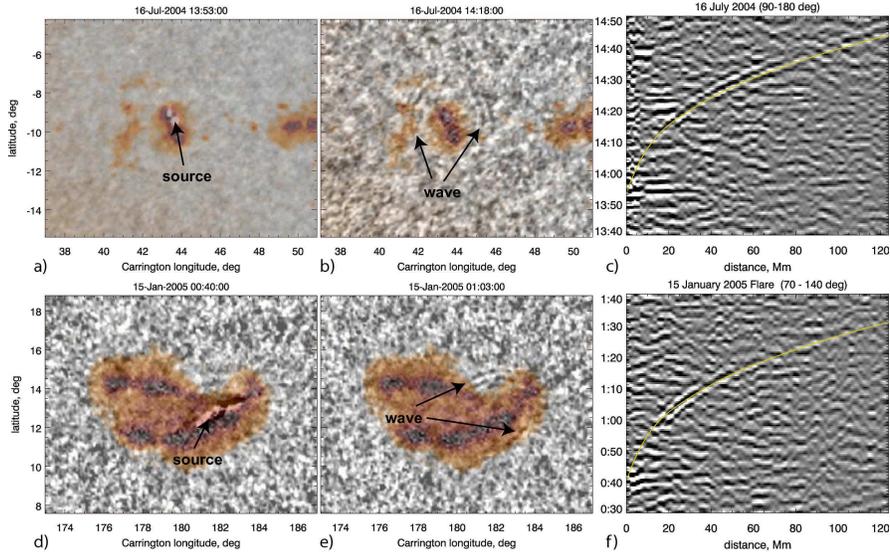}
  \caption{Observations of the seismic response of the Sun (``sunquakes'')
to two solar flares: a-c)  X3 of 16 July, 2004, and d-f) X1 flare of
15 January, 2005. The left panels show a superposition of MDI
white-light images of the active regions and locations of the
sources of the seismic waves determined from MDI Dopplergrams,  the
middle column shows the seismic waves, and the right panels show the
time-distance diagrams of these events. The thin yellow curves in
the right panels represent a theoretical time-distance relation for
helioseismic waves for a standard solar
model.\cite{Kosovichev2006c}}\label{fig14}
\end{center}
\end{figure}

A characteristic feature of the seismic response in this flare and
several others \cite{Kosovichev2006a,Kosovichev2006b,Kosovichev2006c} is anisotropy of the wave front:
the observed wave amplitude is much stronger in one direction than
in the others. In particular, the seismic waves excited during the
October 28, 2003, 16 July, 2004,  flare of 15 January, 2005 flare
had the greatest amplitude in the direction of the expanding flare
ribbons (Fig.~\ref{fig14}). The wave anisotropy can be attributed to
the moving source of the hydrodynamic impact, which is located in
the flare ribbons \cite{Kosovichev2006a,Kosovichev2006c,Kosovichev2007}. The motion of flare ribbons is
often interpreted as a result of the magnetic reconnection processes
in the corona. When the reconnection region moves up it involves
higher magnetic loops, the footpoints of which are further apart. The motion of the footpoints of impact of the high-energy particles is particularly well observed in the SOHO/MDI magnetograms showing magnetic transients moving with supersonic speed, in some cases \cite{Kosovichev2006b}. Of
course, there might be other reasons for the anisotropy of the wave
front, such as inhomogeneities in temperature, magnetic field, and
plasma flows. However, the source motion seems to be a key factor.

Therefore, we conclude that the seismic wave was generated not by a
single impulse  but by a series of impulses, which produce the
hydrodynamic source moving on the solar surface with a supersonic
speed. The seismic effect of the moving source can be easily
calculated by convolving the wave Green's function with a moving
source function. The results of these calculations a strong
anisotropic wavefront, qualitatively similar to the observations
\cite{Kosovichev2007}. Curiously, this effect is quite similar to
the anisotropy of seismic waves on Earth, when the earthquake
rupture moves along the fault. Thus, taking into account the effects
of multiple impulses of accelerated electrons and moving source is
very important for sunquake theories. The impulsive sunquake
oscillations provide unique information about interaction of
acoustic waves with sunspots. Thus, these effects must be studied in
more detail.
\section{Global helioseismology}
\label{global}

\subsection{Basic equations}

A simple theoretical model of solar oscillations can be derived
using the following assumptions:
\begin{enumerate}
\item
linearity: ${\vec v}/c << 1$, where ${\vec v}$ is velocity of
oscillating elements, $c$ is the speed of sound;
\item
adiabaticity: $dS/dt=0$, where $S$ is the specific entropy;
\item
spherical symmetry of the background state;
\item
magnetic forces and Reynolds stresses are negligible.
\end{enumerate}
The basic governing equations are derived from the conservation of
mass, momentum, energy and the Newton's gravity law. The
conservation of mass (continuity equation) assumes that the rate of
mass change in a fluid element of volume $V$ is equal to the mass
flux through the  surface of this element (of area $A$):
\begin{equation}
\frac{\partial}{\partial t}\int_V\rho dV=-\int_A \rho{\vec v}d{\vec
a}= -\int_V\nabla(\rho{\vec v})dV,
\end{equation}
where $\rho$ is the mass density. Then,
\begin{equation}
\frac{\d\rho}{\d t}+\nabla(\rho\vec v)=0,
\end{equation}
or in terms of the material derivative
$d\rho/dt=\partial\rho/\partial t+\vec v\cdot\nabla\rho$:
\begin{equation}
\frac{d\rho}{dt}+\rho\nabla\vec v=0.
\end{equation}
The momentum equation (conservation of momentum of a fluid element)
is:
\begin{equation}
\rho\frac{d\vec v}{dt}=-\nabla P +\rho \vec g,
\end{equation}
where $P$ is pressure, $\vec g$ is the gravity acceleration, which
can be expressed in terms of gravitational potential $\Phi$: $\vec
g=\nabla\Phi$, $d\vec v/dt=\d\vec v/\d t+\vec v\cdot\nabla\vec v$ is
the material derivative for the velocity vector. The adiabaticity
equation (conservation of energy) for a fluid element is:
\begin{equation}
\frac{dS}{dt}=\frac{d}{dt}\left(\frac{P}{\rho^\gamma}\right)=0,
\end{equation}
or
\begin{equation}\frac{dP}{dt}=c^2\frac{d\rho}{dt},\end{equation}
where $c^2=\gamma P/\rho$ is the squared adiabatic sound speed. The
gravitational potential is calculated from the Poisson equation:
\begin{equation}\nabla^2\Phi=4\pi G\rho.\end{equation}


Now, we consider small perturbations of a stationary spherically
symmetrical star in hydrostatic equilibrium:
\[v_0=0, \; \rho=\rho_0(r),\; P=P_0(r).\]
If $\vec\xi(t)$ is a vector of displacement of a fluid element then
velocity $\vec v$ of this element:
\begin{equation}
\vec v = \frac{d\xi}{dt} \approx \frac{\d\vec\xi}{\d
t}.\end{equation} Perturbations of scalar variables, $\rho, P, \Phi$
can be of two general types: Eulerian (denoted with prime symbol),
at a fixed position $\vec r$:
\[\rho(\vec r,t)=\rho_0(r)+\rho'(\vec r,t),\]
and Lagrangian, measured in the moving element (denoted with $\delta$):
\begin{equation}
\delta\rho(\vec r+\vec\xi)=\rho_0(r)+\delta\rho(\vec r,t).
\end{equation}

The Eulerian and Lagrangian perturbations are related to each other:
\begin{equation}
\delta\rho=\rho'+(\vec\xi\cdot\nabla\rho_0)=\rho'+(\vec\xi\cdot\vec
e_r)\frac{d\rho_0}{dr}= \rho'+\xi_r\frac{d\rho_0}{dr},
\label{eq-lagr}
\end{equation}
where $\vec e_r$ is the radial unit vector.

In terms of the Eulerian perturbations and the displacement vector,
$\vec\xi$ the linearized mass, momentum and energy equations can be
expressed in the following form:
\begin{eqnarray}
\rho'  + \nabla(\rho_0\vec\xi)=0,\\
\rho_0\frac{\d\vec v}{\d t}  =  -\nabla P'-g_0\vec e_r\rho'+\rho_0\nabla\Phi',\\
P'  +  \xi_r\frac{dP_0}{dr}=c_0^2(\rho'+\xi_r\frac{d\rho_0}{dr}),\\
\nabla^2\Phi'  =  4\pi G\rho'.
\end{eqnarray}

The equation of solar oscillations can be further simplified by
neglecting the perturbations of the gravitational potential, which
give relatively small corrections to theoretical oscillation
frequencies. This is so-called  Cowling approximation: $\Phi'=0.$

Now, we consider the linearized equations in the spherical
coordinate system, $r,\theta,\phi$. In this system, the displacement
vector has the following form:
\begin{equation}
\vec\xi=\xi_r\vec e_r+\xi_\theta\vec e_\theta+\xi_\phi\vec
e_\phi\equiv \xi_r\vec e_r+\vec\xi_h,\label{eq-xi}
\end{equation}
where $\vec\xi_h=\xi_\theta\vec e_\theta+\xi_\phi\vec e_\phi$ is the
horizontal component of displacement. Also, we use the equation for
divergence of the displacement (called dilatation):
\begin{align}
\nabla\vec\xi\equiv{\rm div}\xi= & \frac{1}{r^2}\frac{\d}{\d
r}(r^2\xi_r)+
\frac{1}{r\sin\theta}\frac{\d}{\d\theta}(\sin\theta\xi_\theta)+
\frac{1}{r\sin\theta}\frac{\d\xi_\phi}{\d\phi}\nonumber =\\ &
=\frac{1}{r^2}\frac{\d}{\d
r}(r^2\xi_r)+\frac{1}{r}\nabla_h\vec\xi_h.
\end{align}
We consider periodic perturbations with frequency $\omega$:
$\vec\xi \propto \exp({i\omega t}), ... $. Here, $\omega$ is the angular frequency measured in rad/sec; it relates to the cyclic frequency, $\nu$, which measures the number of oscillation cycles per sec, as: $\omega=2\pi\nu$.

Then, in the Cowling
approximation, we obtained the following system of the linearized
equations (omitting subscript $0$ for unperturbed variables):
\begin{eqnarray}
\rho'+\frac{1}{r^2}\frac{\d}{\d r}(r^2\rho\xi_r)+\frac{\rho}{r}\nabla_h\vec\xi_h=0,\label{eq-osc1}\\
-\omega^2\rho\xi_r =-\frac{\d P'}{\d  r}+g\rho',\label{eq-osc2}\\
-\omega^2\rho\vec\xi_h=-\frac{1}{r}\nabla_hP',\label{eq-osc3}\\
\rho'=\frac{1}{c^2}P'+\frac{\rho N^2}{g}\xi_r,\label{eq-osc4}
\end{eqnarray}
where
\begin{equation}
N^2=g\left(\frac{1}{\gamma
P}\frac{dP}{dr}-\frac{1}{\rho}\frac{d\rho}{dr}\right)\label{eq-brunt}
\end{equation}
 is {\it the Br\"unt-V\"ais\"al\"a  (or buoyancy) frequency}.

For the boundary conditions, we assume that the solution is regular
at the Sun's center. This correspond to the zero displacement,
$\xi_r=0$ at $r=0$, for all oscillation modes except of the dipole modes of
angular degree $l=1$. In the dipole-mode oscillations the center of a star oscillates (but not the center of mass), and the boundary condition at the center is replaced by a regularity condition. At the surface, we assume that the Lagrangian
pressure perturbation is zero: $\delta P=0$ at $r=R$. This is equivalent
to the absence of external forces. Also, we assume that the solution
is regular at the poles $\theta=0, \pi$.

We seek a solution of Eqs~(\ref{eq-osc1}-\ref{eq-osc4}) by separation of the radial and angular
variables in the following form:
\begin{align}
\rho'(r,\theta,\phi)=\rho'(r)\cdot f(\theta,\phi),\label{eq-osc1a}\\
P'(r,\theta,\phi)=P'(r)\cdot f(\theta,\phi),\label{eq-osc2a}\\
\xi_r(r,\theta,\phi)=\xi_r(r)\cdot f(\theta,\phi),\label{eq-osc3a}\\
\vec\xi_h(r,\theta,\phi)=\xi_h(r)\nabla_hf(\theta,\phi).\label{eq-osc4a}
\end{align}

Then, in the continuity equation:
\begin{equation}
\left[\rho'+\frac{1}{r^2}\frac{\d}{\d r}(r^2\rho\xi_r)\right]
f(\theta,\phi)+ \frac{\rho}{r}\xi_h\nabla_h^2f=0.
\end{equation}
the radial and angular variables can be separated if
\begin{equation}
\nabla_h^2f=\alpha f,
\label{eq-osc1b}
\end{equation}
where $\alpha$ is a constant.

It is well-known that this equation has a non-zero solution regular at the poles
($\theta=0, \pi$) only when
\begin{equation}
\alpha=-l(l+1),
\end{equation}
where $l$ is an integer. This non-zero solution is:
\begin{equation}
f(\theta,\phi)=Y_l^m(\theta,\phi) \propto P_l^m(\theta)e^{im\phi},\label{eq-osc5a}
\end{equation}
where $P_l^m(\theta)$ is the associated Legendre function of angular
degree $l$ and order $m$.

Then, the continuity equation for the radial dependence of the Eulerian density
perturbation, $\rho'(r)$, takes the form:
\begin{equation}
\rho'+\frac{1}{r^2}\frac{\d}{\d
r}\left(r^2\rho\xi_r\right)-\frac{l(l+1)}{r^2}\rho\xi_h=0. \label{eq-osc5}
\end{equation}
The horizontal component of displacement $\xi_h$ can be determined
from the horizontal component of the momentum equation:
\begin{equation}
-\omega^2\rho\xi_h(r)=-\frac{1}{r}P'(r),
\end{equation}
or
\begin{equation}
\xi_h=\frac{1}{\omega^2\rho r}P'.
\end{equation}
Substituting this into the continuity equation (\ref{eq-osc5}) we get:
\begin{equation}
\rho\frac{d\xi_r}{dr}+\xi_h\frac{d\rho}{dr}+\frac{2}{r}\rho\xi_r+\frac{P'}{c^2}+
\frac{\rho N^2}{g}\xi_r-\frac{L^2}{r^2\omega^2\rho}P'=0,
\end{equation}
where we define $L^2=l(l+1)$.

Using the hydrostatic equation for the background (unperturbed)
state, ${dP}/{dr}=-g\rho,$ we finally obtain:
\begin{equation}
\frac{d\xi_r}{dr}+\frac{2}{r}\xi_r-\frac{g}{c^2}\xi_r+\left(1-\frac{L^2c^2}{r^2\omega^2}\right)
\frac{P'}{\rho c^2}=0,
\end{equation}
or
\begin{equation}
\frac{d\xi_r}{dr}+\frac{2}{r}\xi_r-\frac{g}{c^2}\xi_r+\left(1-\frac{S_l^2}{\omega^2}\right)
\frac{P'}{\rho c^2}=0,\label{eq-osc6}
\end{equation}
where
\begin{equation}
S_l^2=\frac{L^2c^2}{r^2}\label{eq-lamb}
\end{equation}
is the {\it Lamb frequency}.

Similarly, for the momentum equation we obtain:
\begin{equation}
\frac{dP'}{dr}+\frac{g}{c^2}P'+(N^2-\omega^2)\rho\xi_r=0.\label{eq-osc7}
\end{equation}

The inner boundary condition at the Sun's center is:
\begin{equation}
\xi_r=0,\label{eq-osc8}
\end{equation}
or a regularity condition for $l=1$.

The outer boundary condition at the surface ($r=R$) is:
\begin{equation}
\delta P=P'+\frac{dP}{dr}\xi_r=0.
\end{equation}
Applying the hydrostatic equation, we get:
\begin{equation}
P'-g\rho\xi_r=0.\label{eq-osc9}
\end{equation}

Using the horizontal component of the momentum equation:
$P'=\omega^2\rho r\xi_h$, the outer boundary condition (\ref{eq-osc9}) can be
written in the following form:
\begin{equation}
\frac{\xi_h}{\xi_r}=\frac{g}{\omega^2r},\label{eq-ocs9}
\end{equation}
that is the ratio of the horizontal and radial components of
displacement is inverse proportional to the squared oscillation
frequency. However, observations show that this relation is only
approximate, presumably, because of the external force caused by the
solar atmosphere.
\begin{figure}
\begin{center}
\includegraphics[width=0.8\linewidth]{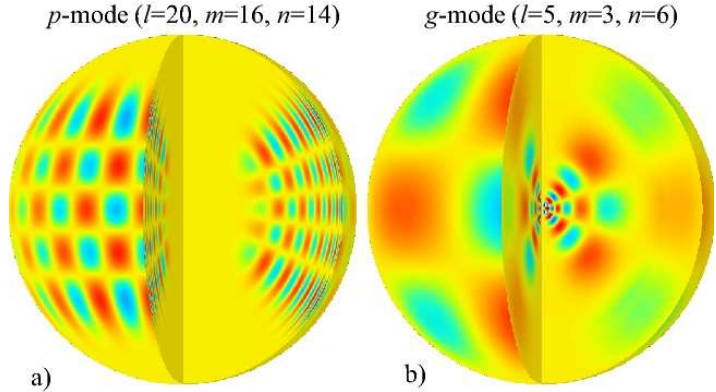}
\caption{Eigenfunctions (\ref{eq-osc9a}) of two normal oscillation modes of the Sun: a) p-mode of angular degree $l=20$, angular degree $m=16$, and radial order $n=16$, b) g-mode of $l=5$, $m=3$, and $n=5$. Red and blue-green colors correspond to positive and negative values.}
\label{fig15}\end{center}
\end{figure}

Equations (\ref{eq-osc6}) and (\ref{eq-osc7}) with boundary
conditions (\ref{eq-osc8})-(\ref{eq-osc9}) constitute an eigenvalue
problem for solar oscillation modes. This eigenvalue problem can be
solved numerically for any solar or stellar model. The solution
gives the frequencies, $\omega_{nl}$, and the radial eigenfunctions, $\xi_r^{(n,l)}(r)$ and ${P'}^{(n,l)}(r)$, of the normal modes.

The radial eigenfunctions multiplied by the angular eigenfunctions
(\ref{eq-osc1a})-(\ref{eq-osc4a}) represented by the spherical
harmonics (\ref{eq-osc5a}) give three-dimensional oscillation
eigenfunctions of the normal modes, e.g.:
\begin{equation}
\xi_r(r,\theta,\phi,\omega)=\xi_r^{(n,l)}(r)Y_l^m(\theta,\phi).\label{eq-osc9a}
\end{equation}
 Examples of such two
eigenfunctions for p- and g-modes are shown  in Fig.~\ref{fig15}. It
illustrates the typical behavior of the modes: the p-modes are
concentrated (have the strongest amplitude) in the outer layers of
the Sun, and g-modes are mostly confined in the central region.
\subsection{JWKB solution}
The basic properties of the oscillation modes can be investigated analytically using an asymptotic  approximation. In this approximation,
we assume that only density $\rho(r)$ varies significantly among the
solar properties in the oscillation equations, and seek for an
oscillatory solution in the JWKB form:
\begin{eqnarray}
\xi_r = A\rho^{-1/2}e^{ik_rr},\\
P'=B\rho^{1/2}e^{ik_rr},
\end{eqnarray}
where the {\it radial wavenumber} $k_r$ is a slowly varying function of
$r$; $A$ and $B$ are constants.

Then, substituting these in Eqs (\ref{eq-osc6}) and (\ref{eq-osc7}) we obtain:
\begin{eqnarray}
\frac{d\xi_r}{dr}=-A\rho^{-1/2}\left(-ik_r+\frac{1}{H}\right)e^{ik_rr},\label{eq-osc10}\\
\frac{dP'}{dr}=-B\rho^{1/2}\left(-ik_r-\frac{1}{H}\right)e^{ik_rr}\label{eq-osc11},
\end{eqnarray}
where
\begin{equation}
H=\left(\frac{d\log\rho}{dr}\right)^{-1},
\end{equation}
is the {\it density scale height}.

From (\ref{eq-osc10}-\ref{eq-osc11}) we get a linear system for the constant, $A$, and $B$:
\begin{equation}
\left(-ik_r+\frac{1}{H}\right)A-\frac{g}{c^2}A+
\frac{1}{c^2}\left(1-\frac{S_l^2}{\omega^2}\right)B=0,
\end{equation}
\begin{equation}
\left(-ik_r-\frac{1}{H}\right)B+\frac{g}{c^2}B+ (N^2-\omega^2)A=0.
\end{equation}
It has a non-zero solution when the determinant is equal zero, that is when
\begin{equation}
k_r^2=\frac{\omega^2-\omega_c^2}{c^2}+\frac{S_l^2}{c^2\omega^2}\left(N^2-\omega^2\right),\label{eq-osc12}
\end{equation}
where
\begin{equation}
\omega_c=\frac{c}{2H}
\label{eq-cutoff}
\end{equation}
is {\it the acoustic cut-off frequency}. Here, we used the relation:
$N^2=g/H-g^2/c^2$.

\begin{figure}
\begin{center}
\includegraphics[width=0.7\linewidth]{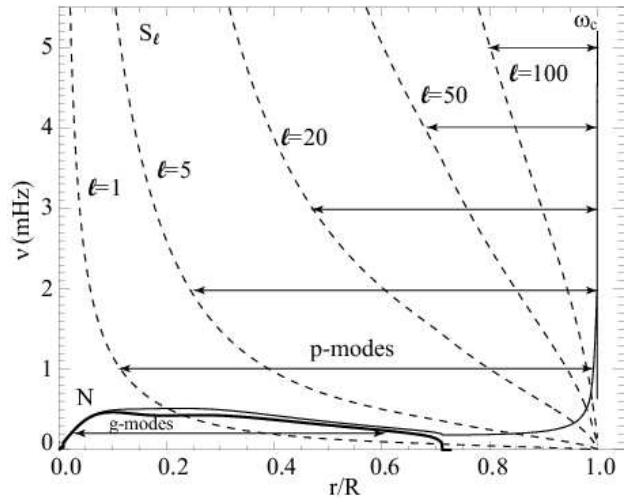}
\caption{ Buoyancy (Br\"unt-V\"ais\"al\"a) frequency $N$ (thick
curve), acoustic cut-off frequency, $\omega_c$ (thin curve) and Lamb
frequency $S_l$ for $l$=1, 5, 20, 50, and 100 (dashed curves) vs.
fractional radius $r/R$ for a standard solar model. The horizontal
lines with arrows indicate the trapping regions for a g mode with
frequency $\nu=0.2$ mHz, and for a sample of five p modes: $l=1$,
$\nu=1$ mHz; $l=5$, $\nu=2$ mHz; $l=20$, $\nu=3$ mHz; $l=50$,
$\nu=4$ mHz; $l=100$, $\nu=5$ mHz.}
\label{fig16}\end{center}
\end{figure}

The frequencies of solar modes depend on the sound speed, $c$, and three characteristic frequencies: acoustic cut-off frequency, $\omega_c$ (\ref{eq-cutoff}), Lamb frequency, $S_l$ (\ref{eq-lamb}), and Br\"unt-V\"ais\"al\"a frequency, $N$ (\ref{eq-brunt}).
These frequencies calculated for a standard solar model are shown in Fig.~\ref{fig16}. The acoustic cut-off and Br\"unt-V\"ais\"al\"a frequencies depend only on the solar structure, but the Lamb frequency depends also on the mode angular degree, $l$. This diagram is very useful for determining the regions of mode propagation.
The waves propagate in the regions where the radial wavenumber is real, that $k_r^2 > 0$. If $k_r^2 < 0$ then the waves exponentially decay with distance (become `evanescent').
The characteristic frequencies define the boundaries of the propagation regions, also called the wave {\it turning points}. The region of propagation for p- and g-modes are indicated in Fig.~\ref{fig16}, and are discussed in the following sections.

We define a horizontal wavenumber as
\begin{equation}
k_h\equiv \frac{L}{r},
\end{equation}
where $L=\sqrt{l(l+1)}$. This definition follows from  the angular
part of the wave equation (\ref{eq-osc1b}):
\begin{equation}
\frac{1}{r^2}\nabla_h^2Y_l^m+\frac{l(l+1)}{r^2}Y_l^m=0,
\end{equation}
where $\nabla_h$ is the horizontal component of gradient.
It can be rewritten in terms of a horizontal wavenumber, $k_h$,
$\frac{1}{r^2}\nabla_h^2Y_l^m+k_h^2Y_l^m=0$ if
$k_h^2=l(l+1)/r^2$.

In term of $k_h$ the Lamb frequency is $S_l=k_hc$, and
Eq.~\ref{eq-osc12} takes the form:
\begin{equation}
k_r^2=\frac{\omega^2-\omega_c^2}{c^2}+k_h^2\left(\frac{N^2}{\omega^2}-1\right),\label{eq-osc13}
\end{equation}

The frequencies of normal modes are determined for the Borh
quantization rule (resonant condition):
\begin{equation}
\int_{r_1}^{r_2} k_rdr=\pi(n+\alpha),
\label{eq-osc14}
\end{equation}
where $r_1$ and $r_2$ are the radii of the inner and outer turning
points where $k_r$=0, $n$ is a radial order -integer number, and
$\alpha$ is a phase shift which depends on properties of the
reflecting boundaries.

\subsection{Dispersion relations for p- and g-modes}

For high-frequency oscillations, when $\omega^2>>N^2$, the
dispersion relation (\ref{eq-osc12})-(\ref{eq-osc13}) can be written as:
\begin{equation}
k_r^2=\frac{\omega^2-\omega_c^2}{c^2}-\frac{S_l^2}{c^2}=\frac{\omega^2-\omega_c^2}{c^2}-k_h^2.
\end{equation}
Then, we obtain:
\begin{equation}
\omega^2=\omega_c^2+(k_r^2+k_h^2)c^2\equiv\omega_c^2+k^2c^2.\label{eq-osc15}
\end{equation}
This is a dispersion relation for acoustic (p) modes, $\omega_c$ is
the acoustic cut-off frequency. The wave with frequencies less than
$\omega_c$ (or wavelength $\lambda > 4\pi H$) do not propagate.
These waves exponentially decay, and called `evanescent'.

For low-frequency perturbations, when $\omega^2<<S_l^2$, one gets:
\begin{equation}
k_r^2=\frac{S_l^2}{c^2\omega^2}(N^2-\omega^2)=\frac{k_h^2}{\omega^2}(N^2-\omega^2),\label{eq-osc16}
\end{equation}
and
\begin{equation}
\omega^2=\frac{k_h^2N^2}{k_r^2}\equiv N^2\cos^2\theta,
\end{equation}
where $\theta$ is the angle between the wavevector, $k$, and
horizontal surface.

These waves are called internal gravity waves or g-modes. They
propagate mostly horizontally, and only if $\omega^2< N^2$. The
frequency of the internal gravity waves does not depend on the
wavenumber, but on the direction of propagation. These waves are
evanescent if $\omega^2> N^2$.


\subsection{Frequencies of p- and g-modes}

Now, we use the Borh quantization rule (\ref{eq-osc14}) and the
dispersion relations for the p- and g-modes (\ref{eq-osc15}-\ref{eq-osc16})
to derive the mode frequencies.

\paragraph{p-modes:} The modes propagate in the region where $k_r^2>0$;
and the radii of the turning points, $r_1$ and $r_2$,
are determined from the relation $k_r^2=0$:
\begin{equation}
\omega^2=\omega_c^2+\frac{L^2c^2}{r^2}=0.
\end{equation}
The acoustic cut-off is only significant near the Sun's surface.
The lower turning point is located in the interior where
$\omega_c<<\omega$ (Fig.~\ref{fig16}. Then, at the lower turning point, $r=r_1$: $\omega\approx
{Lc}/{r}$, or
\begin{equation}
\frac{c(r_1)}{r_1}=\frac{\omega}{L}
\end{equation}
represents the equation for the radius of the lower turning point,
$r_1$. The upper turning point is determined by the acoustic frequency term:
$\omega_c(r_2)\approx \omega$. Since
$\omega_c(r)$ is a steep function of $r$ near the surface, then
\begin{equation}
r_2\approx R.
\end{equation}
The p-mode propagation region is illustrated in Fig.~\ref{fig16}
Thus, the resonant condition for the p-modes is:
\begin{equation}
\int_{r_1}^R \sqrt{\frac{\omega^2}{c^2}-\frac{L^2}{r^2}}dr =
\pi(n+\alpha)
\label{eq-osc17}
\end{equation}
In the case, of the low-degree ``global'' modes, for which $l<<n$,
the lower turning point is almost at the center, $r_1\approx 0$, and
we obtain \cite{Vandakurov1968}:
\begin{equation}
\omega\approx \frac{\pi(n+L/2+\alpha)}{\int_0^R{dr/c}}.
\label{eq-osc18}
\end{equation}
This relation shows  is the spectrum of low-degree p-modes is
approximately equidistant with the frequency spacing:
\begin{equation}
\Delta\nu=\left(4\int_0^R\frac{dr}{c}\right)^{-1}.
\label{eq-osc19}
\end{equation}
\begin{figure}
\begin{center}
\includegraphics[width=0.8\linewidth]{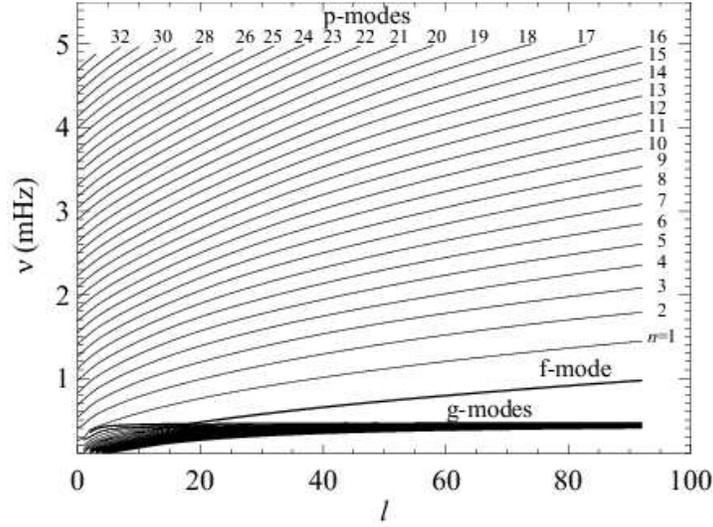}
\caption{Spectrum of normal modes calculated for a standard solar
model. The thick gray curve shows $f$-mode. Labels $p_1$-$p_{33}$
mark p-modes of the radial order $n=1, \ldots, 33$.
}
\label{fig17}\end{center}
\end{figure}

This corresponds very well to the observational power spectrum shown
in Fig.~\ref{fig4}. According to this relation, the frequencies of
mode pairs, $(n,l)$ and $(n-1,l+2)$, coincide. However, calculations
to the second-order shows that the frequencies in these pairs are
separated by the amount \cite{Tassoul1980,Gough1993}:
\begin{equation}
\delta\nu_{nl}=\nu_{nl}-\nu_{n-1,l+2}\approx
-(4l+6)\frac{\Delta\nu}{4\pi^2\nu_{nl}}\int_0^R\frac{dc}{dr}\frac{dr}{r}.
\label{eq-osc19a}
\end{equation}
This is so-called ``small separation". For the Sun, $\Delta\nu\approx
136 \mu$Hz, and $\delta\nu\approx 9\mu$Hz. The $l-\nu$ for the p-modes is illustrated in Fig.~\ref{fig17}.

\begin{figure}
\begin{center}
\includegraphics[width=0.7\linewidth]{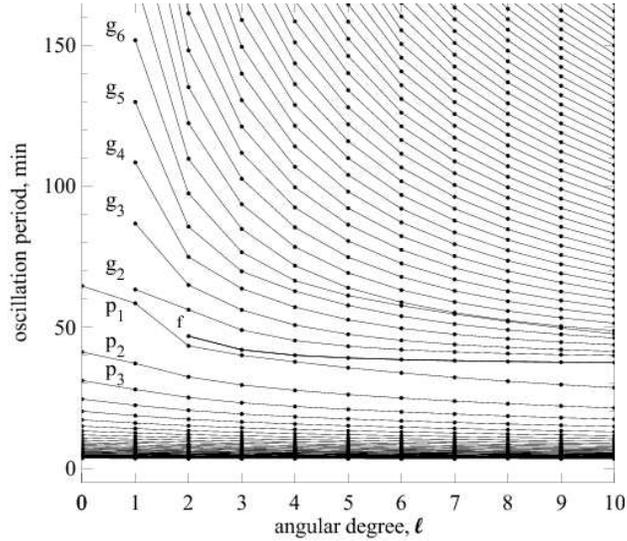}
\caption{Periods of solar oscillation modes in the angular degree range, $l=0-10$. Labels $g_1$-$g_{6}$
mark g-modes of the radial order $n=1, \ldots, 6$}
\end{center}\label{fig18}
\end{figure}

\paragraph{g-modes:} The turning points, $k^r=0$, are determined from equation (\ref{eq-osc16}):
\begin{equation}
N(r)=\omega.
\end{equation}
In the propagation region, $k_r>0$, (see Fig.~\ref{fig16}), far from the turning points
($N>>\omega$):
\begin{equation}
k_r \approx \frac{LN}{r\omega}.
\end{equation}
Then, from the resonant condition:
\begin{equation}
\int_{r_1}^{r_2}\frac{L}{\omega}N\frac{dr}{r}=\pi(n+\alpha).
\end{equation}
we find an asymptotic formula for the g-mode frequencies:
\begin{equation}
\omega\approx \frac{L\int_{r_1}^{r_2}N\frac{dr}{r}}{\pi(n+\alpha)}.
\label{eq-osc20}
\end{equation}
It follows that for a given $l$ value the oscillation periods form a regular equally spaced pattern:
\begin{equation}
P=\frac{2\pi}{\omega}=\frac{\pi(n+\alpha)}{L\int_{r_1}^{r_2}N\frac{dr}{r}}.
\label{eq-osc21}
\end{equation}
The distribution of numerically calculated g-mode periods is shown in Fig.~\ref{fig18}.

\subsection{Asymptotic ray-path approximation}
The asymptotic approximation provides an important representation of solar oscillations in terms of the ray theory.
Consider the wave path equation in the ray approximation:
\begin{equation}
\frac{\partial\vec r}{\partial t}=\frac{\partial\omega}{\partial\vec
k}.
\label{eq-osc22}
\end{equation}
Then, the radial and angular components of this equation are:
\begin{eqnarray}
\frac{dr}{dt}=\frac{\partial\omega}{\partial k_r},\label{eq-osc23}\\
r\frac{d\theta}{dt}=\frac{\partial\omega}{\partial k_h}.\label{eq-osc24}
\end{eqnarray}
Using the dispersion relation for acoustic (p) modes:
\begin{equation}
\omega^2=c^2(k_r^2+k_h^2),
\end{equation}
in which we neglected the $\omega_c$ term. (It can be neglected everywhere except near
the upper turning point, $R$), we get
\begin{equation}
dt=\frac{dr}{c\left(1-{k_h^2c^2}/{\omega^2}\right)^{1/2}}.
\end{equation}
Thus,
is the travel time from the lower turning point to the surface.

\begin{figure}
\begin{center}
\includegraphics[width=1.0\linewidth]{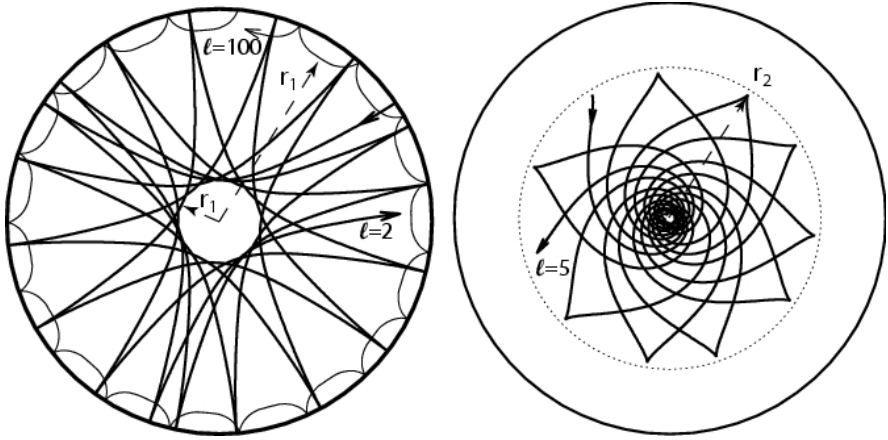}
\caption{Ray paths for a) two solar p-modes of angular degree $l=2$,
frequency $\nu=1429.4\;\mu$Hz (thick curve), and $l=100$,
$\nu=3357.5\;\mu$Hz (thin curve); b) g-mode of $l=5$,
$\nu=192.6\;\mu$Hz (the dotted curve indicates the base of the
convection zone). The lower turning points, $r_1$ of the p-modes are
shown by arrows. The upper turning points of these modes are close
to the surface and not shown. For the g-mode, the upper turning
point, $r_2$, is shown by arrow. The inner turning point is close to
the center and not shown.}
\label{fig19}\end{center}
\end{figure}

The equation for the acoustic ray path is given by the ratio of equations (\ref{eq-osc24}) and (\ref{eq-osc23}):
\begin{equation}
r\frac{d\theta}{dr}=\left(\frac{\partial\omega}{\partial
k_h}\right)/ \left(\frac{\partial\omega}{\partial
k_r}\right)=\frac{k_h}{k_r},
\end{equation}
or
\begin{equation}
r\frac{d\theta}{dr}=\frac{k_h}{k_r}=\frac{L/r}{\sqrt{{\omega^2}/{c^2}-{L^2}/{r^2}}}.
\label{eq-osc25}
\end{equation}
For any given values of $\omega$ and $l$, and initial coordinates, $r$ and $\theta$,  this equation gives trajectories of ray paths of p-modes
inside the Sun. The ray paths calculated for two solar p-modes are shown
in Fig.~\ref{fig19}a. They illustrate an important property that the acoustic waves excited by a source near the solar surface travel into the interior and come back to surface. The distance, $\Delta$, between the surface points for one skip can be calculated as the integral:
\begin{equation}
\Delta=2\int_{r_1}^R d\theta =2\int_{r_1}^R \frac{L/r}{\sqrt{\omega^2/c^2-L^2/r^2}}dr\equiv 2\int_{r_1}^R \frac{c/r}{\sqrt{\omega^2/L^2-c^2/r^2}}dr.
\label{eq-dist}
\end{equation}
The corresponding travel time is calculated by integrating equation (\ref{eq-osc23}):
\begin{equation}
\tau = 2 \int_{r_1}^R dt = \int_{r_1}^R \frac{dr}{c\left(1-{k_h^2c^2}/{\omega^2}\right)^{1/2}}\equiv \int_{r_1}^R \frac{dr}{c\left(1-{L^2c^2}/{r^2\omega^2}\right)^{1/2}}.
\label{eq-time}
\end{equation}
These equations give a {\it time-distance} relation, $\tau-\Delta$, for acoustic waves traveling between two surface points through the solar interior. The ray representation of the solar modes and the time-distance relation provided a motivation for developing {\it time-distance helioseismology} (Sec.~\ref{tomography}), a local helioseismology method \cite{Duvall1993}.

The ray paths for g-modes are calculated similarly.
For the g-modes, the dispersion relation is:
\begin{equation}
\omega^2=\frac{k_h^2N^2}{k_r^2+k_h^2}.
\end{equation}
Then, the corresponding ray path equation:
\begin{equation}
r\frac{d\theta}{dr}=-\frac{k_r}{k_h}=-\sqrt{\frac{N^2}{\omega^2}-1}.
\end{equation}

The solution for a g-mode of $l=5, \nu=192.6\mu$Hz is shown in
Fig.~\ref{fig19}b. Note that the g-mode travels mostly in the
central region. Therefore, the frequencies of g-modes are mostly
sensitive to the central conditions.

\subsection{Duvall's law}
\begin{figure}[h]
\begin{center}
\includegraphics[width=0.6\linewidth]{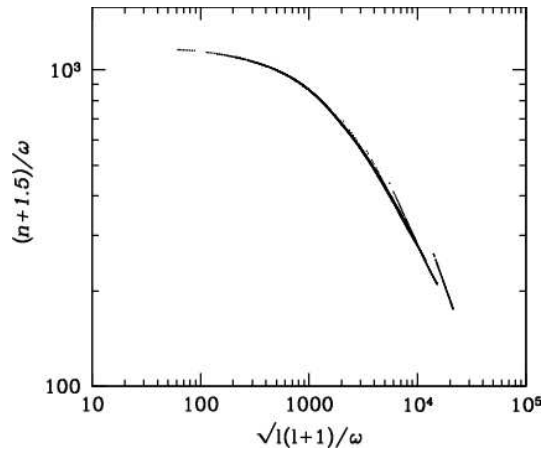}
\caption{The observed Duvall's law relation for modes of $l=0-250$.}
\label{fig20}\end{center}
\end{figure}

The solar p-modes, observed in the period range of 3--8 minutes, can be considered as high-frequency modes and described by the asymptotic theory quite accurately.
Consider the resonant condition (\ref{eq-osc17}) for p-modes:
\begin{equation}
\int_{r_1}^R
\left(\frac{\omega^2}{c^2}-\frac{L^2}{r^2}\right)^{1/2}dr=\pi(n+\alpha),
\end{equation}
Dividing both sides by $\omega$ we get:
\begin{equation}
\int_{r_1}^R
\left(\frac{r^2}{c^2}-\frac{L^2}{\omega^2}\right)^{1/2}\frac{dr}{r}=\frac{\pi(n+\alpha)}{\omega}.
\label{eq-osc26}
\end{equation}
Since the lower integral limit, $r_1$ depends only on the ratio
$L/\omega$, then the whole left-hand side is a function of only one parameter,
$L/\omega$, that is:
\begin{equation}
F\left(\frac{L}{\omega}\right)=\frac{\pi(n+\alpha)}{\omega}.
\end{equation}
This relation represents so-called Duvall's law \cite{Duvall1982}.
It means that a 2D dispersion relation $\omega=\omega(n,l)$ is
reduced to the 1D relation between two ratios $L/\omega$ and
$(n+\alpha)/\omega$. With an appropriate choice of parameter $\alpha$ (e.g. 1.5)
these ratios can easily calculated from a table of observed solar frequencies.
An example of such calculations shown in Fig.~\ref{fig20}) illustrates
that the Duvall's law holds quite well for the observed solar modes.
The short bottom branch that separates from the main curve correspond to f-modes.

\subsection{Asymptotic sound-speed inversion}

The Duvall's law demonstrate that the asymptotic theory provides a rather accurate description of the observed solar p-modes. Thus, it can be used for solving the inverse problem of helioseismology - determination of the internal properties from the observed frequencies.
Theoretically, the internal structure of the Sun is described by the stellar evolution theory
\cite{Christensen-Dalsgaard1996}. This theory calculates the thermodynamic structure of the Sun during the evolution on the Main Sequence. The evolutionary model of the current age, $\approx 4.6\times 10^9$ years, is called the standard solar model. Helioseismology provides estimates of the interior properties, such as the sound-speed profiles, that can be compared with the predictions of the standard model.

Our goal is to find corrections to a solar model from the observed
frequency differences between the Sun and the model using the
asymptotic formula for the Duvall's law \cite{Christensen-Dalsgaard1988}.

We consider a small perturbation of the sound-speed, $c\rightarrow
c+\Delta c$, and the corresponding perturbation of frequency:
$\omega\rightarrow \omega+\Delta\omega$.
Then, from equation (\ref{eq-osc26}) we obtain:
\begin{equation}
\int_{r_t}^R \left[\frac{(\omega+\Delta\omega)^2}{(c+\Delta
c)^2}-\frac{L^2}{r^2}\right]^{1/2}dr=\pi(n+\alpha).
\end{equation}
Expanding this in terms of $\Delta c/c$ and $\Delta\omega/\omega$
and keeping only the first-order terms we get:
\begin{equation}
\frac{\Delta\omega}{\omega}\int_{r_t}^R
\frac{dr}{c\left(1-{L^2c^2}/{r^2\omega^2}\right)^{1/2}}=\int_{r_t}^R
\frac{\Delta
c}{c}\frac{dr}{c\left(1-{L^2c^2}/{r^2\omega^2}\right)^{1/2}} .
\end{equation}

If we introduce a new variable:
\begin{equation}
T=\int_{r_t}^R
\frac{dr}{c\left(1-{L^2c^2}/{r^2\omega^2}\right)^{1/2}},
\end{equation}
then
\begin{equation}
\frac{\Delta\omega}{\omega}=\frac{1}{T}\int_{r_t}^R \frac{\Delta
c}{c}\frac{dr}{c\left(1-{L^2c^2}/{r^2\omega^2}\right)^{1/2}}.
\label{eq-osc27}
\end{equation}

This equation has a simple physical interpretation: $T$ is the
travel time of acoustic waves to travel along the acoustic ray path
between the lower and upper turning points (Fig.~\ref{fig19}). The
right-hand side integral is an average of the sound-speed
perturbations along this ray path (compare with Eq.(\ref{eq-time})).

Equation (\ref{eq-osc27}) can be reduced to {\it the Abel integral equation} by
making a substitution of variables. The new variables are:
\begin{eqnarray}
x=\frac{\omega^2}{L^2},\\
y=\frac{c^2}{r^2},
\end{eqnarray}
where $x$ is a measured quantity, and $y$ is associated with the
sound-speed distribution of an unperturbed solar model.

Then, we obtain an equation for $x$ and $y$:
\begin{equation}
F(x)=\int_0^x\frac{f(y)dy}{\sqrt{x-y}},
\end{equation}
where
\[F(x)=T\frac{\Delta\omega}{\omega}\frac{1}{\sqrt{x}},\]
\[f(y)=\frac{\Delta c}{c}\frac{1}{2y^{3/2}\left(\dfrac{d\log c}{d\log r}+1\right)}.\]

To solve for $f(y)$ we multiply both sides of Eq.(12) by
${dx}/{\sqrt{z-x}}$
and integrate with respect to $x$ from 0 to $z$:
\[
\int_0^z\frac{F(x)dx}{\sqrt{z-x}}=\int_0^z\frac{dx}{\sqrt{z-x}}
\int_0^x\frac{f(y)dy}{\sqrt{x-y}}=\]
\[
=\int_0^x f(y)dy\int_y^z\frac{dx}{\sqrt{(z-x)(x-y)}}.
\]
Here we changed the order of integration.

Note that
\[\int_y^z\frac{dx}{\sqrt{(z-x)(x-y)}}=\pi,\]
then
\[\int_0^z\frac{F(x)dx}{\sqrt{z-x}}=\pi\int_0^x f(y)dy.\]
Differentiating with respect to $x$, we obtain the final solution:
\begin{equation}
f(y)=\frac{1}{\pi}\frac{d}{dx}\int_0^z\frac{F(x)dx}{\sqrt{z-x}}.
\end{equation}
Then, from $f(y)$ we find the sound-speed correction $\Delta c/c$.

 This method based on linearization of the asymptotic Abel integral
 is called  "differential asymptotic sound-speed inversion"
 \cite{Christensen-Dalsgaard1988}. It provides estimates of the sound-speed
 deviations from a reference solar model.

 Alternatively, the
 sound-speed profile inside the Sun can be found from a implicit
 solution of the Abel obtained by differentiating the Duvall's law
 equation (\ref{eq-osc26}) with respect to variable $y=L/\omega$. Then, this
 equation can be solved analytically. The solution provides an
 implicit relationship between the solar radius and sound speed
 \cite{Gough1986}:

 \begin{equation} \ln(r/R)=\int_{r/c}^{R/c_s}
 \frac{dF}{dy}\left(y^2-\frac{r^2}{c^2}\right)^{-1/2}dy,
 \end{equation}

 where $c_s$ is the sound speed at the solar surface
 $r=R$. The calculation of  the derivative, $dF/dy$, is essentially
 differentiation of a smooth function approximating the Duvall's
 law, that is differentiating $\pi(n+\alpha)/\omega$ with respect to
 $L/\omega$. Both of these quantities are obtained from the observed
 frequency table, $\omega(n,l)$.

 The first inversion results using
 this approach was published by Christensen-Dalsgaard et al
 \cite{Christensen-Dalsgaard1988}. These technique can be
 generalized by including the Br\"unt-V\"ais\"al\"a frequency term
 in the p-mode dispersion relation, and also taking into account the
 frequency dependence of the phase shift, $\alpha$
 \cite{Kosovichev1988}. The results  show that this inversion
 procedure  provides a good agreement with the solar models, used
 for testing, except the central core, where the asymptotic and
 Cowling approximations become inaccurate.

\begin{figure}
\begin{center}
\includegraphics[width=1.0\linewidth]{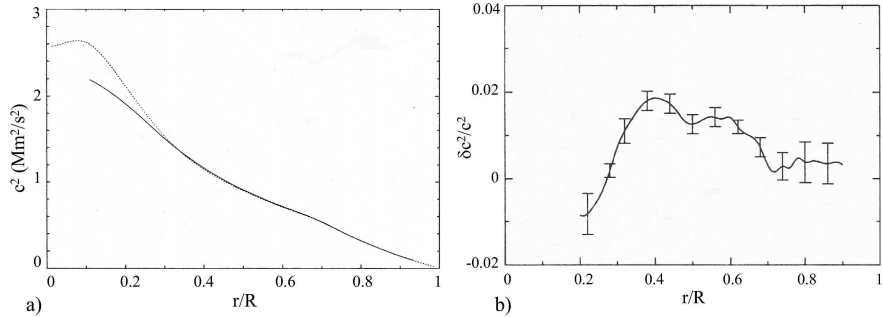}
\caption{a) Result of the asymptotic sound inversion (solid curve)
\cite{Kosovichev1988e} for the p-mode frequencies \cite{Duvall1988}.
It confirmed the standard solar model (model 1)
\cite{Christensen-Dalsgaard1982} (dots). The
large discrepancy in the central region is due inaccuracy of the
data and the asymptotic approximation. b) The relative difference
in the squared sound speed between the asymptotic inversions
of the observed and theoretical frequencies. }\label{fig21}
\end{center}
\end{figure}

 Figure~\ref{fig21} shows
 the inversion results \cite{Kosovichev1988e} for the p-mode frequencies measured by Duvall
 et al. \cite{Duvall1988}. The deviation of the sound speed from a
 standard solar model is about 1\%. Later, the agreement between the solar model and
 and the helioseismic inversions was
 improved by using more precise opacity tables and including element
 diffusion in the model calculations
 \cite{Christensen-Dalsgaard1996}. Also, a more accurate inversion method was developed by using a perturbation theory based on a variational principle for the normal mode frequencies (Sec.~\ref{inverse}).

\subsection{Surface gravity waves (f-mode)}

The surface gravity (f-mode) waves are similar in nature to the surface ocean waves.
 They are driven by the buoyancy force, and exist because of the sharp
 density decrease at the solar surface. These waves are missing in the JWKB solution.
These waves propagate at the surface boundary where Lagrangian
pressure perturbation $\delta P\sim 0$.

To investigate these waves we consider the oscillation equations in terms of $\delta P$ by making
use of the relation between Eulerian and Lagrangian variables (\ref{eq-lagr}):
\[P'=\delta P+g\rho\xi_r.\]
The oscillation equations (\ref{eq-osc6}) and (\ref{eq-osc7}) in terms of $\xi_r$ and $\delta P$ are:
\begin{equation}
\frac{d\xi_r}{dr}-\frac{L^2g}{\omega^2r^2}\xi_r+\left(1-\frac{L^2c^2}{\omega^2r^2}\right)\frac{\delta
P}{\rho c^2}=0,
\end{equation}
\begin{equation}
\frac{d\delta P}{dr}+\frac{L^2g}{\omega^2r^2}\delta P-\frac{g\rho
f}{r}\xi_r=0,
\end{equation}
where
\begin{equation}
f\approx \frac{\omega^2r}{g}-\frac{L^2g}{\omega^2 r}.
\end{equation}
These equations have a peculiar solution:
\[\delta P=0,\;\; f=0.\]
For this solution:
\begin{equation}
\omega^2=\frac{Lg}{R}=k_hg
\end{equation}
-dispersion relation for f-mode.

The eigenfunction equation:
\begin{equation}
\frac{d\xi_r}{dr}-\frac{L}{r}\xi_r=0
\end{equation}
has a solution
\begin{equation}
\xi_r \propto e^{k_h(r-R)}
\end{equation}
exponentially decaying with depth.

These waves are similar in nature to water waves which have the same
dispersion relation: $\omega=gk_h$. The f-mode waves are
incompressible: $\nabla\vec v=0.$ These waves are not sensitive to
the sound speed but are sensitive to the density gradient at the
solar surface. They are used for measurements of the `seismic
radius' of the Sun.

\subsection{The seismic radius}

The frequencies of f-modes:
\begin{equation}
\omega^2=gk_h\equiv\frac{GM}{R^2}\frac{L}{R}\equiv L\frac{GM}{R^3}.
\label{eq-osc28}
\end{equation}
If the frequencies are determined in observations for given $l$,
then we can define the `seismic radius', $R$, as
\begin{equation}
R=\left(\frac{LGM}{\omega^2}\right)^{1/3}.
\label{eq-osc29}
\end{equation}

\begin{figure}
\begin{center}
\includegraphics[width=0.85\linewidth]{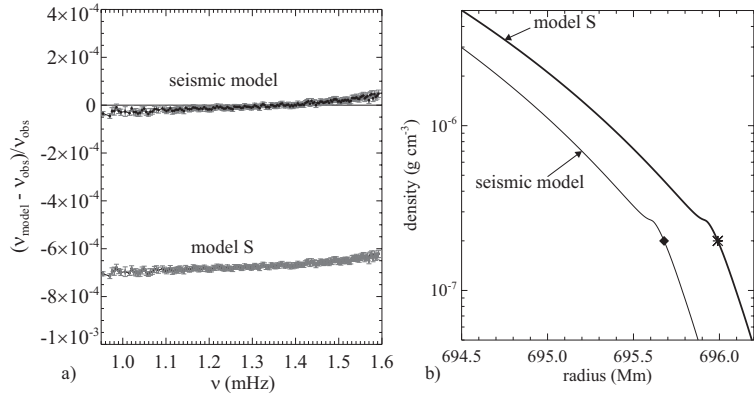}
\caption{a) Relative differences between the f-mode frequencies of
$l=88-250$ computed for a standard solar model (Model S) and the
observed frequencies. The `seismic model' frequencies are obtained
by scaling the frequencies of model S with factor 1.00066 which
corresponds to scaling down the model radius with $(1.00066)^{2/3}
\approx 1.00044$. The error bars are $3\sigma$ error estimates of
the observed frequencies. b) Density as a function of radius
near the surface for the standard and seismic models. The star
indicates the photospheric radius. The diamond shows the seismic
radius, $695.68$ Mm. }\label{fig22}
\end{center}
\end{figure}

The procedure of measuring the solar seismic radius is simple \cite{Schou1997}.
The lower curve in Figure~\ref{fig22}a shows the relative difference
between the f-mode frequencies of
$l=88-250$ calculated for a standard solar model (Model S) and the
frequencies obtained from the SOHO/MDI observations. This difference shows
that the model frequencies are systematically, by $\approx 6.6\times 10^{-4}$,
lower than the observed frequencies.
Then from equation (\ref{eq-osc28}):
\begin{equation}
\frac{\Delta R}{R}=-\frac{2}{3}\frac{\Delta \nu}{\nu} \approx 4.4\times 10^{-4},
\label{eq-osc30}
\end{equation}
This means that the seismic radius is approximately equal to
695.68 Mm, which is about 0.3 Mm less than the standard radius,
695.99 Mm, used for calibrating the model calculation.
This radius is usually measured astrometrically as a position of the
 inflection point in the solar limb profile. However, in the model
 calculations it is considered as a height where the optical depth
 of continuum radiation is equal 1. The difference between this height and
 the height of the inflection point can explain the discrepancy between the
 model and seismic radius.

\begin{figure}
\begin{center}
\includegraphics[width=0.6\linewidth]{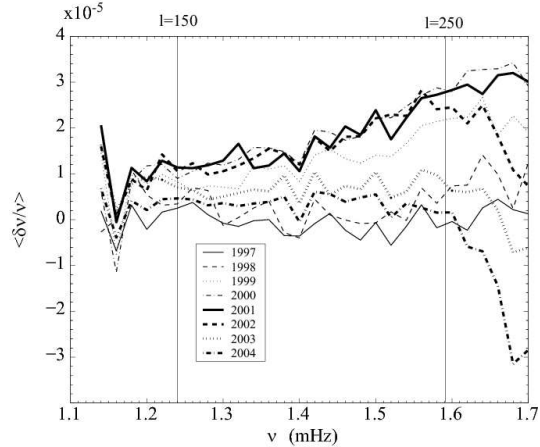}
\caption{Average relative frequency differences in f-mode
$\left\langle \delta\nu/\nu \right\rangle$ as a function of
$\left\langle \nu \right\rangle$, average frequencies binned every
20 $\mu$Hz. The reference year is 1996. }\label{fig23}
\end{center}
\end{figure}

 Figure~\ref{fig22}b illustrates the density profiles
 in the standard solar model (model S \cite{Christensen-Dalsgaard1996})
 and a `seismic' model, calibrated to the seismic radius. The f-mode frequencies
 of the seismic model match the observations.

Since the f-mode frequencies provide an accurate estimate of the seismic radius, then it is interesting to investigate the variations of the solar radius during the solar activity cycle, which are quite important for understanding physical mechanisms of solar variability (e.g. \cite{Rozelot2004}). Figure~\ref{fig23} shows the f-mode frequency variations during the solar cycle 24, in 1997-2004, relative to the f-mode frequencies observed in 1996 during the solar minimum \cite{Lefebvre2005}.

The results show a systematic increase of the f-mode frequency with the increased solar activity, which means a decrease of the seismic radius. However, the variations of the f-mode frequencies are not constant as this is expected from equation (\ref{eq-osc30}
for a simple homologous change of the solar structure. A detailed investigation of these variations showed that the frequency dependence can be explained if the variations of the solar structure are not homologous and if the deeper subsurface layers expand but the shallower layers shrink with the increased solar activity \cite{Lefebvre2005,Lefebvre2007}.

\section{General helioseismic inverse problem}
\label{inverse}

In the asymptotic (high-frequency of short-wavelength)
approximation (\ref{eq-osc26}), the oscillation frequencies depend
only on the sound-speed profile. This dependence is expressed
in terms of the Abel integral equation (\ref{eq-osc27}), which can be solved
analytically.

In the general case, the relation between the frequencies
and internal properties is more complicated, the frequencies depend
not only on the sound speed, but also on other internal properties,
and there is no analytical solution.
Generally, the frequencies determined from the oscillation
equations (\ref{eq-osc6}) and (\ref{eq-osc7})
depend on the density, $\rho(r)$, the pressure, $P(r)$, and the
adiabatic exponent, $\gamma(r)$. However, $\rho$ and $P$ are not
independent, and related to each other through the hydrostatic
equation:
\begin{equation}
\frac{dP}{dr}=-g\rho,
\end{equation}
where $g={Gm}/{r^2},\;\;m=4\pi\int_0^r \rho r'\,^2dr'.$
Therefore, only two thermodynamic (hydrostatic) properties
of the Sun are independent, e.g. pairs of $(\rho,  \gamma)$,
$(P, \gamma)$, or their combinations: $(P/\rho, \gamma)$,
$(c^2, \gamma)$, $(c^2, \rho)$ etc.

The general inverse problem of helioseismology is formulated in
terms of small corrections to the standard solar model because
the differences between the Sun and the standard model are
typically $1\%$ or less. When necessary the corrections can be
applied repeatedly, using an iterative procedure.

\subsection{Variational principle}
We consider the oscillation equations as a formal operator equation
in terms of the vector displacement, $\vec\xi$:
\begin{equation}
\omega^2\vec\xi={\cal L}(\vec\xi),\label{eq-gl1}
\end{equation}
where ${\cal L}$ in the general case is an integro-differential
operator. If we multiply this equation by $\vec\xi^*$ and integrate over
the mass of the Sun we get:
\begin{equation}
\omega^2\int_V \rho\vec\xi^*\cdot\vec\xi dV=
\int_V\vec\xi^*\cdot{\cal L}\vec\xi \rho dV,\label{eq-gl2}
\end{equation}
where $\rho$ is the model density, $V$ is the solar volume.

Then, the oscillation frequencies can be determined as a ratio of two integrals:
\begin{equation}
\omega^2=\frac{\int_V\vec\xi^*\cdot{\cal L}\vec\xi \rho dV}{\int_V \rho\vec\xi^*\cdot\vec\xi dV}.\label{eq-gl3}
\end{equation}
The frequencies are expressed in terms of eigenfunctions $\vec\xi$ and the solar properties
properties represented by coefficients of the operator ${\cal L}$.
For small perturbations of solar parameters the frequency change will
depend on these perturbations and the corresponding perturbations of the eigenfunctions, e.g.
\begin{equation}
\delta\omega^2=\Psi[\delta\rho,\delta\gamma,\delta\vec\xi].
\end{equation}

The variational principle states that the perturbation of the eigenfunctions
constitute second-order corrections, that is to the first-order approximation the frequency variations depend only on variations of the model properties:
\begin{equation}
\delta\omega^2\approx\Psi[\delta\rho,\delta\gamma].\label{eq-gl4}
\end{equation}
The variational principle allows us to neglect the perturbation of the eigenfunctions in the
first-order perturbation theory. This was first established by Rayleigh.
Thus, equation (\ref{eq-gl3}) is called the Rayleigh's Quotient, and the variational principle is called the Rayleigh's Principle. The original formulation of this principle is: for an oscillatory system the averaged over
period kinetic energy is equal the averaged potential energy. In our case, the left-hand side of equation (\ref{eq-gl2}) is proportional to the mean kinetic energy, and the right-hand side is proportional to the potential energy of solar oscillations.

\subsection{Perturbation theory}
We consider a small perturbation of the operator ${\cal L}$ caused by variations of the solar structure properties:
\[{\cal L}(\vec\xi)={\cal L}_0(\vec\xi)+{\cal L}_1(\vec\xi).\]
Then, the corresponding frequency perturbations are
determined from the following equation:
\[\delta\omega^2=\frac{\int_V\vec\xi^*\cdot{\cal L}_1\vec\xi \rho dV}{\int_V \rho\vec\xi^*\cdot\vec\xi dV},\]
or
\begin{equation}
\frac{\delta\omega}{\omega}=\frac{1}{2\omega_0I}{\int_V\vec\xi^*\cdot{\cal L}_1\vec\xi \rho dV},\label{eq-gl5}
\end{equation}
where
\begin{equation}
I=\int_V \rho\vec\xi^*\cdot\vec\xi dV\label{eq-gl6}
\end{equation}
is so-called {\it mode inertia} or {\it mode mass}. The {\it mode energy} is $E=I\omega_0^2a^2$,
where $a$ is the amplitude of the surface displacement. The mode eigenfunctions are
usually normalized such that $\xi_r(R)=1$.

Using explicit formulations for operator ${\cal L}_1$ Eq.~\ref{eq-gl5} can be reduced
to a system of integral equations for a chosen pair of independent
variables \cite{Dziembowski1990,Gough1988,Gough1990a,Kosovichev1999},
e.g. for $(\rho,\gamma)$
\begin{equation}
\frac{\delta\omega^{(n,l)}}{\omega^{(n,l)}}=\int_0^R K_{\rho,\gamma}^{(n,l)}\frac{\delta\rho}{\rho}dr+
\int_0^R K_{\gamma,\rho}^{(n,l)}\frac{\delta\gamma}{\gamma}dr,
\label{eq-gl7}
\end{equation}
where $K_{\rho,\gamma}^{(n,l)}(r)$ and $ K_{\gamma,\rho}^{(n,l)}(r)$
are sensitivity (or `seismic') kernels. They are calculated using the
initial solar model parameters, $\rho_0$, $P_0$, $\gamma$, and the oscillation
eigenfunctions for these model, $\vec\xi$.


\subsection{Kernel transformations}
The sensitivity kernels for various pairs of solar parameters can  be obtained
by using the relations among these parameters, which follows from
the equations of solar structure (`stellar evolution theory').

A general procedure for calculating the sensitivity kernels
developed by Kosovichev \cite{Kosovichev1999} can be illustrated in an operator form.
Consider two pairs of solar variables, $\vec X$ and $\vec Y$, e.g.
\[\vec X=\left(\frac{\delta\rho}{\rho},\frac{\delta\gamma}{\gamma}\right);\;\;
\vec X=\left(\frac{\delta u}{u},\frac{\delta Y}{Y}\right),\]
where $u=P/\rho$, $Y$ is the helium abundance.

The linearized structure equations (the hydrostatic equilibrium equation
and the equation of state) that relates these variables can be written
symbolically:
\begin{equation}
{\cal A}\vec X=\vec Y.\label{eq-gl8}
\end{equation}

Let $\vec K_X$ and $\vec K_Y$ be the sensitivity kernels for $X$ and $Y$,
then the frequency perturbation is:
\begin{equation}
\frac{\delta\omega}{\omega}=\int_0^R \vec K_X\cdot\vec X dr\equiv\left<\vec K_X\cdot\vec X\right>,\label{eq-gl9}
\end{equation}
where $<\cdot>$ denotes the inner product.
Similarly,
\begin{equation}
 \frac{\delta\omega}{\omega}=\left<\vec K_Y\cdot\vec Y\right>.\label{eq-gl10}
\end{equation}

Then from equations (\ref{eq-gl9}) and (\ref{eq-gl10}) we obtain the following relation:
\begin{equation}
\left<\vec K_Y\cdot\vec Y\right>=\left<\vec K_Y\cdot{\cal A}\vec X\right>=\left<{\cal A}^*\vec K_Y\cdot\vec X\right>,\label{eq-gl11}
\end{equation}
where ${\cal A}^*$ is an adjoint operator. This operator is adjoint to the stellar structure operator, ${\cal A}$. The second part of equation (\ref{eq-gl11}) represent a formal definition of this operator.

From Eq.(\ref{eq-gl9}) and (\ref{eq-gl11}) we get:
\[\left<{\cal A}^*\vec K_Y\cdot\vec X\right>=\left<\vec K_X\cdot\vec X\right>.
\]
This equation is valid for any $\vec X$ only if
\begin{equation}
{\cal A}^*\vec K_Y=\vec K_X.\label{eq-gl12}
\end{equation}
That means that the equation for the sensitivity kernels is adjoint
to the stellar structure equations. The explicit formulation of the adjoint equations for the sensitivity kernels for various pairs of variables is given in \cite{Kosovichev1999}.
\begin{figure}[h]
\begin{center}
\includegraphics[width=0.95\linewidth]{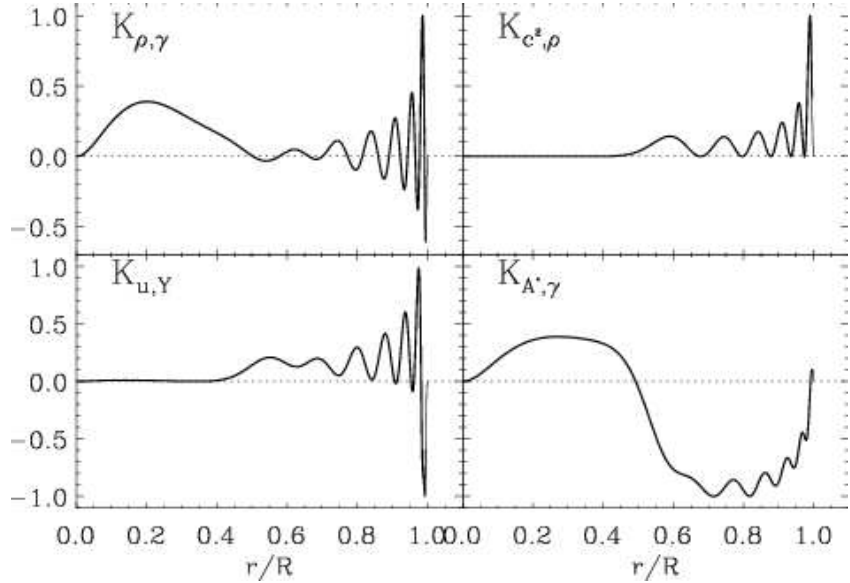}
\caption{Sensitivity kernels for the acoustic mode of the angular degree, $l$=10, and the
radial order, $n$=6. $K_{\rho,\gamma}$ is the kernel for density, $\rho$, at constant
adiabatic exponent, $\gamma$; $K_{c^2,\rho}$ is the kernel for the squared sound speed, $c^2$,
at constant $\rho$; $K_{u,Y}$ is the kernel for function $u$, - the ratio pressure, $p$, to density
at constant helium abundance, $Y$;
and $K_{A^*,\gamma}$ is the kernel for the parameter of convective stability,
$A^*=rN^2/g$, at constant $\gamma$.}\label{fig24}
\end{center}
\end{figure}

\begin{figure}[h]
\begin{center}
\includegraphics[width=0.75\linewidth]{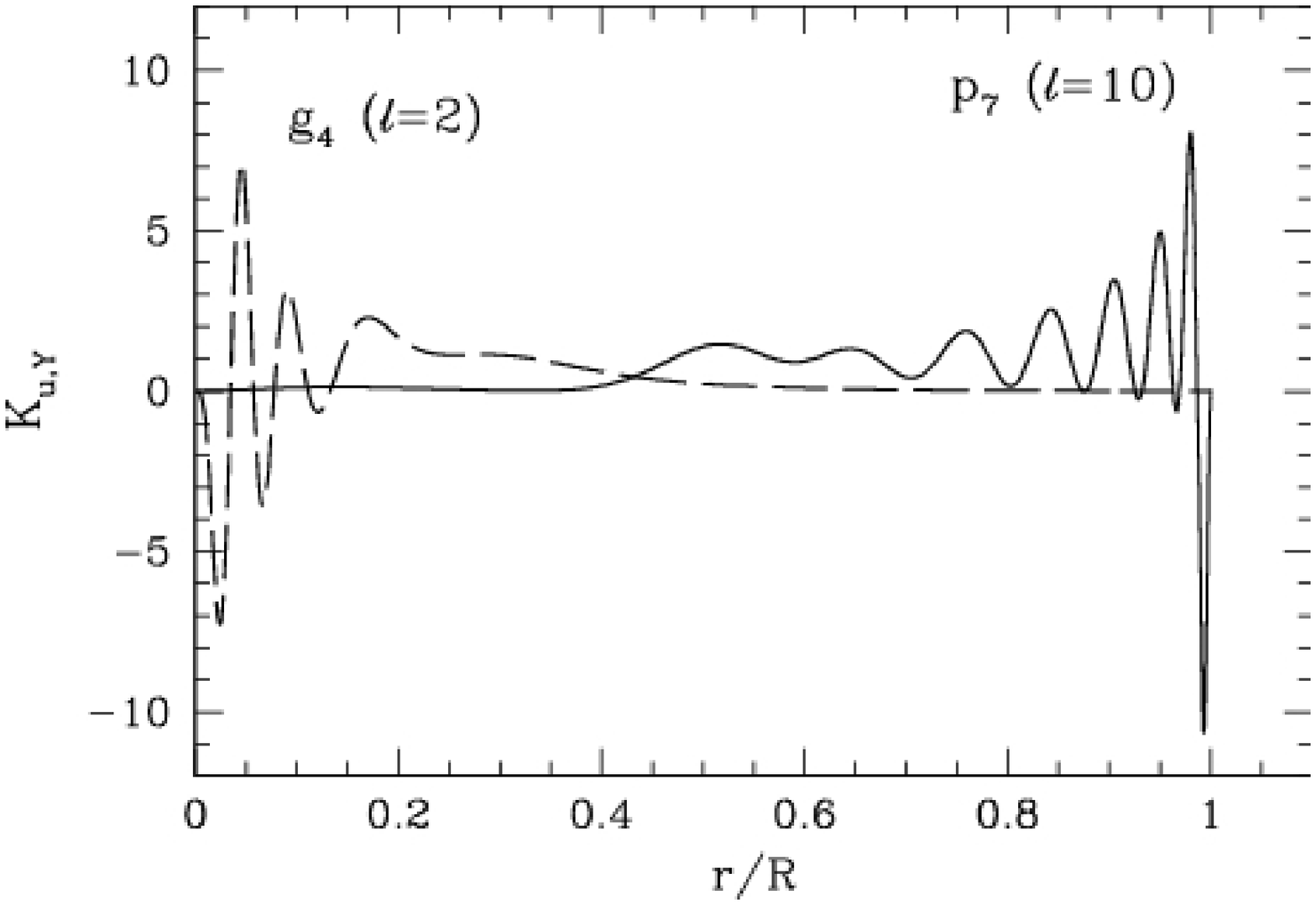}
\caption{Sensitivity kernels for p- and g-modes for
$u=P/\rho$ and helium abundance $Y$.}\label{fig25}
\end{center}
\end{figure}

Examples of the sensitivity kernels for solar properties are shown in Figures~\ref{fig24}. Figure~\ref{fig25} illustrates the difference in sensitivities of the p- and g-modes. The frequencies of solar p-modes are mostly sensitive to properties of the outer layers of the Sun while the frequencies of g-modes have the greatest sensitivity to the parameters of the solar core.

\subsection{Solution of inverse problem}

The variation formulation provides us with a system integral equations (\ref{eq-gl7}) for a set of observed mode
frequencies. Typically, the number of observed frequencies, $N \simeq 2000$.
Thus, we have a problem of determining two functions from this finite set of measurements.
In general, it is impossible to determine these functions precisely.
We can always find some rapidly oscillating functions, $f(r)$, that being added
to the unknowns, $\delta\rho/\rho$ and $\delta\gamma/\gamma$, do not
change the values of the integrals, e.g.
\[\int_0^R  K_{\rho,\gamma}^{(n,l)}(r)f(r)dr=0.\]

Such problems without a unique solution are called "ill-posed".
The general approach is to find a smooth solution that satisfies the
integral equations (\ref{eq-gl7}) by applying some smoothness constraints to the
unknown functions. This is called a {\it regularization procedure}.

There are two basic methods for the helioseismic inverse problem:
\begin{enumerate}
\item
Optimally Localized Averages (OLA) method - (Backus-Gilbert method) \cite{Backus1968}
\item
Regularized Least-Squares (RLS) method - (Tikhonov method) \cite{Tikhonov1977}
\end{enumerate}

\subsection{Optimally localized averages method}

The idea of the OLA method is to find a linear combination
of data such as the corresponding linear combination of the
sensitivity kernels for one unknown has an isolated peak at a given
radial point, $r_0$, (resembling a $\delta$-function), and the combination
for the other unknown is close to zero. Then, this linear
combination provides an estimate for the first unknown at $r_0$.

Indeed, consider a linear combination of (\ref{eq-gl7}) with some
unknown coefficient $a^{(n,l)}$:
\begin{equation}
\sum a^{(n,l)}\frac{\delta\omega^{(n,l)}}{\omega^{(n,l)}}=
=\int_0^R \sum a^{(n,l)}K_{\rho,\gamma}^{(n,l)}\frac{\delta\rho}{\rho}dr+
\int_0^R \sum a^{(n,l)}K_{\gamma,\rho}^{(n,l)}\frac{\delta\gamma}{\gamma}dr.\label{eq-gl10a}
\end{equation}
If in the first term the linear combination of the kernels is close to a $\delta$-function at $r=r_0$, that is
\begin{equation}
\sum a^{(n,l)}K_{\rho,\gamma}^{(n,l)}(r) \simeq \delta(r-r_0),\label{eq-gl11a}
\end{equation}
and the linear combination in the second term vanishes:
\begin{equation}
\sum a^{(n,l)}K_{\gamma,\rho}^{(n,l)}(r) \simeq 0,\label{eq-gl12a}
\end{equation}
then equation (\ref{eq-gl10a}) gives an estimate of the density perturbation, $\delta\rho/\rho$, at $r=r_0$:
\begin{equation}
\sum a^{(n,l)}\frac{\delta\omega^{(n,l)}}{\omega^{(n,l)}} \approx \int_0^R \delta(r-r_0)\frac{\delta\rho}{\rho}dr = \overline{\left(\frac{\delta\rho}{\rho}\right)}_{r_0}.\label{eq-gl13}
\end{equation}
Of course, the coefficients, $a^{(n,l)}$, of equation (\ref{eq-gl13}) must be calculated from conditions (\ref{eq-gl11a}) and (\ref{eq-gl12a}) for various target radii $r_0$.

The functions,
\begin{equation}
\sum a^{(n,l)}K_{\rho,\gamma}^{(n,l)}(r)\equiv A(r_0,r),\label{eq-gl14}
\end{equation}
\begin{equation}
\sum a^{(n,l)}K_{\gamma,\rho}^{(n,l)}(r)\equiv B(r_0,r),\label{eq-gl15}
\end{equation}
are called the {\it averaging kernels}. They play a fundamental role in the helioseismic inverse theory for determining the resolving power of helioseismic data.

\begin{figure}[h]
\begin{center}
\includegraphics[width=0.75\linewidth]{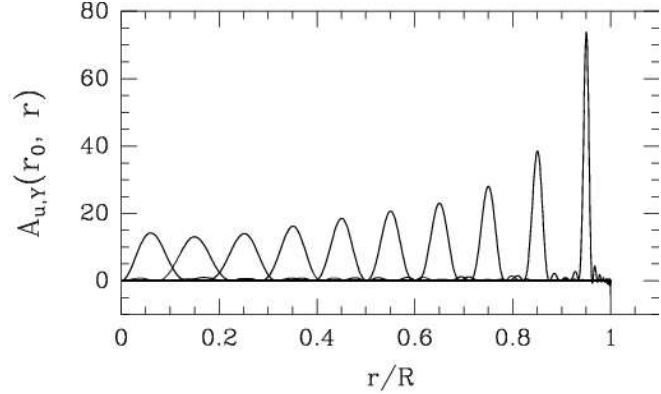}
\caption{A sample of the optimally localized
averaging kernels for the structure function, $u$,
the ratio of pressure, $P$, to density, $\rho$, $u=P/\rho$. The second, eliminated, parameter in these kernels is the helium abundance, $Y$.
}\label{fig26}
\end{center}
\end{figure}

The coefficients, $a^{n,l}$,
are determined my
minimizing a quadratic form :
\begin{eqnarray}
M(r_0, A, \alpha, \beta) = \int_0^R J(r_0,r) \left[ A(r_0,r)\right]^2 dr+\\ \nonumber
+ \beta \int_0^R \left[ B(r_0,r)\right]^2 dr
+ \alpha \sum_{i,j} E_{n,l;n',l'}a^{n,l}a^{n',l'},\label{eq-gl16}
\end{eqnarray}
where  function $J(r_0,r) = 12(r - r_0)^2$ provides a localization of the averaging kernels $A(r,r_0)$ at $r=r_0$,  $E_{n,l;n',l'}$ is a covariance matrix
of observational errors, $\alpha$ and $\beta$ are  {\it regularization parameters}.
The first integral in eq.~(\ref{eq-gl16}) represents the Backus-Gilbert
criterion of $\delta$-ness for $A(r_0,r)$; the second term minimizes
the contribution from $B(r_0,r)$, thus, effectively eliminating
the second unknown function, ($\delta\gamma/\gamma$ in this case);
and the last term minimizes the
errors. A practical minimization algorithm is presented in \cite{Kosovichev1999}.
An example of the averaging kernels is shown in Fig.~\ref{fig26}

\subsection{Inversion results for solar structure}

As an example, consider the results of inversion of the recent
data obtained from the MDI instrument on board the SOHO space observatory.
The data represent 2176 frequencies of solar oscillations of the angular degree, $l$,
from 0 to 250. These frequencies were obtained by fitting peaks in the oscillation
power spectra from a 360-day observing run,  between May 1, 1996 and April 25, 1997.\\

\begin{figure}[h]
\begin{center}
\centerline{\epsfxsize=3.6in \epsfbox{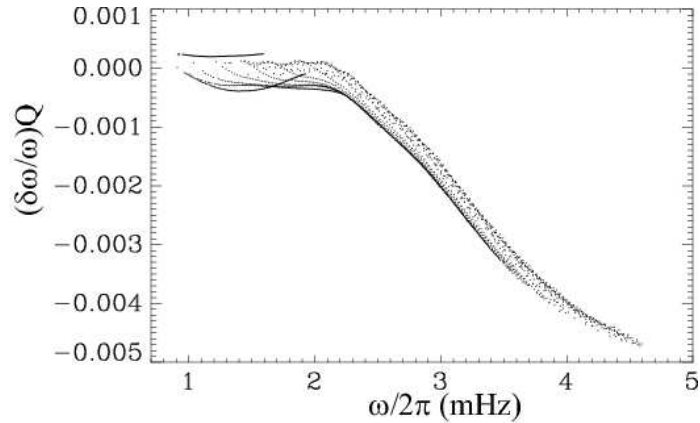}}
\caption{ The relative frequency difference, scaled with the relative mode inertia factor, $Q=I/I_0$ (\ref{eq-gl6}), between the Sun and the standard solar model.
}\label{fig27}
\end{center}
\end{figure}

Figure~\ref{fig27} shows the relative frequency difference, $\delta\omega/\omega$, between the observed frequencies and the corresponding frequencies calculated for the standard model S
\cite{Christensen-Dalsgaard1996}. The frequency difference is scaled with a factor $Q\equiv I(\omega)/I_0(\omega)$,
where $I(\omega)$ is the mode inertia, and $I_0(\omega)$
is the mode inertia of radial modes ($l=0$), calculated at
the same frequency.

This scaled frequency difference depends mainly on the frequency
alone meaning that most of the difference between the Sun and the reference
solar model is in the near-surface layers. Physically, this follows from the fact that
the p-modes of different $l$ behave
similarly near the surface where they propagate almost vertically. This behavior is illustrated by the p-mode ray paths in Fig.~\ref{fig19}a, which become almost radial near the surface. In the inversion procedure, this frequency dependence is eliminated by adding an additional ``surface term" in equation (\ref{eq-gl7}) \cite{Kosovichev1999}.
However, there is also a significant
scatter along the general frequency trend. This scatter is due to the variations
of the structure in the deep interior, and it is the basic task of the inversion
methods to uncover the variations.

\begin{figure}[h]
\begin{center}
\includegraphics[width=0.85\linewidth]{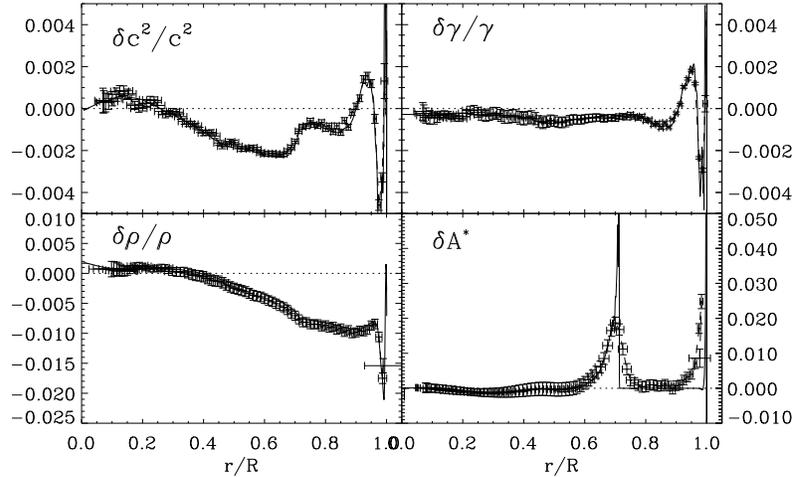}
\caption{The results of test inversions (points with the error bars, connected with dashed curves)
of frequency differences between two solar models
for the squared sound speed, $c^2$, the adiabatic exponent, $\gamma$, the density, $\rho$, and the
parameter of convective stability, $A^*$. The solid curves show the actual differences between
the two models. Random Gaussian noise was added to the frequencies of a test solar model.
The vertical bars show the formal error estimates, the horizontal bars show the
characteristic width of the localized averaging kernels. The central points of the averages are
plotted at the centers of gravity of the averaging kernels.
}\label{fig28}
\end{center}
\end{figure}

First, we test the inversion procedure by considering the frequency difference
for two solar models and trying to recover the differences between model
properties. Results of the test
inversion (Fig.~\ref{fig28}) show  good agreement with the actual
differences. However, the sharp variations, like a peak in the parameter of convective stability, $A^* \equiv rN^2/g$, at the base of the convection zone, are smoothed.
Also, the
inner 5\% of the Sun and the subsurface layers (outer 2-3\%) are
not resolved.

\begin{figure}[h]
\begin{center}
\includegraphics[width=0.85\linewidth]{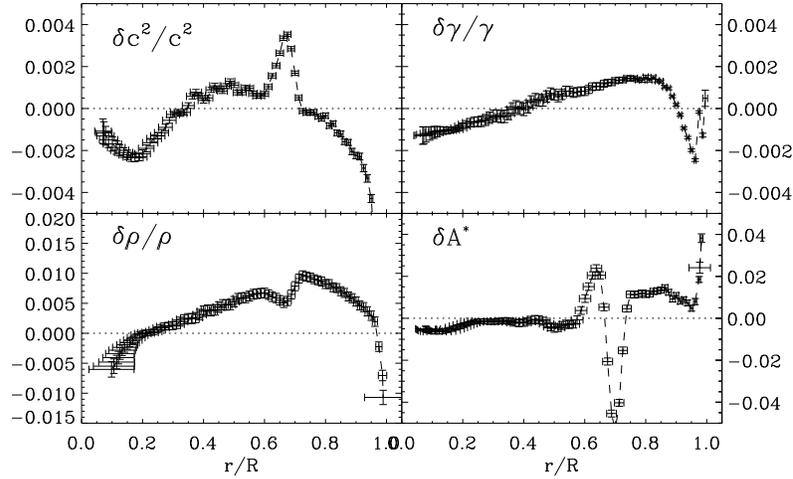}
\caption{ The relative differences between the Sun and the standard solar model
\cite{Christensen-Dalsgaard1996}in
the squared sound speed, $c^2$, the adiabatic exponent, $\gamma$, the density,
$\rho$, and the
parameter of convective stability, $A^*$, inferred from the solar frequencies
determined
from the 360-day series of SOHO MDI data.
}\label{fig29}
\end{center}
\end{figure}

Then, we apply this procedure to the real solar data.
The results (Fig.~\ref{fig29}) show that the differences between the inferred structure and the reference
solar model (model S)  are quite small, generally less than 1\%. The small differences provide
a justification for the linearization procedure, based on the variational principle.
This also means that the modern standard model of the Sun \cite{Christensen-Dalsgaard1996}
provides an accurate description
of the solar properties compared to the earlier solar model \cite{Christensen-Dalsgaard1982}, used for the asymptotic inversions (Fig.~\ref{fig21}).
A significant improvement in the solar modeling was achieved by using more accurate radiative opacity data and by including the effects of gravitational settling of heavy elements and element diffusion. However, recent spectroscopic estimates of the heavy element abundance on the Sun, based on radiative hydrodynamics simulations of
solar convection, indicated that the heavy element abundance on the Sun may be lower than the value used in the standard model \cite{Asplund2009}. The solar model with a low heavy element abundance do not agree with the helioseismology measurements (e.g. \cite{Bahcall2005}). This problem in the solar modeling has not been resolved.
Thus, the helioseismic inferences of the solar structure  lead to better understanding of the structure and evolution
of the star, and have important applications in other fields of astrophysics.

The prominent peak of the squared sound speed, $\delta c^2/c^2$, at the
base of the convection zone, $r/R\approx 0.7$,
 indicates on additional mixing
which may be caused by rotational shear flows or by convective
overshoot. The variation in the sound speed in the energy-generating
core at $r/R <0.2$ might be also caused by a partial mixing.

The monotonic decrease of the adiabatic exponent, $\gamma$, in the
core was recently explained by the relativistic corrections to the
equation of state \cite{Elliott1998}.
Near surface variations of
$\gamma$, in the zones of ionization of helium and hydrogen, and
below these zones, are most likely caused by deficiencies in the
theoretical models of the weakly coupled plasma employed in
the equation of
state calculations \cite{Dappen2007}.

The monotonic decrease of the squared
sound speed variation in the convection zone ($r/R > 0.7$) is
partly due to an error in the solar seismic radius used to calibrate the
standard model \cite{Schou1997}, and partly due to the inaccurate description of the subsurface layers by the standard solar model, based on a mixing-length convection theory.

\subsection{Regularized least-squares method}
The Regularized Least-Squares (RLS) method \cite{Tikhonov1977} is based on minimization of the quantity
\begin{eqnarray}
 {\cal E} \equiv \sum_{n,l} \frac{1}{\sigma^2_{n,l}} \left[\frac{\delta\omega^{(n,l)}}{\omega^{(n,l)}} -
\int_0^R \left( K_{(f,g)}^{(n,l)} \frac{\delta f}{f}
+ K_{(g,f)}^{(n,l)} \frac{\delta g}{g}\right) dr \right]^2 + \nonumber\\ + \int_0^R
\left[\alpha_1\left(L_1\frac{\delta f}{f}\right)^2 +
\alpha_2\left(L_2\frac{\delta g}{g}\right)^2\right]
dr\label{eq-gl16a},
\end{eqnarray}
in which the unknown structure correction functions, ${\delta
f}/{f}$ and  ${\delta g}/{g}$, are both represented by
piece-wise linear functions  or by cubic splines. The second integral specifies
smoothness constraints for the unknown functions, in which $L_1$
and $L_2$ are linear differential operators, e.g.
$L_{1,2}={d^2}/{d^2r}$; $\sigma_i$ are error estimates of the
relative frequency differences.

In this inversion method, the estimates of the
structure corrections are, once again, linear combinations
of the frequency differences obtained from observations,
and corresponding averaging kernels exist too.
However, unlike the OLA kernels $A(r_0;r)$, the RLS averaging kernels
may have negative sidelobes and significant peaks near the surface,
thus making interpretation of the inversion results to some extent ambiguous.
Nevertheless, it works well in most cases, and may provide a higher resolution
compared to the OLA method.

\subsection{Inversions for solar rotation}

The eigenfrequencies of a spherically-symmetrical static star are degenerate with respect
to the azimuthal number $m$. Rotation breaks the symmetry and splits each mode of
radial order, $n$, and angular degree, $l$, into $(2l+1)$ components of $m=-l,...,l$
({\it mode multiplets}). The rotational frequency splitting can be computed using a more
general variational principle derived by Lynden-Bell and Ostriker \cite{Lynden-Bell1967}. From this
variational principle, one can obtain mode frequencies $\omega_{nlm}$ relative to
the degenerate frequency $\omega_{nl}$ of the non-rotating star:
\begin{equation}
\Delta\omega_{nlm}\equiv\omega_{nlm}-\omega_{nl}=\frac{1}{I_{nl}}\int_V\left[m\vxi\cdot\vxi^*+
i\vece_\Omega(\vxi\times\vxi^*)\right]\Omega\rho dV,\label{eq-gl17}
\end{equation}
where $\vece_\Omega$ is the unit vector defining the rotation axis, and $\Omega=\Omega(r,\theta)$
is the angular velocity which is a function of radius $r$ and co-latitude $\theta$, and $I_{nl}$ is the mode inertia.

Equation~(\ref{eq-gl17}) can be rewritten
as a two-dimensional integral equation for $\Omega(r,\theta)$:
\begin{equation}
\Delta\omega_{nlm}=\int_0^R\int_0^\pi K^{(\Omega)}_{nlm}(r,\theta) \Omega(r,\theta)
 d\theta dr.\label{eq-gl18}
\end{equation}
where $K^{(\Omega)}_{nlm}(r,\theta)$ represent the rotational splitting kernels:
\begin{eqnarray}
K^{(\Omega)}_{nlm}(r,\theta) =  \frac{m}{I_{nl}} 4\pi\rho r^2\left\{
(\xi_{nl}^2-2\xi_{nl}\eta_{nl})(P_l^m)^2 +
\eta_{nl}^2\left[\left(\frac{dP_l^m}{d\theta}\right)^2- \right.\right.
 \nonumber \\
\vspace*{-2cm} \left.\left. -2P_l^m\frac{dP_l^m}{d\theta}
\frac{\cos\theta}{\sin\theta}+\frac{m^2}{\sin^2\theta}(P_l^m)^2\right]\right\}\sin\theta.\label{eq-gl18a}
\end{eqnarray}
Here $\xi_{nl}$ and $\eta_{nl}$ are the radial and horizontal
components of eigenfunctions
of the mean spherically
symmetric structure of the Sun, $P_l^m(\theta)$ is an
associated normalized Legendre function ($\int_0^\pi (P_l^m)^2\sin\theta d\theta=1$).
The kernels are symmetric relative to the equator, $\theta=\pi/2$. Therefore,
the frequency splittings are sensitive only to the symmetric component of rotation
in the first approximation. The non-symmetric component can, in principle, be determined
from the second-order correction to the frequency splitting, or from
local helioseismic techniques, such as time-distance seismology.

For a given set of observed frequency splitting, $\Delta\omega_{nlm}$, eq.~(\ref{eq-gl17})
constitutes a two-dimensional linear inverse problem for the angular velocity,
$\Omega(r,\theta)$, which can be solved by the OLA or RLS techniques.

\subsection{Results for Solar Rotation}
\begin{figure}
\begin{center}
\includegraphics[width=0.95\linewidth]{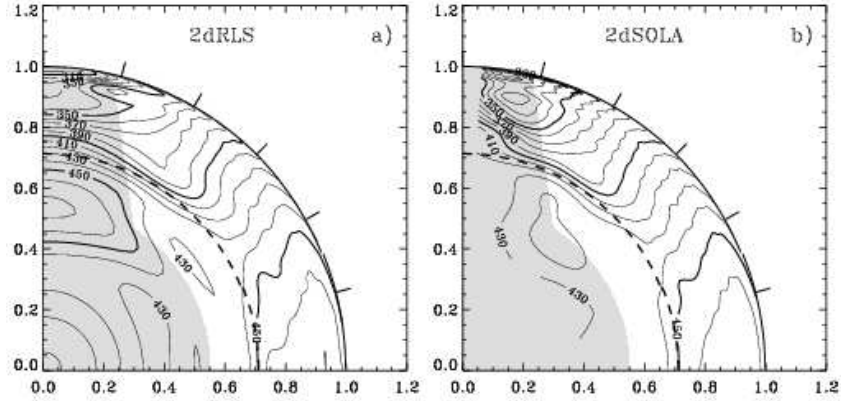}
\caption{Contour lines of the rotation rate (in nHz) inside the Sun obtained by
inverting the rotational frequency splittings from a 144-day
observing run from SOHO MDI by the RLS and SOLA methods.
The shaded areas are the areas where the localized averaging kernels substantially
deviate from the target positions.}\label{fig30}
\end{center}
\end{figure}
As an example, we present the inversion results for solar rotation
obtained from SOHO/MDI
data. The frequency splitting data were obtained from the 144-day MDI time series by J.
Schou for $j=1,...,36$ and $1\leq l\leq 250$ \cite{Schou1998}. The total number of
measurements in this data set was $M=37366$.

Figure~\ref{fig30} shows results of inversion of the SOI-MDI data by
the two methods. The results are generally in  good agreement in
most of the area where good averaging kernels were obtained.
However, the results differ in the high-latitude region. In
particular, a prominent feature of the RLS inversion at coordinates (0.2, 0.95) in Fig.~\ref{fig30}a, which
can be interpreted as a `polar jet', is barely visible in Fig.~\ref{fig30}b, showing the OLA inversion of the same data.
Therefore, obtaining reliable inversion results in this region and
also in the shaded area is one of the main current goals of
helioseismology. This can be achieved by obtaining more accurate
measurements of rotational frequency splitting and improving
inversion techniques. Of course, the radical improvement can be made by observing the polar regions of the Sun. These measurements can be done by using spacecraft with an orbit highly inclined to the ecliptic plane, such as a proposed Solar Polar Imager (SPI) and POLARIS missions \cite{Appourchaux2009}.

\begin{figure}[t]
 \begin{center}
 \includegraphics[width=0.6\textwidth]{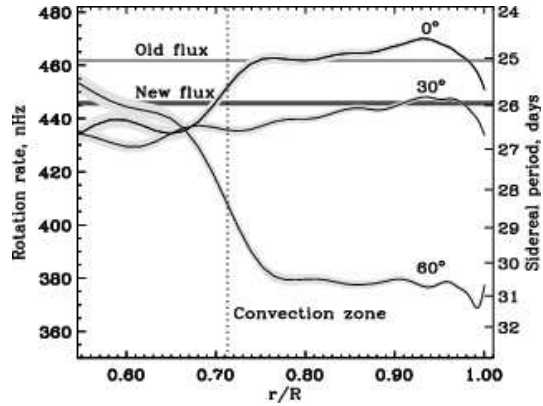}
  \caption{ The solar rotation rate as a function of radius at three
  latitudes. The horizontal lines indicate the rotation rate of the
  surface magnetic flux at the end of solar cycle 22 ("old magnetic flux")
  and at the beginning of cycle 23 ("new magnetic flux")
  \cite{Benevolenskaya1999}}. \label{fig31}
  \end{center}
\end{figure}

The most characteristic feature of solar rotation is the
differential rotation of the convection zone, which occupies the our
30\% of the solar radius. While the radiative core rotates almost
uniformly, the equatorial regions of the convection zone rotate
significantly faster than the polar regions. The main interest is in
understanding the role of Sun's internal rotation in the dynamo
process of generation of solar magnetic fields and the origin of the
11-year sunspot cycle. The results of these measurements
(Fig.~\ref{fig31}a) reveal two radial shear layers
at the bottom of the convection zone (so-called tachocline) and in
the upper convective boundary layer. A common assumption is that the
solar dynamo operates in the tachocline area (interface dynamo)
where it is easier to explain storage of magnetic  flux than in the
upper convection zone because of the flux buoyancy. However, there
are theoretical and observational difficulties with this concept.
First, the magnetic field in the tachocline must be quite strong,
$\sim 60-160$ kG, to sustain the action of the Coriolis force
transporting the emerging flux tubes into high-latitude regions
\cite{DSilva1993}. The magnetic energy of such field is above the
equipartition level of the turbulent energy. Second, the
back-reaction such strong field should suppress turbulent motions
affecting the Reynolds stresses. Since these turbulent stresses
support the differential rotation one should expect significant
changes in the rotation rate in the tachocline. However, no
significant variations with the 11-year solar cycle are detected.
Third, magnetic fields often tend to emerge in compact regions on
the solar surface during long periods lasting several solar
rotations. This effect is known as "complexes of activity" or
"active longitudes". However, the helioseismology observations show
that the rotation rate of the solar tachocline is significantly
lower than the surface rotation rate. Thus, magnetic flux emerging
from the tachocline should be spread over longitudes (with new flux
lagging the previously emerged flux) whether it remains connected to
the dynamo region or disconnected. It is well-known that sunspots
rotate faster than surrounding plasma. This means that the magnetic
field of sunspots is anchored in subsurface layers. Observations
show that the rotation rate of magnetic flux matches the internal
plasma rotation in the upper shear layer
(Fig.~\ref{fig31}b) indicating that this layer is
playing an important role in the solar dynamo, and causing a shift
in the dynamo paradigm \cite{Brandenburg2005}.

\begin{figure}[t]
 \begin{center}
 \includegraphics[width=0.7\textwidth]{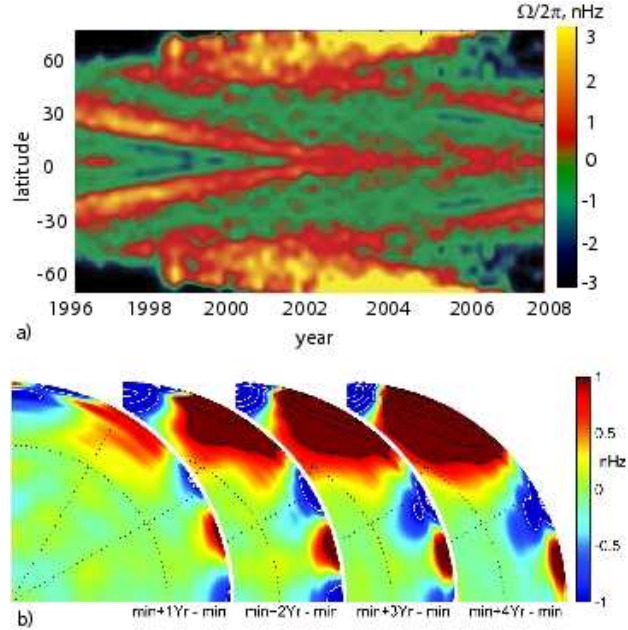}
  \caption{a) Migration of the subsurface zonal flows with latitude
  during solar cycle 23 from SOHO/MDI data \cite{Howe2008}. Red
  shows zones of faster rotation, green and blue show slower
  rotation. b) Variations of the zonal flows with depth and latitude
  during the first 4 years after the solar minimum.
  \cite{Vorontsov2002}}\label{fig32}
  \end{center}
\end{figure}

Variations in solar rotation clearly related to the 11-year sunspot
cycle are observed in the upper convection zone. These are so-called
`torsional oscillations' which represent bands of slower and faster
rotation, migrating towards the equator as the solar cycle
progresses (Fig.~\ref{fig32}). The torsional oscillations were
first discovered on the Sun's surface \cite{Howard1980}, and then
were found in the upper convection zone by helioseismology
\cite{Kosovichev1997b,Howe2000}. The depth of these evolving zonal
flows is not yet established. However, there are indications that
they may be persistent   through most of the convection zone, at
least, at high latitudes \cite{Vorontsov2002}. The physical
mechanism is not understood. Nevertheless, it is clear that these
zonal flows are closely related to the internal dynamo mechanism
that produces toroidal magnetic field. On the solar surface, this
field forms sunspots and active regions which tend to appear in the
areas of shear flows at the outer (relative to the equator) part of
the faster bands. Thus, the torsional flows are an important key to
understanding the solar dynamo, and one of the challenges is to
establish their precise depth and detect corresponding variations in
the thermodynamic structure of the convection zone. Recent modeling
of the torsional oscillations by the Lorentz force feedback on
differential rotation showed that the poleward-propagating
high-latitude branch of the torsional oscillations can be explained
as a response of the coupled differential rotation/meridional flow
system to periodic forcing in midlatitudes of either mechanical
(Lorentz force) or thermal nature \cite{Rempel2007}. However, the
main equatorward-propagating branches cannot be explained by the
Lorenz force, but maybe driven by thermal perturbations caused by
magnetic field \cite{Spruit2003}. It is intriguing that starting
from 2002, during the solar maximum, the helioseismology
observations show new branches of "torsional oscillations" migrating
from about $45^\circ$ latitude towards the equator
(Fig.~\ref{fig32}a). They indicate the start of the next solar
cycle, number 24, in the interior, and are obviously related to
magnetic processes inside the Sun. However, magnetic field of the
new cycle appeared on the surface only in 2008.


\section{Local-area helioseismology}
\label{local}

\subsection{Basic principles}
In the previous sections we discussed methods of global
helioseismology, which are based on inversions of accurately measured
frequencies and frequency splitting of normal oscillation modes of
the Sun. The frequencies are measured from long time series of
observations of the Doppler velocity of the solar disk. These
time series are much longer than the mode lifetimes, typically, two
or three 36-day-long `GONG months', that is 72 or 108 days. The long
time series allow us to resolve individual mode peaks in the power
spectrum, and accurately measure the frequencies and other
parameters of these modes. However, because of the long integration
times global helioseismology cannot capture the fast evolution of
magnetic activity in subsurface layers of the Sun. Also, it provides
only information about the axisymmetrical structure of the Sun and
the differential rotation (zonal flows).

Local helioseismology attempts to determine the subsurface structure
and dynamics of the Sun in local areas by analyzing local
characteristics of solar oscillations, such as frequency and phase
shifts and variations in wave travel times. This is a relatively new
and rapidly growing field. It takes advantage of high-resolution
observations of solar oscillations, currently available from the
GONG+ helioseismology network and the space mission SOHO, and
are anticipated from the SDO mission.

\subsection{Ring-diagram analysis}

Local helioseismology was pioneered by Douglas Gough and Juri Toomre \cite{Gough1983} first proposed to
measure oscillation frequencies of solar modes as a function of the
wavevector, $\omega(\veck)$, ({\it the dispersion relation}) in local areas,
and use these measurements for diagnostics of the local flows and thermodynamic properties.
They noticed that subsurface variations of temperature cause change
in the  frequencies, and that subsurface flows result in distortion of
the dispersion relation because of the advection effect.

\begin{figure}
\begin{center}
\includegraphics[width=0.6\linewidth]{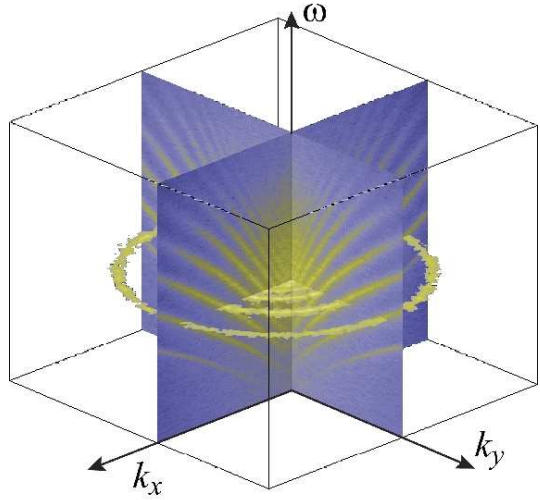}
\caption{Three-dimensional power spectrum of solar oscillations,
$P(k_x,k_y,\omega)$. The vertical panels with blue background
show the mode ridge structure similar to
the global oscillation spectrum shown in Fig.~\ref{fig3}.
The horizontal cut with transparent background shows the ring structure
of the power spectrum at a given frequency (courtesy of Amara Graps).}\label{fig33}
\end{center}
\end{figure}

This idea was implemented by Frank Hill \cite{Hill1988} in the form
of a {\it ring-diagram analysis}. The name of this technique comes from
the ring appearance of the 3D dispersion relation,
$\omega=\omega(k_x,k_y)$, in the $(k_x, k_y)$ plane, where $k_x$ and
$k_y$ are $x$- and $y$-components of the wave vector, $\veck$ (Fig.~\ref{fig33}). The ridges in the vertical cuts represent the same mode ridges as in Fig.~\ref{fig3}, corresponding to the normal oscillation modes of different radial orders $n$.

In the presence of a horizontal flow field, $\vecU=(U_x,U_y)$ the
dispersion relation has the form:
\begin{equation}
\omega=\omega_0(k)+\veck\cdot\vecU\equiv \omega_0+(U_xk_x+U_yk_y),\label{eq-loc1}
\end{equation}
where $\omega_0(k)$ is the symmetrical part of the dispersion
relation in the $(k_x,k_y)$-plane. It depends only on the magnitude of
the wave vector, $\veck$. The power spectrum, $P(\omega,\veck)$ for each $\veck$ is fitted with a Lorentian profile \cite{Haber2002}:
\begin{equation}
P(\omega,\veck)=\frac{A}{(\omega - \omega_0+k_xU_x+k_yU_y)^2+\Gamma^2}+\frac{b_0}{k^3},\label{eq-loc2}
\end{equation}
where $A, \omega_0, \Gamma$, and $b_0$ are respectively the amplitude, central frequency, line width and a background noise parameter.

In some realizations, the fitting formula includes the line asymmetry (Sec.~\ref{basic}). Also, the central frequency can be
fitted by assuming a power-law relation: $\omega_0=ck^p$, where $c$
and $p$ are constants \cite{Basu2004,Hill1988}. This relationship is
valid for a polytropic adiabatic stratification, where $p=1/2$
\cite{Gough1983}.  If the flow velocity changes with depth then the
parameter, $\vecU$, represent a velocity, averaged with the depth
with a weighting factor proportional to the kinetic energy density
of the waves, $\rho\vecxi\cdot\vecxi$ \cite{Gough1976}:
\begin{equation}
\vecU=\frac{\int \vecu(z)\rho\vecxi\cdot\vecxi dz}{\int
\rho\vecxi\cdot\vecxi dz},\label{eq-loc3}
\end{equation}
where $\vecxi(z)=(\xi_r,\xi_h)$ is the wave amplitude, given by the
mode displacement  eigenfunctions (\ref{eq-xi}. The integral is
taken over the entire extent of the solar envelope. Equation
(\ref{eq-loc3}) is solved by the RLS or OLA techniques (Sec.~\ref{inverse}).

The ring-diagram method has provided important results about
the structure and evolution of large-scale and meridional flows and dynamics of active regions
\cite{Haber2000,Haber2002,Haber2004,Howe2008,Komm2008}. In
particular, large-scale patterns of subsurface flows converging around
magnetic active regions were discovered \cite{Haber2004}. These
flows cause variations of the mean meridional circulation with the
solar cycle \cite{Haber2002}, which may affect transport of magnetic
flux of decaying active regions from low latitudes to the polar
regions, and thus change the duration and magnitude of the solar
cycles \cite{Dikpati2006}.

However, the ring-diagram technique in the present formulation has
limitations in terms of the spatial and temporal resolution and the
depth coverage. The local oscillation power spectra are typically calculated for regions with the horizontal size covering 15 heliographic degrees ($\simeq 180$ Mm). This is significantly larger than the typical size of supergranulation and active regions ($\simeq 30$ Mm). There have been attempts to increase the resolution by doing the measurements in overlapping regions (so-called "dense-packed diagrams"). However, since such measurements are not independent, their resolution is unclear. The measurements of the power spectra calculated for smaller regions (2-4 degrees in size) increase the spatial resolution but decrease the depth coverage \cite{Hindman2006}.

\subsection{Time-distance helioseismology (Solar tomography)}

Further developments of local seismology led to the
idea to perform  measurements of local wave distortions in the
time-distance space instead of the traditional frequency-wavenumber Fourier space
\cite{Duvall1993}. In this case, the wave distortions can be
measured as perturbations of wave travel times. However, because
of the stochastic nature of solar waves it is impossible to track
individual wave fronts. Instead, it was suggested to use a
cross-covariance (time-distance) function that provides a
statistical measure of the wave distortion.
Indeed, by cross-correlating solar oscillation signals at two points one
may expect that the main contribution to this cross-correlation will
be from the waves traveling between these points along the acoustic ray paths \cite{Claerbout1968,Rickett2000}. Thus, the cross-ccovariance function calculated for oscillation signals measured at two points separated by a distance, $\Delta$, for various time lags, $\tau$, has a peak when the time lag is equal to the travel time of acoustic waves between these points. Physically, the cross-covariance function corresponds to the Green's function of the wave equation, representing the wave signal from a point source. Of course, in reality, because of the finite wavelength effects, non-uniform distribution of acoustic sources, and complicated wave interaction with turbulence and magnetic fields the interpretation
of the travel-time measurements is extremely challenging. Various approximations
are used to relate the observed perturbations of the travel times to
the internal properties such as sound-speed perturbations and flow
velocities. We discuss the basic principles and the current status of the time-distance helioseismology method in Sec.~\ref{tomography}.

\subsection{Acoustic holography and imaging}

The acoustic holography \cite{Lindsey1997} and acoustic imaging \cite{Chang1997} techniques are developed on the principles of day-light imaging by collecting over large areas on the solar surface coherent acoustic signals emitted from selected target points of the interior. The idea is that the constructed this way signals contain information about objects located below the surface  because of wave absorption or scattering at the target points. The phases of individual signals are calculated by using the time-distance relation, $\tau(\Delta)$, f   or acoustic waves traveling along the ray paths. The constructed signals, $\psi_{\rm out,in}(t)$, are calculated using the following relation \cite{Chou1999}:
\begin{equation}
\psi_{\rm out,in}(t)=\sum_{\tau_1}^{\tau_2} W \overline\psi(\Delta,t\pm \tau),\label{eq-loc4}
\end{equation}
where $\overline\psi(\Delta,t+\tau)$ is the azimuthal-averaged signal at a distance $\Delta$ from a target point at time $t \pm \tau(\Delta)$. The summation variable $\tau$ is equally spaced in the interval $(\tau_1,\tau_2)$; and the weighting factor, $W \propto (\sin\Delta/\tau^2)^{1/2}$, describes the geometrical spreading of acoustic waves with distance. The positive sign in equation (\ref{eq-loc4}) corresponds to $\psi_{\rm out}$ constructed with waves traveling outward from a target point ("egression signal" \cite{Lindsey1997}), while the negative sign provides $\psi_{\rm in}$ constructed with the incoming waves ("ingression signal").

The amplitude and phase of the constructed signals contain information about subsurface perturbation. A practical approach to extract this is to cross-correlate the outgoing and incoming signals \cite{Chen1998,Braun2000}:
\begin{equation}
C(t)=\int \psi_{\rm in}(t')\psi_{\rm out}(t'+t)dt',\label{eq-loc4a}
\end{equation}
and then to measure time shifts of this function for various target positions relative to the corresponding quiet Sun values. These measurements correspond to the travel-time variations obtained by time-distance helioseismology \cite{Chou2000a,Chou2007}. Further analysis of the travel-time variations is similar to the time-distance helioseismology method \cite{Kosovichev1997}. The advantages and disadvantages of the time-distance helioseismology and acoustic holography/imaging are not clear. Both, approaches are being tested using various types of artificial data and applied for measuring subsurface structures and flows. Most of the current inferences of subsurface structures and flows have been obtained using the time-distance approach \cite{Duvall1993,Kosovichev1997}. The time-distance helioseismology method, also called {\it solar tomography} is described in more detail in the following section.

\section{Solar tomography}
\label{tomography}

\subsection{Time-distance diagram}
\begin{figure}
\begin{center}
\includegraphics[width=0.4\linewidth]{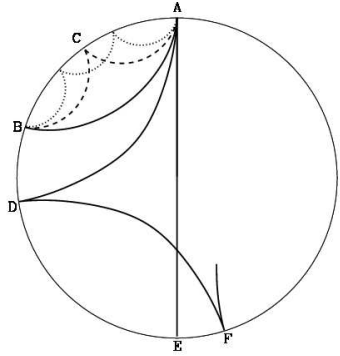}
\caption{A cross-section diagram through the solar interior showing
a sample of wave paths inside the Sun. }\label{fig34}
\end{center}
\end{figure}

Solar acoustic waves (p-modes) are excited by turbulent convection near the solar surface
and travel through the interior with the speed of sound. Because the sound speed
increases with depth the waves are refracted and reappear on the surface at some
distance from the source. The wave propagation is illustrated in Figure~\ref{fig34}.
Waves excited at point A will reappear at the surface points B, C, D, E, F,
and others after propagating along the ray paths indicated by the curves connecting
these points.

The basic idea of {\it time-distance helioseismology}, or {\it helioseismic tomography}, is to measure the
acoustic travel time between different points on the solar surface,
and then to use these measurements for inferring variations of wave-speed perturbations
and flow velocities in the interior by inversion \cite{Duvall1993}. This idea is similar to seismology of  Earth.
However, unlike in Earth, the solar waves are generated stochastically
by numerous acoustic sources in a subsurface layer of turbulent
convection.

\begin{figure}
\begin{center}
\includegraphics[width=\linewidth]{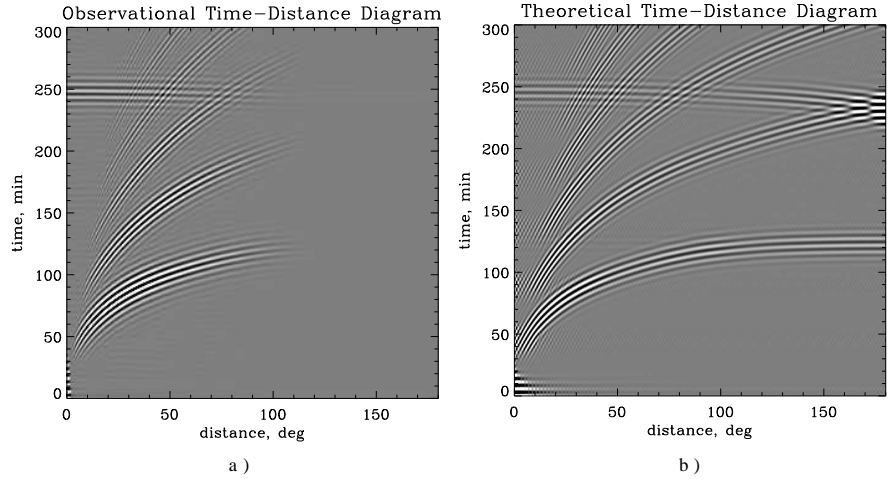}
\caption{The observational (a) and theoretical (b)
cross-covariance functions (time-distance diagrams) as a function of distance on
the solar surface, $\Delta$, and the delay time, $\tau$.
The lowest set of ridges (`first bounce')
corresponds to waves propagated to the
distance, $\Delta$, without additional reflections from the solar surface.
The second from the bottom ridge (`second bounce') is produced by the waves
arriving to the same distance
after one reflection from the surface, and the third ridge (`third bounce')
results from the waves arriving after two bounces from the surface. The backward
ridge at $\tau \approx 250$ min is a continuation of the second-bounce ridge
 due to the choice of the angular
distance range from 0 to 180 degrees (that is, the counterclockwise distance ADF
in Fig.\ref{fig34} is substituted with the clockwise distance AF).
Because of foreshortening close to the
solar limb the observational cross-covariance function covers only $\sim 110$ degrees
of distance.}\label{fig35}
\end{center}
\end{figure}

Therefore, the wave travel time is determined from the cross-covariance
function, $\Psi(\tau, \Delta)$, of the oscillation signal, $f(t,\vecr)$:

\begin{equation}
\Psi(\tau,\Delta) = \int_0^T f(t,\vecr_1) f^*(t+\tau,\vecr_2) dt,\label{eq-tom1}
\end{equation}
where $\Delta$ is the horizontal distance
between two points with coordinates $\vecr_1$ and $\vecr_2$, $\tau$ is the
lag time, and $T$ is the total time of the observations. The normalized cross-covariance function is called cross-correlation. The time-distance analysis is based on non-normalized cross-covariance. Because of the
stochastic nature of solar oscillations, function $\Psi$ must be
averaged over some areas  to achieve a good signal-to-noise
ratio sufficient for measuring the travel times. The oscillation signal,
$f(t,\vecr)$, is  measured from the Doppler shift or intensity of a spectral line.
A typical cross-covariance function obtained from full-disk solar observations of the Doppler shift shown in  Fig.~\ref{fig35}a displays a set
of ridges. The ridges correspond to acoustic wave packets traveling between two points on the surface directly through the interior or with intermediate reflections (bounces) from the surface as illustrated in Figure \ref{fig34}

The waves originated at point A may reach point B directly (solid curve) forming the
first-bounce ridge, or after one bounce at point C (dashed curve) forming the
second-bounce ridge, or after two bounces (dotted curve) - the third-bounce
ridge and so on. Because the sound speed is higher in the deeper layers the
direct waves arrive first, followed by the second-bounce and higher-bounce
waves.

\begin{figure}
\begin{center}
\includegraphics[width=0.7\linewidth]{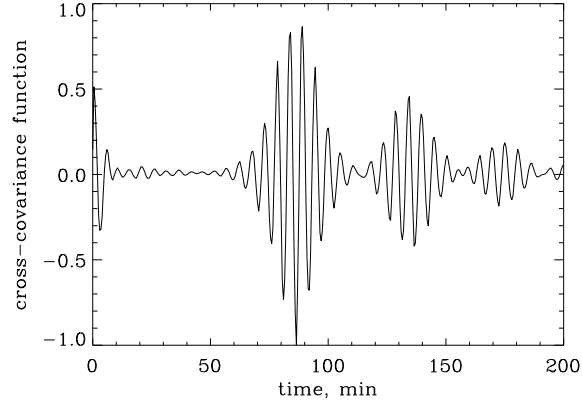}
\caption{The observed cross-covariance signal as a function of time at the distance
of 30 degrees. }\label{fig36}
\end{center}
\end{figure}

The cross-covariance function represents a {\it time-distance diagram}, or a solar `seismogram'.
Figure \ref{fig36} shows the cross-covariance signal as a function of time
for the travel distance, $\Delta$, of 30 degrees. It consists of three wave packets
corresponding to the first, second and third bounces.
Ideally, like in Earth seismology,
the seismogram can be inverted to infer the structure and flows
using a wave theory. However, in practice, modeling the wave fronts is a computationally intensive task. Therefore, the analysis is performed by measuring and inverting the phase and group travel times of the wave packets employing various approximations, the most simple and powerful of
which is the ray-path approximation.

Generally, the observed  solar oscillation signal corresponds to displacement
or pressure perturbation, and can be represented in terms of the normal modes eigenfunctions.  Therefore, the cross-covariance function also can be expressed in
terms of the normal modes. In addition, it can  be represented as a superposition of traveling
wave packets, as we show in the next subsection \cite{Kosovichev1997}.
An example of the theoretical cross-covariance function calculated using normal p-modes of the standard solar model is shown in
Fig.~\ref{fig35}b. This model reproduces the observational cross-covariance
function very well in the observed range of distances, from 0 to 90 degrees.
The theoretical model was calculated for larger distances than the corresponding observational diagram in Fig.~\ref{fig35}a, including points on
the far side of the Sun, which is not accessible for measurements.  A backward
propagating ridge originating from the second-bounce ridge at 180 degrees is a
geometrical effect due to the choice of the range of the angular distance from
0 to 180 degrees.  In the theoretical diagram (Fig.~\ref{fig35}b) one can
notice a very weak backward ridge between 30 and 70 degrees and at 120 min. This
ridge is due to reflection from the boundary between the convection and radiative zones. However, this signal has not
been detected in observations.

\subsection{Wave travel times}

For simplicity we consider solar oscillation signals observed not far from the disk center and describe these in terms of the radial displacement neglecting the horizontal displacement. The general theory was developed by Nigam and Kosovichev \cite{Nigam2010}.
In the simple case, the solar oscillation signal can be represented in terms of the radial eigenfunctions (\ref{eq-osc9a}):
\begin{equation}
f(t,r,\theta,\phi) = \sum_{nlm} a_{nlm}\xi_r^{(n,l,m)}(r,\theta,\phi)
\exp(i\omega_{nlm}t + i\phi_{nlm}),\label{eq-tom2}
\end{equation}
where $n, l$ and $m$ are the radial order, angular degree and
angular order of a normal mode respectively, $\xi_{nlm}(r,\theta,\phi)$
is a mode eigenfunction in the spherical coordinates, $r,\theta$ and $\phi$,
$\omega_{nlm}$ is the eigenfrequency, and $\phi_{nlm}$ is an initial
phase of the mode. Using equation (\ref{eq-tom2}), we calculate
the cross-covariance function, and express it  as a
superposition of traveling wave packets.
Such a representation is important for interpretation of the time-distance
data.
A similar correspondence between the normal modes and the wave packets has been
discussed for surface oscillations in Earth's seismology
\cite{Ben-Menahem1964} and also for ocean waves \cite{Tindle1974}.

To simplify the analysis, we consider the spherically symmetrical case. In this case, the mode eigenfrequencies do not depend on the azimuthal order $m$.
For a radially stratified sphere, the eigenfunctions can be represented
in terms of spherical harmonics $Y_{lm}(\theta,\phi)$ (\ref{eq-osc9a}):
\begin{equation}
\xi_r^{(n,l,m)}(r,\theta,\phi) =  \xi_r^{(n,l)}(r)Y_{lm}(\theta,\phi),\label{eq-tom3}
\end{equation}
where $\xi_r^{(n,l)}(r)$ is the radial eigenfunction \cite{Unno1989}.

Using, the convolution theorem \cite{Bracewell1986} we express the cross-covariance function in terms of a Fourier intergral:
\begin{equation}
\Psi(\tau,\Delta) = \int_{-\infty}^\infty F(\omega,\vecr_1) F^*(\omega,\vecr_2)
\exp(i\omega\tau)d\omega,\label{eq-tom4}
\end{equation}
where $F(\omega,\vecr)$ is Fourier transform of the oscillation signal
$f(t,\vecr)$.

The oscillation signal is considered as band-limited and
filtered to select a p-mode frequency range using a Gaussian
transfer function:
\begin{equation}
G(\omega)=\exp\left[-\frac{1}{2}\left(\frac{\omega-\omega_0}{\delta\omega}\right)^2\right],\label{eq-tom5}
\end{equation}
where $\omega$ is the cyclic frequency, $\omega_0$ is the central frequency
and $\delta\omega$ is the characteristic bandwidth of the filter.
The cross-covariance function in Fig. 1 displays three sets of ridges
which correspond to the first, second
and third bounces of packets of acoustic wave packets from the surface.

The time series used in our analysis are considerably longer than the
travel time $\tau$, therefore, we can neglect the effect of the window function,
and represent $F(\omega,\vecr)$ in the form
\begin{equation}
F(\omega,r,\theta,\phi) \approx A\sum_{nlm} \xi_r^{(n,l)}(r)Y_{lm}(\theta,\phi)
\delta(\omega-\omega_{nl})\exp\left[-\frac{1}{2}\left(\frac{\omega-\omega_0}{\delta\omega}\right)^2\right],
\label{eq-tom6}
\end{equation}
where $\delta(x)$ is the delta-function, $\omega_{nl}$ are frequencies
of the normal modes, and $A$ is the amplitude of the Gaussian envelope of the amplitude spectrum at $\omega=\omega_0$.
In addition, we assume the normalization conditions:  $\xi_r^{(n,l)}(R)=1$,
$a_{nl}=A\,G(\omega)$.
Then, the cross-covariance function is
\begin{equation}
\Psi(\tau,\Delta)=A^2\sum_{nl}\exp\left[-\left(\frac{\omega_{nl}-\omega_0}
{\delta\omega}\right)^2 + i\omega_{nl}\tau\right]\sum_{m=-l}^{l}
Y_{lm}(\theta_1,\phi_1)Y^*_{lm}(\theta_2,\phi_2),\label{eq-tom7}
\end{equation}
where $\theta_1,\phi_1$ and $\theta_2,\phi_2$ are the spherical heliographic coordinates of the two observational points.
The sum of the spherical function products
\begin{equation}
\sum_{m=-l}^{l}
Y_{lm}(\theta_1,\phi_1)Y^*_{lm}(\theta_2,\phi_2)=\alpha_lP_l(\cos\Delta),\label{eq-tom8}
\end{equation}
where $P_l(\cos\Delta)$ is the Legendre polynomial, $\Delta$ is the angular
distance between points 1 and 2 along the great circle on the sphere,
$\cos\Delta=\cos\theta_1\cos\theta_2+\sin\theta_1\sin\theta_2\cos(\phi_2-\phi_1)$,
and $\alpha_l=\sqrt{4\pi/(2l+1)}$.
Then, the cross-covariance function is:
\begin{equation}
\Psi(\tau,\Delta) \approx A^2\sum_{nl}\alpha_lP_l(\cos\Delta)
\exp\left[-\left(\frac{\omega_{nl}-\omega_0}
{\delta\omega}\right)^2+i\omega_{nl}\tau\right].\label{eq-tom9}
\end{equation}
For large values of $l\Delta$, but when $\Delta$ is small,
\begin{equation}
P_l(\cos\Delta)
\simeq \sqrt{\frac{2}{\pi L\Delta}} \cos\left(L\Delta - \frac{\pi}{4}\right).\label{eq-tom10}
\end{equation}
 Thus,
\begin{equation}
\Psi(\tau,\Delta) = A^2\sum_{nl}\frac{2}{L\sqrt{\Delta}}\exp\left[-\frac{(\omega_{nl}-\omega_0)^2}
{\delta\omega^2}\right]\cos(\omega_{nl}\tau)\cos(L\Delta).\label{eq-tom10a}
\end{equation}
Now the double sum can be reduced to a convenient sum of integrals
if we regroup the modes so that the outer sum is over the ratio
$v = \omega_{nl}/L$ and the inner sum is over $\omega_{nl}$.

According to the ray-path theory, the travel distance $\Delta$ of an acoustic
wave is determined by the ratio $v$, which represent the horizontal angular phase velocity ($v=\omega_{nl}/L\equiv (\omega_{nl}/k_h)/r$). Because of the band-limited nature of the function $G$, only values of $L$ which are close to
$L_0 \equiv \omega_0 / v$ contribute to the sum. We consider the relation $L$ vs $\omega_{nl}$ as a continuous function along the mode ridges (Fig.~\ref{fig3}), and expand $L$ near the central frequency $\omega_0$:
\begin{equation}
L \simeq  L_0 +
\frac{\partial L}{\partial\omega_{nl}}(\omega_{nl} - \omega_0)  =
\frac{\omega_0}{v} + \frac{\omega_{nl} - \omega_0}{u},\label{eq-tom11}
\end{equation}
where $u \equiv \partial\omega_{nl}/\partial L$. Furthermore,
\begin{equation}
\cos(\omega_{nl})\tau)\cos(L\Delta)=\cos\left[\left(\tau - \frac{\Delta}{u}\right)\omega_{nl} +
\left(\frac{1}{u} - \frac{1}{v}\right)\Delta\omega_0\right],\label{eq-tom12}
\end{equation}
and the other term is identical except that $\tau$ has been replaced
with $-\tau$ (negative time lag).  The result is that the double sum
in equation~(\ref{eq-tom12}) becomes
\begin{equation}
\Psi(\tau,\Delta) \simeq A^2\sum_{v} \frac{2}{L_0\sqrt{\Delta}}
\sum_{\omega_{nl}} \exp\left[-\frac{(\omega-\omega_0)^2}
{\delta\omega^2}\right] \cos\left[\left(\pm\tau - \frac{\Delta}{u}\right)
+ \left(\frac{1}{u} - \frac{1}{v}\right)\Delta\omega_0 \right].\label{eq-tom13}
\end{equation}

The inner sum can be approximated by an integral, considering $\omega_{nl}$ as a continuous variable along the mode ridges:
\begin{eqnarray}
\lefteqn{\int_{-\infty}^{\infty} d\omega\,\exp\left[-\frac{(\omega-\omega_0)^2}
{\delta\omega^2}\right] \cos\left[\left(\tau -
\frac{\Delta}{u}\right)\omega - \left(\frac{1}{u} - \frac{1}{v}\right)\Delta\omega_0 \right] = }
\nonumber\\
& &
\sqrt{\pi\,\delta\omega^2}\exp\left[-\frac{\delta\omega^2}{4}\left(\tau -
\frac{\Delta}{u}\right)^2\right] \cos\left[\omega_0 \left(\tau -
\frac{\Delta}{v}\right)\right].\label{eq-tom13a}
\end{eqnarray}
The integration limits reflect the fact that the amplitude
function $G(\omega)$ is essentially zero for very large and very small frequencies.
Finally,  the cross-covariance is expressed in the following form \cite{Kosovichev1997}:
\begin{equation}
\Psi(\tau,\Delta) = B\sum_{v} \cos\left[\omega_0\left(\tau-
\tau_{\rm ph}\right)\right]\exp\left[-\frac{\delta\omega^2}{4}\left(\tau-
\tau_{\rm gr}\right)^2\right],\label{eq-tom14}
\end{equation}
where $B$ is constant, $\tau_{\rm ph}=\Delta/v$ and $\tau_{\rm gr}=\Delta/u$ are the phase and group travel times. Equation (\ref{eq-tom14}) has the form of a Gabor wavelet. The  phase and group travel times are measured by
fitting individual terms of equation ({\ref{eq-tom14}}) to the observed  cross-covariance function using a least-squares technique.

\subsection{Deep- and surface-focus measurement schemes}

\begin{figure}[h]
\begin{center}
\includegraphics[width=0.9\linewidth]{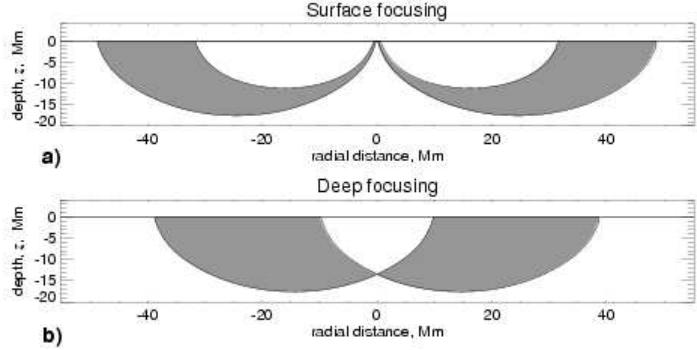}
\caption{The regions of ray propagation (shaded areas) as a function of depth,
$z$,
and the radial distance, $\Delta$, from a point on the surface for two
observing schemes: `surface focusing' (a) and `deep focusing'
(b). The
rays are also averaged over a circular regions on the surface, forming
three-dimensional figures of revolution.}\label{fig37}
\end{center}
\end{figure}

As we have pointed out the travel-time measurements require averaging of the cross-covariance function in order to obtain a good signal-to-noise ratio.
Two typical schemes of the spatial averaging suggested by Duvall \cite{Duvall1995} are shown in Fig.~\ref{fig37}.

 For the so-called `surface-focusing' scheme
(Fig.\ref{fig37}a) the measured travel times are mostly sensitive to the near
surface condition at the central point where the ray paths are focused.
However, by measuring the travel times for several distances and applying an
inversion procedure it is possible to infer the distribution of the variations
of the wave speed and flow velocities with depth.
The averaging also can be done in such a way that the `focus' point is located beneath
the surface. An example of the `deep-focusing' scheme is shown
in Fig.\ref{fig37}b.  In this case the travel times are more sensitive to deep
structures but still inversions are required for correct interpretation.

\subsection{Sensitivity kernels: Ray-path approximation}

The travel-time inversion procedures are based on theoretical relations between the travel-time variations and interior properties constituting the forward problem of local helioseismology. Similarly to global helioseismology, these relations are expressed in the form linear integral equations with sensitivity
kernels. Two basic types of the sensitivity kernels have been used: ray-path kernels  \cite{Kosovichev1997} and Born-approximation kernels \cite{Birch2000,Birch2004,Couvidat2006}. The ray-path kernels are based on a simple and generally robust theoretical ray approximation, but they do not take into account finite wavelength effects and thus are not sufficiently accurate for diagnostics of small-scale structures. For reliable inferences it is important to use both these kernels.

 In the ray approximation, the travel times are
sensitive only to the perturbations along the ray paths  given by
Hamilton's equations (\ref{eq-osc22}). The variations of the phase travel time obey
the Fermat's Principle:
\begin{equation}
\delta\tau = \frac{1}{\omega}\int_\Gamma \delta\veck d\vecr,\label{eq-tom15}
\end{equation}
where $\delta\veck$ is the perturbation of the wave vector, $\veck$, due to the
structural inhomogeneities and flows along the unperturbed ray path, $\Gamma$.
Using the dispersion relation for acoustic waves in the convection zone
the travel-time variations can be expressed in terms of the sound-speed,
magnetic field strength and flow velocity.

The dispersion relation for magnetoacoustic waves in the
convection zone is
\begin{equation}
(\omega - \veck\cdot\vecU)^2 = \omega_c^2 + k^2 c_f^2,\label{eq-tom16}
\end{equation}
where $\vecU$ is the flow velocity, $\omega_c$ is the
acoustic cut-off frequency, $\;\;c_f^2=\frac{1}{2}
\left(c^2+c_A^2+\sqrt{\left(c^2+c_A^2\right)^2-
4c^2(\veck\cdot\veccA)^2/k^2}\right)$ is the fast magnetoacoustic
speed, $\veccA = \vecB/\sqrt{4\pi\rho}\;\;$ is the vector Alfv\'en
velocity, $\vecB$ is the magnetic field strength, $c$ is the
adiabatic sound speed, and $\rho$ is the plasma density. If we
assume that, in the unperturbed state $\vecU=\vecB=0$, then, to
the first-order approximation
\begin{equation}
\delta\tau=-\int_\Gamma\left[\frac{(\vecn\cdot\vecU)}{c^2} + \frac{\delta c}{c}S
+ \left(\frac{\delta\omega_c}{\omega_c}\right)
\frac{\omega_c^2}{\omega^2c^2S} +\frac{1}{2}\left(\frac{c_A^2}{c^2}
-\frac{(\veck\cdot\veccA)^2}{k^2c^2}\right)S\right]ds,\;\;\;\label{eq-tom17}
\end{equation}
where $\vecn$ is a unit vector tangent to the ray, $S=k/\omega$ is the
phase slowness.

Then, we separate the effects of flows and structural
perturbations by measuring the travel times of acoustic waves traveling in opposite directions along the same ray path, and calculating the difference, $\tau_{\rm diff}$  and the mean, $\tau_{\rm mean}$, of these reciprocal travel times:
\begin{equation}
\delta\tau_{\rm diff} = - 2\int_\Gamma\frac{(\vecn\cdot\vecU)}{c^2}ds;\\\label{eq-tom18}
\end{equation}
\begin{equation}
\delta\tau_{\rm mean} =-\int_\Gamma\left[\frac{\delta c}{c}S +
\left(\frac{\delta\omega_c}{\omega_c}\right)
\frac{\omega_c^2}{\omega^2c^2S} +\frac{1}{2}\left(\frac{c_A^2}{c^2}
-\frac{(\veck\cdot\veccA)^2}{k^2c^2}\right)S\right]ds.\label{eq-tom19}
\end{equation}
Anisotropy of the last term of equation (\ref{eq-tom19}) allows us to separate, at least
partly, the magnetic effects from the variations of the
sound speed and the acoustic cut-off frequency. The acoustic cut-off frequency,
$\omega_c$ may be perturbed by surface magnetic fields and by
temperature and density inhomogeneities. The effect of the
cut-off frequency variation depends strongly on the wave frequency,
and, therefore, it results in a frequency dependence in $\tau_{\rm mean}$.

In practice, the travel times are measured for from the cross-covariance functions between selected central  points on the solar surface and surrounding quadrants symmetrical
relative to the North, South, East and West directions.
In each quadrant, the travel times are averaged over narrow ranges of
the travel distance, $\Delta$.
The travel times of the northward-directed waves are subtracted from
the times of the south-directed waves to yield the time, $\tau_{\rm diff}^{\rm NS}$, which is predominantly sensitive to subsurface north-south flows. Similarly, the time
differences, $\tau_{\rm diff}^{\rm EW}$, between westward- and eastward directed
waves yields a measure of the east-ward flows.
The time,  $\tau_{\rm diff}^{\rm oi}$, between the outward- and inward-directed waves, averaged over the full annuli, is mainly sensitive to vertical flows and divergence of the horizontal flows. This represents a cross-talk effect between the vertical flows and horizontal flows, which is difficult to resolve when the vertical flows are weak \cite{Zhao2003a}.

Thus, the effects of flows and structural
perturbations are separated from each other
by taking the difference and the mean of the reciprocal travel
times:
\begin{equation}
\delta\tau_{\rm diff} \approx - 2\int_\Gamma\frac{(\vecn\vecU)}{c^2}ds;\\\label{eq-tom20}
\end{equation}
\begin{equation}
\delta\tau_{\rm mean} \approx -\int_\Gamma\frac{\delta w}{c}Sds,\label{eq-tom21}
\end{equation}
where $c$ is the adiabatic sound speed,
$\vecn$ is a unit vector tangent to the ray, $S=k/\omega$ is the
phase slowness, $\delta w$ is the local wave speed perturbation:
\begin{equation}
\frac{\delta w}{c}=\frac{\delta c}{c} +\mbox{$\frac{1}{2}$}\left(\frac{c_A^2}{c^2}-\frac{(\veck\veccA)^2}{k^2c^2}\right).\label{eq-tom22}
\end{equation}
Magnetic field causes anisotropy of the mean travel times, which
 allows us to separate, in
principle, the magnetic effects from the variations of the
sound speed (or temperature). So far, only a combined effect of the magnetic
fields and temperature variations has been measured reliably.

\subsection{Born approximation}

The development of a more accurate theory for the travel times,
based on the Born approximation is currently under way \cite{Birch2000,Birch2001,Gizon2001,Birch2004,Couvidat2006}.

\begin{figure}
\begin{center}
\includegraphics[width=0.75\linewidth]{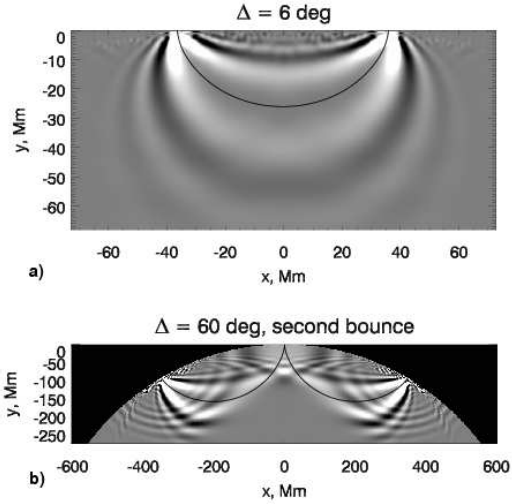}
\caption{Travel-time sensitivity kernels
in the first Born approximation  for sound-speed variations
as a function of the horizontal, $x$, and vertical, $y$, coordinates
for: a) the first-bounce signal for distance $\Delta=6$ degrees, b)
the second-bounce signal for $\Delta=60$ degrees. The solid curves show
the corresponding ray paths at frequency $\nu=3$ mHz \cite{Kosovichev2003}.
 }\label{fig38}
\end{center}
\end{figure}

One unexpected feature of the single-source travel-time kernels calculated in the Born
approximation is that these kernels have zero value along the ray path
(called `banana-doughnut kernels').
Examples of the Born kernels for the first and the second bounces are shown
in Fig.\ref{fig38}. The kernels are mostly sensitive to perturbations
within the first Fresnel zone.

\begin{figure}
\begin{center}
\includegraphics[width=0.65\linewidth]{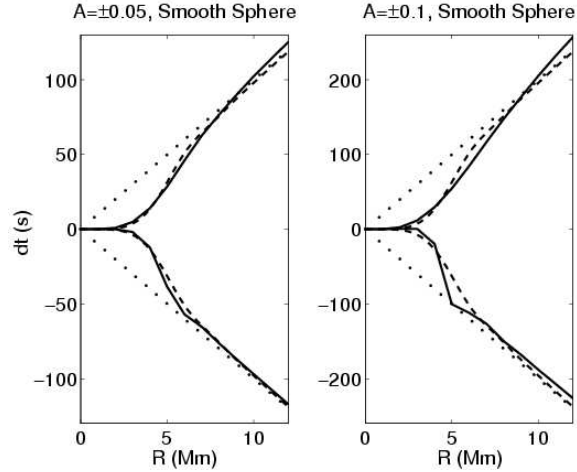}
\caption{Tests of the ray and Born approximations:
travel times for smooth spheres as functions of sphere radius at half
maximum.  The solid lines are the numerical results.
The dashed curves are the Born approximation
travel times and the dotted lines are the first order ray approximation.  The left
panel shows the two perturbations of the relative amplitude,
$A=\pm0.05$. The right panel is for the cases $A=\pm0.1$. \cite{Birch2001}}
\label{fig39}
\end{center}
\end{figure}

Figure \ref{fig39} shows the test results for both  the ray and Born
approximations for a simple model of a smooth sphere in an uniform medium
by comparing with precise numerical results \cite{Birch2001}.
These results show that for typical perturbations in the solar interior
the Born approximation is sufficiently accurate, while the ray approximation
significantly overestimates the travel times for perturbations smaller than
the size of the first Fresnel zone. That means that the inversion results
based on the ray theory may underestimate the strength of
the small-scale perturbations. The comparison of the inversion results for sub-surface sound-speed structures beneath sunspots have showed a very good agreement between the ray-paths and Born theories \cite{Couvidat2006}.

\section{Inversion results of solar acoustic tomography}

The results of test inversions (e.g. \cite{Kosovichev1997,Zhao2003a,Zhao2007})
demonstrate an accurate reconstruction of sound-speed variations and
the horizontal
components of subsurface flows. However, vertical flows in deep layers
are not resolved because of the predominantly horizontal propagation
of the rays in these layers. The vertical velocities are also systematically
underestimated in the upper layers. When the vertical flow is weak, e.g. such as in supergranulation, the vertical velocity is not estimated correctly, because the trave-time signal is dominated by the horizontal flow divergence. In such situation, it is difficult to determine even the direction of the vertical flow \cite{Zhao2007}.  Similarly, the sound-speed
variations are underestimated in the deep layers and close to the surface.
These limitations of the solar tomography should be taken into account
in interpretation of the inversion results.

Here, I briefly present some examples of the local helioseismology inferences obtained by inversion of acoustic travel times.


\subsection{Diagnostics of supergranulation.}

\begin{figure}
\begin{center}
\includegraphics[width=\linewidth]{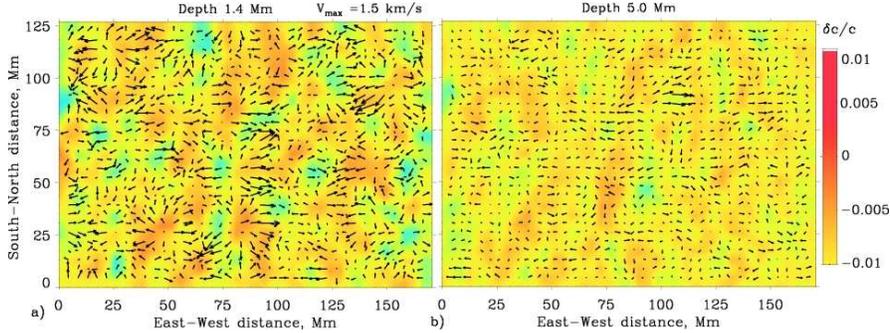}
\caption{The supergranulation horizontal flow velocity field (arrows)
and the sound-speed perturbation  (color background) at the depths
of 1.4 Mm (a) and 5.0 Mm (b), as inferred from
the SOHO/MDI high-resolution data of
27 January 1996. \cite{Kosovichev1997}}\label{fig40}
\end{center}
\end{figure}

The data used were for 8.5 hours on 27 January, 1996 from the
high resolution mode of the MDI instrument.
 The results of inversion of these data are shown in Figure \ref{fig40} \cite{Kosovichev1997}.
It has been found that, in the upper layers, 2-3 Mm deep, the horizontal flow
is organized in supergranular cells, with outflows from the center of
the supergranules. The characteristic size of the cells is 20-30 Mm.
Comparing with MDI magnetograms, it was found  that the cell
boundaries coincide with the areas of enhanced magnetic field.
These results are consistent with the observations of supergranulation
on the solar surface.
 However, in the layers deeper than
$\sim 5$ Mm, the supergranulation pattern disappears.
The inversions show an evidence of reverse converging flows
at the depth of $\sim 10$ Mm \cite{Zhao2003a}. This means that supergranulation is a relatively shallow phenomenon.

\subsection{Structure and dynamics of sunspot}

\begin{figure}
\begin{center}
\includegraphics[width=\linewidth]{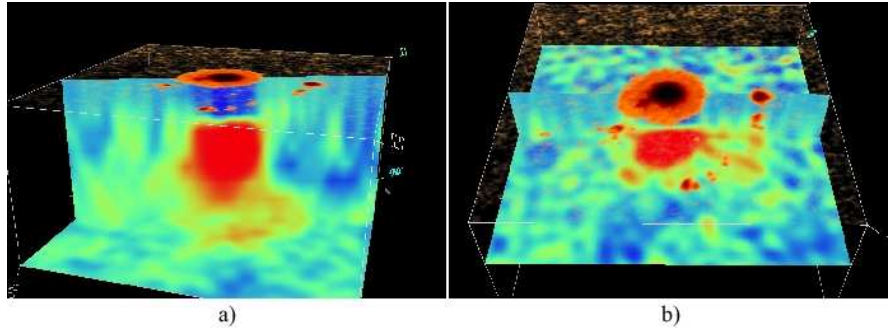}
\caption{The sound-speed perturbation in a large sunspot observed
on June 20, 1998, are shown as vertical and horizontal cuts.
The horizontal size of the box is 13 degrees (158 Mm), the depth is 24 Mm.
The positive variations of the sound speed are shown in red, and the negative variations (just beneath the sunspot)are in blue. The upper
semitransparent panel is the surface intensity image (dark color shows
umbra, and light color shows penumbra). In panel b) the horizontal
sound-speed plane is located at the depth of 4 Mm, and shows long narrow
structures (`fingers') connecting the main sunspot structure with surrounding
pores  the same magnetic polarity as the spot \cite{Kosovichev2000}.
}\label{fig41}
\end{center}
\end{figure}

The high-resolution data from the SOHO and Hinode space missions have allowed us to investigate the structure and dynamics beneath sunspots.
Figure \ref{fig41} shows an example of the internal structure of a large
sunspot observed on June 17, 1998 \cite{Kosovichev2000}. An image of the spot taken in
the continuum is shown at the top. The wave-speed perturbations
under the sunspot are much stronger than these of the emerging flux, and can reach
$\sim 3$ km/s. It is interesting that beneath the spot
the perturbation is negative in the subsurface layers and becomes
positive in the deeper interior.
One can suggest that the negative perturbations beneath the spot are,
probably, due to the lower temperature.
It follows that
magnetic inhibition of convection that makes sunspots cooler
is most effective  within the top 2-3 Mm of the convection zone.
The strong positive perturbation below suggests that the deep sunspot
structure is hotter than the surrounding plasma.
 However, the effects of temperature and magnetic field have not been separated
in these inversions. Separating these effects
 is an important problem of solar tomography.
These data also show at a depth of $\sim 4$ Mm connections
to the spot of small pores, which have the same magnetic polarity
as the main spot. The pores of the opposite polarity are not connected to the main sunspot.
This suggests that sunspots represent a tree-like structure in the
upper convection zone.

\begin{figure}
\begin{center}
\includegraphics[width=0.7\linewidth]{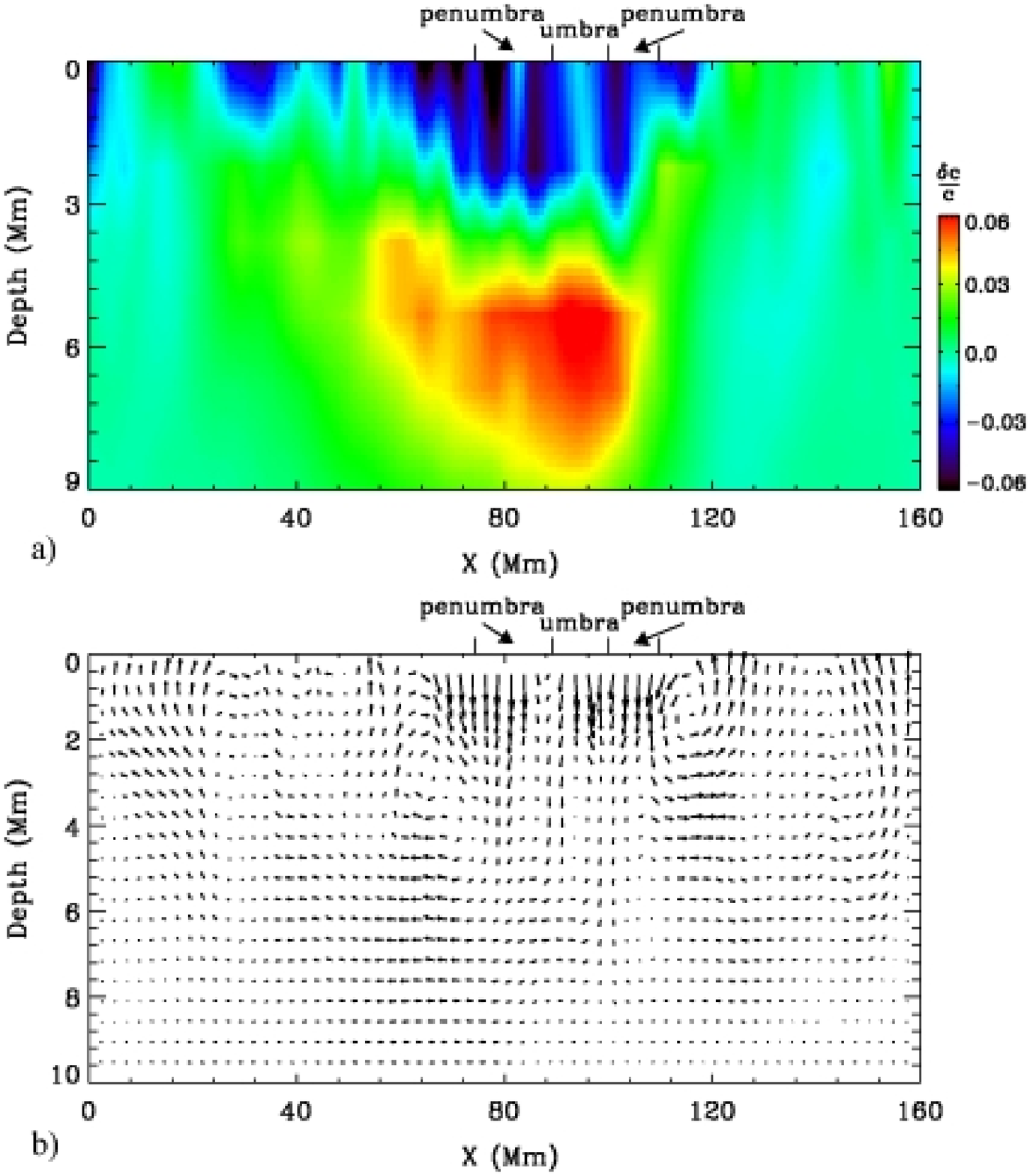}
\caption{Wave speed perturbation  and flow velocities beneath
sunspots from Hinode data\cite{Zhao2010}} \label{fig42}
\end{center}
\end{figure}
Figure~\ref{fig42} shows the subsurface structures and flows beneath
a sunspot obtained from Hinode \cite{Zhao2010}. A vertical cut  along
the East-West direction approximately in the middle of a large
sunspot observed in AR 10953, May 2, 2007, (Fig.~\ref{fig42}a), shows that
the wave speed anomalies extend about half of the sunspot size
beyond the sunspot penumbra into the plage area. In the vertical
direction, the negative wave speed perturbation extends to a depth
of 3--4 Mm. The positive perturbation is about 9 Mm deep, but it is
not clear whether it extends further, because our inversion cannot
reach deeper layers because of the small field of view. Similar
two-layer sunspot structures were observed before from SOHO/MDI
\cite{Kosovichev2000}(Fig.~\ref{fig41}). But, it is striking that the new
images strongly indicate on the cluster structure of the sunspot \cite{Parker1979}.
This was not previously seen in the tomographic images of sunspots
obtained with lower resolution.

The high-resolution flow field below
the sunspot is also significantly more complicated than the
previously inferred from SOHO/MDI \cite{Zhao2001}, but
reveals the same general converging downdraft pattern. A vertical
view of an averaged flow field (Fig.~\ref{fig42}b) shows nicely the flow
structure beneath the active region. Strong downdrafts are seen
immediately below the sunspot's surface, and extends up to 6 Mm in
depth. A little beyond the sunspot's boundary, one can find both
upward and inward flows. Clearly, large-scale mass circulations form
outside the sunspot, bringing plasma down along the sunspot's
boundary, and back to the photosphere within about twice of the
sunspot's radius. It is remarkable that such an apparent mass
circulation is obtained directly from the helioseismic inversions
without using any additional constraints, such as forced mass
conservation. Previously, the circulation pattern was not that
clear.
\begin{figure}
\begin{center}
\includegraphics[width=\linewidth]{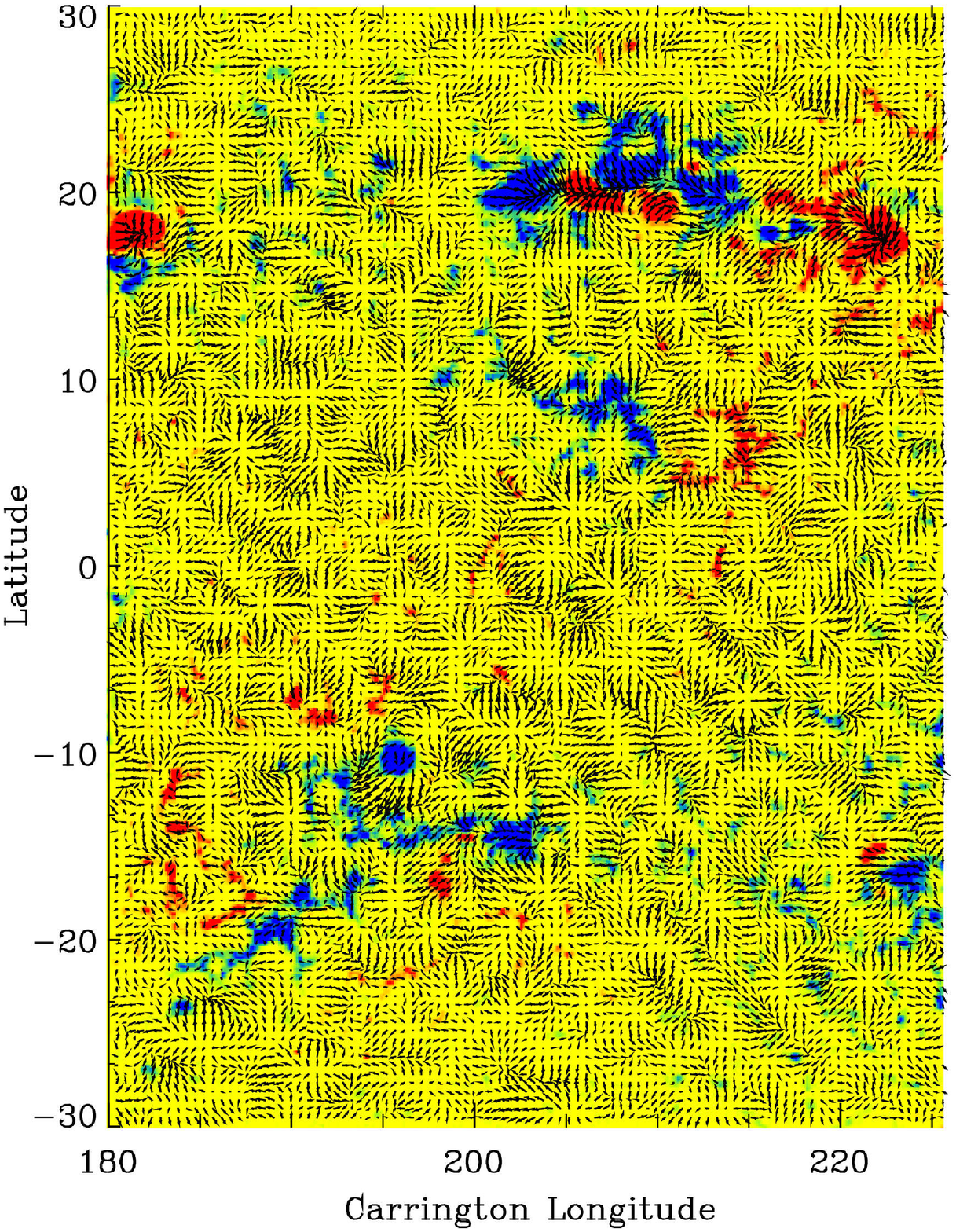}
\caption{A portion of a synoptic subsurface flow map at depth of 2 Mm. The color background shows the distribution of magnetic field on the surface \cite{Zhao2004}.
\label{fig43}}
\end{center}
\end{figure}

\subsection{Large-scale and meridional flows}

\begin{figure}
\begin{center}
\includegraphics[width=\linewidth]{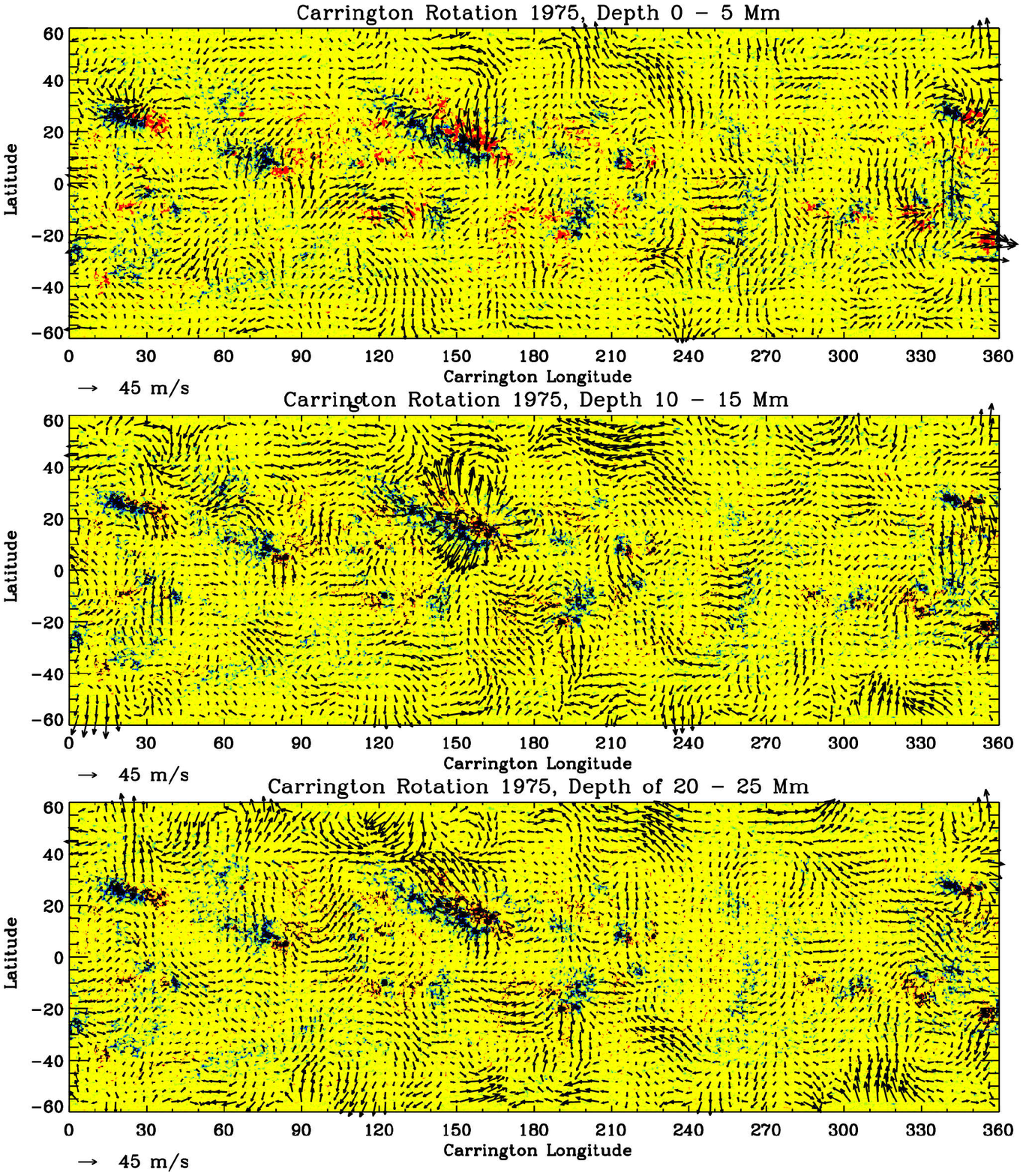}
\caption{Subsurface synoptic flow maps at three depths. The color background shows the distribution of magnetic field on the surface \cite{Zhao2004}.}
\label{fig44}
\end{center}
\end{figure}

Time-distance helioseismology \cite{Zhao2004} and also local
measurements of the p-mode frequency shifts by the `ring-diagram'
analysis \cite{Haber2002,Haber2000,Haber2004}, have provided  synoptic maps of
subsurface flows over the whole surface of the Sun. Figure~\ref{fig43} shows a portion
of a high-resolution synoptic flow map at the depth of 2 Mm below the surface.
In addition, to the supergranulation pattern these maps reveal large-scale converging plasma flow around the active regions where  magnetic field is concentrated. These flows are particularly well visible in low-resolution synoptic flow maps (Fig.~\ref{fig44}).  The characteristic speed of these flows is about 50 m/s.

\begin{figure}
\begin{center}
\includegraphics[width=\linewidth]{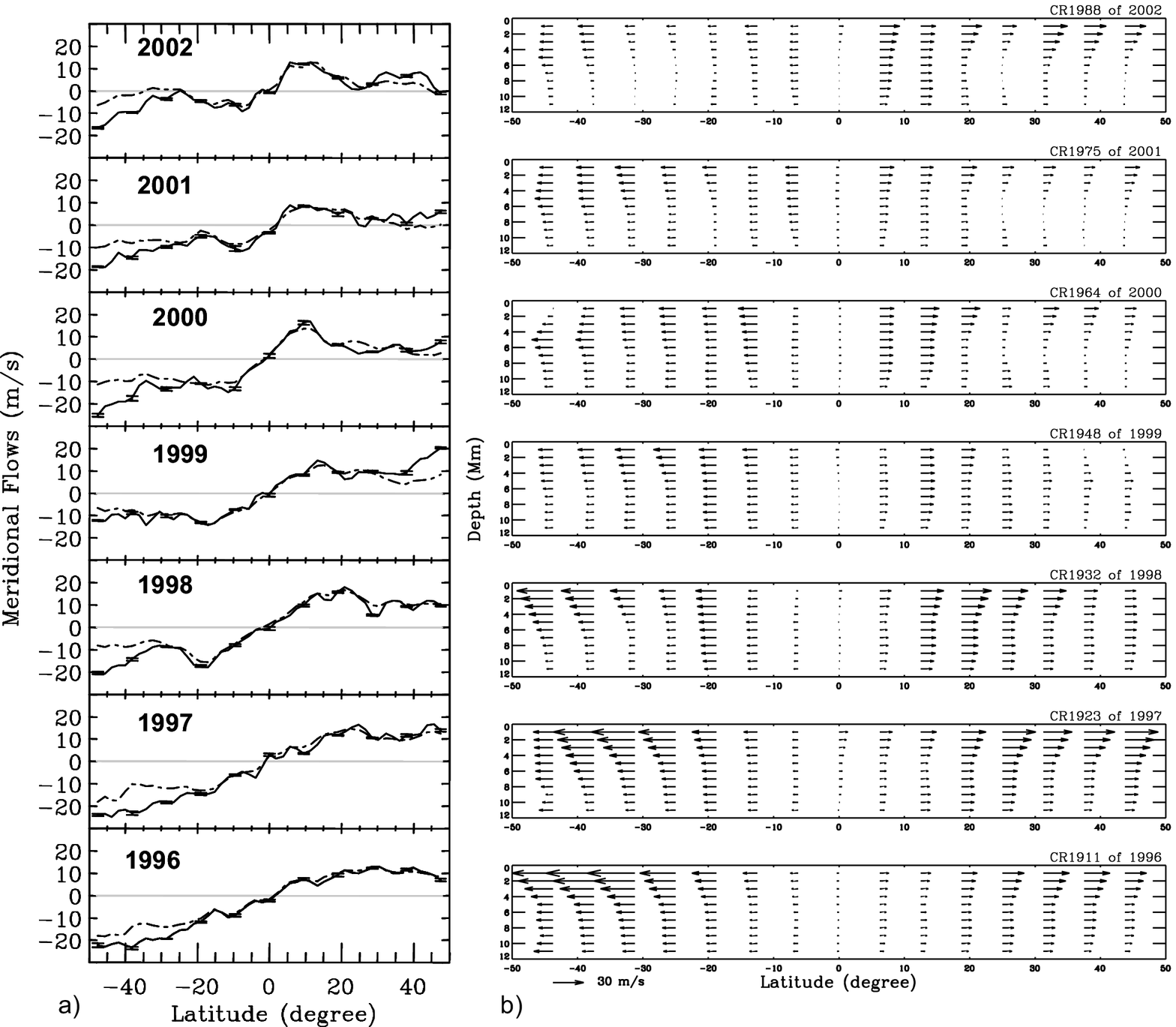}
\caption{Evolution of subsurface meridional flow during 1996-2002 for various
    Carrington rotations \cite{Zhao2004}.}
\label{fig45}
\end{center}
\end{figure}

These stable
long-living flow patterns affect the global circulation in the Sun.
It is particularly important that these flows change the mean
meridional flow from the equator to the poles, slowing it down
during the solar maximum years (Fig.~\ref{fig45}). This may have
important consequences for the solar dynamo theories which invoke
the meridional flow to explain the magnetic flux transport into the
polar regions and the polar magnetic field polarity reversals
usually happening during the period of maximum of solar activity.

\section{Conclusion and outlook}

During the past decade thanks to the long-term continuous
observations from the ground and space the physics of solar
oscillations made a tremendous progress in understanding the mechanism of solar oscillations, and in developing new techniques for
helioseismic diagnostics of the solar structure and dynamics.
However, many problems are still unresolved. Most of them are
related to phenomena in strong magnetic field regions and in the deep interior.
The prime helioseismology tasks are to detect processes
of magnetic field generation and transport in the solar interior,
and formation of active regions and sunspots. This will be help to
understand the physics of the solar dynamo and the cyclic behavior
of solar activity.

For solving these tasks it is very important to continue developing
realistic MHD simulations of solar convection and oscillations and
to obtain continuous high-resolution helioseismology data for the
whole Sun. The recent observations from Hinode have convincingly
demonstrated advantages of high-resolution helioseismology, but
unfortunately such data are available only for small regions and for
short periods of time. A new substantial progress in observations of
solar oscillations is expected from the Solar Dynamics Observatory
(SDO) space mission  launched in February 2010.

The
Helioseismic and Magnetic Imager (HMI) instrument on SDO will
provide uninterrupted Doppler shift measurements over the whole
visible disk of the Sun with a spatial resolution of 0.5 arcsec per
pixel ($4096\times 4096$ images) and 40-50 sec time cadence. The
total amount of data from this instrument will reach 2 Tb per day.
This tremendous amount of data will be processed through a specially
developed data analysis pipeline and will provide high-resolution
maps of subsurface flows and sound-speed structures
\cite{Kosovichev2007a}. These data will enable investigations of the
multi-scale dynamics and magnetism of the Sun and also contribute to
our understanding of the Sun as a star.

The tools that will be used in the HMI program include:
helioseismology to map and probe the solar convection zone where a
magnetic dynamo likely generates this diverse range of activity;
measurements of the photospheric magnetic field which results from
the internal processes and drives the processes in the atmosphere;
and brightness measurements which can reveal the relationship
between magnetic and convective processes and solar irradiance
variability.

Helioseismology, which uses solar oscillations to probe flows and
structures in the solar interior, is providing remarkable new
perspectives about the complex interactions between highly turbulent
convection, rotation and magnetism. It has revealed a region of
intense rotational shear at the base of the convection zone, called
the tachocline, which is the likely seat of the global dynamo.
Convective flows also have a crucial role in advecting and shearing
the magnetic fields, twisting the emerging flux tubes and displacing
the photospheric footpoints of magnetic structures present in the
corona. Flows of all spatial scales influence the evolution of the
magnetic fields, including how the fields generated near the base of
the convection zone rise and emerge at the solar surface, and how
the magnetic fields already present at the surface are advected and
redistributed. Both of these mechanisms contribute to the
establishment of magnetic field configurations that may become
unstable and lead to eruptions that affect the near-Earth
environment.

New methods of local-area helioseismology have begun to reveal the
great complexity of rapidly evolving 3-D magnetic structures and
flows in the sub-surface shear layer in which the sunspots and
active regions are embedded. Most of these new techniques were
developed  during analysis of MDI observations. As useful as they
are, the limitations of MDI telemetry availability and the limited
field of view at high resolution has prevented the full exploitation
of the methods to answer the important questions about the origins
of solar variability. By using these techniques on continuous,
full-disk, high-resolution observations, HMI will enable detailed
probing of dynamics and magnetism within the near-surface shear
layer, and provide sensitive measures of variations in the
tachocline.

\begin{figure}
\begin{center}
\includegraphics[width=\linewidth]{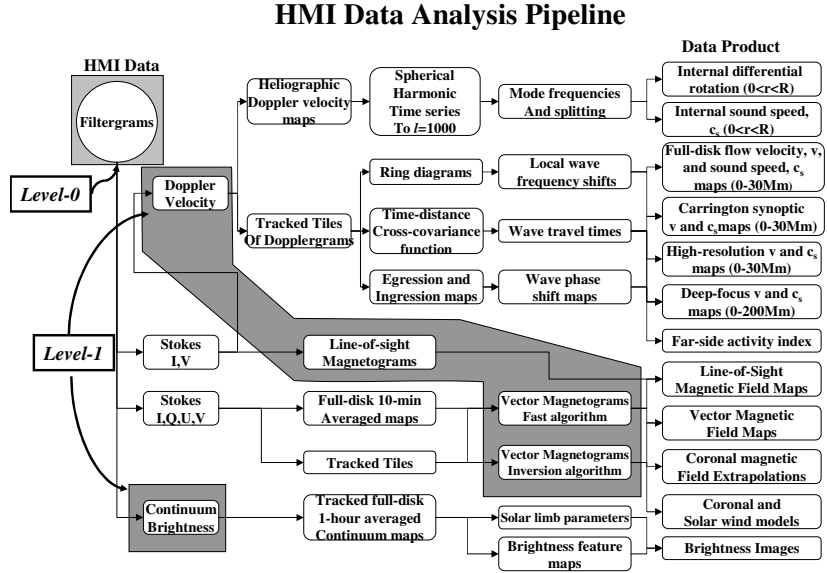}
\caption{A schematic illustration of the Solar Dynamics Observatory
HMI data analysis pipeline and data products.
The dark shaded area indicates Level-1 data
products. The boxes to the right of this area represent intermediate
and final Level-2 data products. The data products are described in
detail in the HMI Science Plan \cite{Kosovichev2007a}.}
\label{fig46}
\end{center}
\end{figure}

The scientific operation modes and data products can be divided into
four main areas: global helioseismology, local-area helioseismology,
line-of-sight and vector magnetography and continuum intensity
studies. The principal data flows and products are summarized in
Figure~\ref{fig46}.

\paragraph{Global Helioseismology:Diagnostics of global changes
inside the Sun. }
The traditional normal-mode method described in Sec.~\ref{global}-\ref{inverse},
will provide
large-scale axisymmetrical distributions of sound speed, density,
adiabatic exponent and flow velocities through the whole solar
interior from the energy-generating core to the near-surface
convective boundary layer. These diagnostics will be based on
frequencies and frequency splitting of modes of angular degree up to
1000, obtained for several day intervals each month and up to
$l$=300 for each 2-month interval. These will be used to produce a
regular sequence of internal rotation and sound-speed inversions to
allow observation of the tachocline and average near surface shear.

\paragraph{Local-Area Helioseismology: 3D imaging of the solar interior.
} The new methods of local-area helioseismology (Sec.~\ref{local}-ref{tomography}), time-distance technique,
ring-diagram analysis and acoustic holography represent powerful
tools for investigating physical processes inside the Sun. These
methods on measuring local properties of acoustic and surface
gravity waves, such as travel times, frequency and phase shifts.
They will provide images of internal structures and flows on various
spatial and temporal scales and depth resolution. The targeted
high-level regular data products include:
\begin{itemize}
\item Full-disk velocity and sound-speed maps of the upper convection zone (covering the top 30 Mm) obtained every 8 hours with the time-distance methods on a Carrington grid;
\item Synoptic maps of mass flows and sound-speed perturbations in the upper convection zone for each Carrington rotation with a 2-degree resolution, from averages of full disk time-distance maps;
\item Synoptic maps of horizontal flows in upper convection zone for each Carrington rotation with a 5 degree resolution from ring-diagram analyses.
\item Higher-resolution maps zoomed on particular active regions, sunspots and other targets, obtained with 4-8-hour resolution for up to 9 days continuously, from the time-distance method;
\item Deep-focus maps covering the whole convection zone depth, 0-200 Mm, with 10-15 degree resolution;
\item Far-side images of the sound-speed perturbations associated with large active regions every 24 hours.
\end{itemize}

The HMI science investigation addresses the
fundamental problems of solar variability with studies in all
interlinked time and space domains, including global scale, active
regions, small scale, and coronal connections. One of the prime
objectives of the Living With a Star program is to understand how
well predictions of evolving space weather variability can be made.
The HMI investigation will examine these questions in parallel with
the fundamental science questions of how the Sun varies and how that
variability drives global change and space weather.

\section*{Acknowledgment}
This work was supported by the CNRS, the
International Space Science Institute (Bern), Nordita (Stockholm) and NASA.


\begin{thebibliography}{100}
\providecommand{\url}[1]{{#1}}
\providecommand{\urlprefix}{URL }
\expandafter\ifx\csname urlstyle\endcsname\relax
  \providecommand{\doi}[1]{DOI \discretionary{}{}{}#1}\else
  \providecommand{\doi}{DOI \discretionary{}{}{}\begingroup
  \urlstyle{rm}\Url}\fi

\bibitem{Eddington1926}
A.S. {Eddington}, \emph{{The Internal Constitution of the Stars}} (1926)

\bibitem{Leighton1962}
R.B. {Leighton}, R.W. {Noyes}, G.W. {Simon}, \apj \textbf{135}, 474 (1962)

\bibitem{Mein1966}
P.~{Mein}, Annales d'Astrophysique \textbf{29}, 153 (1966)

\bibitem{Frazier1968}
E.N. {Frazier}, \apj \textbf{152}, 557 (1968)

\bibitem{Ulrich1970}
R.K. {Ulrich}, \apj \textbf{162}, 993 (1970)

\bibitem{Deubner1975}
F.~{Deubner}, \aap \textbf{44}, 371 (1975)

\bibitem{Rhodes1977}
E.J. {Rhodes}, Jr., R.K. {Ulrich}, G.W. {Simon}, \apj \textbf{218}, 901 (1977)

\bibitem{Ando1977}
H.~{Ando}, Y.~{Osaki}, \pasj \textbf{29}, 221 (1977)

\bibitem{Bahcall1969}
J.N. {Bahcall}, N.A. {Bahcall}, R.K. {Ulrich}, \apj \textbf{156}, 559 (1969)

\bibitem{Hill1975}
H.A. {Hill}, R.T. {Stebbins}, T.M. {Brown}, in \emph{Bulletin of the American
  Astronomical Society}, \emph{Bulletin of the American Astronomical Society},
  vol.~7 (1975), \emph{Bulletin of the American Astronomical Society}, vol.~7,
  p. 478

\bibitem{Severny1976}
A.B. {Severny}, V.A. {Kotov}, T.T. {Tsap}, \nat \textbf{259}, 87 (1976)

\bibitem{Brookes1976}
J.R. {Brookes}, G.R. {Isaak}, H.B. {van der Raay}, \nat \textbf{259}, 92 (1976)

\bibitem{Scherrer1979}
P.H. {Scherrer}, J.M. {Wilcox}, V.A. {Kotov}, A.B. {Severny}, T.T. {Tsap}, \nat
  \textbf{277}, 635 (1979)

\bibitem{Grec1980}
G.~{Grec}, E.~{Fossat}, M.~{Pomerantz}, \nat \textbf{288}, 541 (1980)

\bibitem{Pall'e1998}
P.L. {Pall{\'e}}, T.~{Roca Cort{\'e}s}, B.~{Gelly}, {the GOLF Team}, in
  \emph{Structure and Dynamics of the Interior of the Sun and Sun-like Stars},
  \emph{ESA Special Publication}, vol. 418, ed. by {S.~Korzennik} (1998),
  \emph{ESA Special Publication}, vol. 418, pp. 291

\bibitem{Claverie1979}
A.~{Claverie}, G.R. {Isaak}, C.P. {McLeod}, H.B. {van der Raay}, T.R. {Cortes},

\bibitem{Vandakurov1968}
Y.V. {Vandakurov}, Soviet Astronomy \textbf{11}, 630 (1968)

\bibitem{Iben1976}
I.~{Iben}, Jr., J.~{Mahaffy}, \apjl \textbf{209}, L39 (1976)

\bibitem{Christensen-Dalsgaard1979}
J.~{Christensen-Dalsgaard}, D.O. {Gough}, J.G. {Morgan}, \aap \textbf{73}, 121
  (1979)

\bibitem{Christensen-Dalsgaard1981}
J.~{Christensen-Dalsgaard}, D.O. {Gough}, \aap \textbf{104}, 173 (1981)

\bibitem{Duvall1983}
T.L. {Duvall}, Jr., J.W. {Harvey}, \nat \textbf{302}, 24 (1983)

\bibitem{Ahmad2002a}
Q.R. {Ahmad}, {et al.}, Physical Review Letters \textbf{89}(1), 011301 (2002)

\bibitem{Pekeris1938}
C.L. {Pekeris}, \apj \textbf{88}, 189 (1938)

\bibitem{Cowling1941}
T.G. {Cowling}, \mnras \textbf{101}, 367 (1941)

\bibitem{Ledoux1958}
P.~{Ledoux}, T.~{Walraven}, Handbuch der Physik \textbf{51}, 353 (1958)

\bibitem{Gough1990}
D.O. {Gough}, M.J. {Thompson}, \mnras \textbf{242}, 25 (1990)

\bibitem{Dziembowski1984}
W.~{Dziembowski}, P.R. {Goode}, Memorie della Societa Astronomica Italiana
  \textbf{55}, 185 (1984)

\bibitem{Dziembowski1989}
W.A. {Dziembowski}, P.R. {Goode}, \apj \textbf{347}, 540 (1989)

\bibitem{Dziembowski2005}
W.A. {Dziembowski}, P.R. {Goode}, \apj \textbf{625}, 548 (2005)

\bibitem{Rhodes1979}
E.J. {Rhodes}, Jr., R.K. {Ulrich}, F.~{Deubner}, \apj \textbf{227}, 629 (1979)

\bibitem{Ulrich1979}
R.K. {Ulrich}, E.J. {Rhodes}, Jr., F.~{Deubner}, \apj \textbf{227}, 638 (1979)

\bibitem{Deubner1979}
F.~{Deubner}, R.K. {Ulrich}, E.J. {Rhodes}, Jr., \aap \textbf{72}, 177 (1979)

\bibitem{Duvall1984}
T.L. {Duvall}, Jr., J.W. {Harvey}, \nat \textbf{310}, 19 (1984)

\bibitem{Duvall1984a}
T.L. {Duvall}, Jr., W.A. {Dziembowski}, P.R. {Goode}, D.O. {Gough}, J.W.
  {Harvey}, J.W. {Leibacher}, \nat \textbf{310}, 22 (1984)

\bibitem{Brown1987}
T.M. {Brown}, C.A. {Morrow}, \apjl \textbf{314}, L21 (1987)

\bibitem{Kosovichev1988}
A.G. {Kosovichev}, Soviet Astronomy Letters \textbf{14}, 145 (1988)

\bibitem{Brown1989}
T.M. {Brown}, J.~{Christensen-Dalsgaard}, W.A. {Dziembowski}, P.~{Goode}, D.O.
  {Gough}, C.A. {Morrow}, \apj \textbf{343}, 526 (1989)

\bibitem{Rosner1985}
R.~{Rosner}, N.O. {Weiss}, \nat \textbf{317}, 790 (1985)

\bibitem{Parker1993}
E.N. {Parker}, \apj \textbf{408}, 707 (1993)

\bibitem{Harvey1988}
J.W. {Harvey}, K.~{Abdel-Gawad}, W.~{Ball}, B.~{Boxum}, F.~{Bull}, J.~{Cole},
  L.~{Cole}, S.~{Colley}, K.~{Dowdney}, R.~{Drake}, in \emph{Seismology of the
  Sun and Sun-Like Stars}, \emph{ESA Special Publication}, vol. 286, ed. by
  {E.~J.~Rolfe} (1988), \emph{ESA Special Publication}, vol. 286, pp. 203--208

\bibitem{Brookes1978}
J.R. {Brookes}, G.R. {Isaak}, H.B. {van der Raay}, \mnras \textbf{185}, 1
  (1978)

\bibitem{Isaak1989}
G.R. {Isaak}, C.P. {McLeod}, P.L. {Palle}, H.B. {van der Raay}, T.~{Roca
  Cortes}, \aap \textbf{208}, 297 (1989)

\bibitem{Domingo1995}
V.~{Domingo}, B.~{Fleck}, A.I. {Poland}, \solphys \textbf{162}, 1 (1995)

\bibitem{Scherrer1995}
P.H. {Scherrer}, R.S. {Bogart}, R.I. {Bush}, J.T. {Hoeksema}, A.G.
  {Kosovichev}, J.~{Schou}, W.~{Rosenberg}, L.~{Springer}, T.D. {Tarbell},
  A.~{Title}, C.J. {Wolfson}, I.~{Zayer}, {MDI Engineering Team}, \solphys
  \textbf{162}, 129 (1995)

\bibitem{Leibacher2003}
J.~{Leibacher}, in \emph{IAU Joint Discussion}, \emph{IAU Joint Discussion},
  vol.~12 (2003), \emph{IAU Joint Discussion}, vol.~12

\bibitem{Gough1983}
D.O. {Gough}, J.~{Toomre}, \solphys \textbf{82}, 401 (1983)

\bibitem{Hill1988}
F.~{Hill}, \apj \textbf{333}, 996 (1988)

\bibitem{Duvall1993}
T.L. {Duvall}, Jr., S.M. {Jefferies}, J.W. {Harvey}, M.A. {Pomerantz}, \nat
  \textbf{362}, 430 (1993)

\bibitem{Kosovichev1996}
A.G. {Kosovichev}, \apjl \textbf{469}, L61 (1996)

\bibitem{Kosovichev1997}
A.G. {Kosovichev}, T.L. {Duvall}, Jr., in \emph{SCORe'96 : Solar Convection and
  Oscillations and their Relationship}, \emph{Astrophysics and Space Science
  Library}, vol. 225, ed. by F.P. {Pijpers}, J.~{Christensen-Dalsgaard}, C.S.
  {Rosenthal} (1997), \emph{Astrophysics and Space Science Library}, vol. 225,
  pp. ~241--260

\bibitem{Chang1997}
H.~{Chang}, D.~{Chou}, B.~{Labonte}, {The TON Team}, \nat \textbf{389}, 825
  (1997)

\bibitem{Lindsey2000}
C.~{Lindsey}, D.C. {Braun}, \solphys \textbf{192}, 261 (2000)

\bibitem{Kosovichev2007a}
A.G. {Kosovichev}, {HMI Science Team}, Astronomische Nachrichten \textbf{328},
  339 (2007)

\bibitem{Stein1989}
R.F. {Stein}, A.~{Nordlund}, \apjl \textbf{342}, L95 (1989)

\bibitem{Zhao2007}
J.~{Zhao}, D.~{Georgobiani}, A.G. {Kosovichev}, D.~{Benson}, R.F. {Stein},
  {\AA}.~{Nordlund}, \apj \textbf{659}, 848 (2007)

\bibitem{Jacoutot2008}
L.~{Jacoutot}, A.G. {Kosovichev}, A.A. {Wray}, N.N. {Mansour}, \apj
  \textbf{682}, 1386 (2008)

\bibitem{Hanasoge2007}
S.M. {Hanasoge}, T.L. {Duvall}, Jr., S.~{Couvidat}, \apj \textbf{664}, 1234
  (2007)

\bibitem{Parchevsky2008}
K.V. {Parchevsky}, J.~{Zhao}, A.G. {Kosovichev}, \apj \textbf{678}, 1498 (2008)

\bibitem{Hartlep2008}
T.~{Hartlep}, J.~{Zhao}, N.N. {Mansour}, A.G. {Kosovichev}, \apj \textbf{689},
  1373 (2008)

\bibitem{Rimmele1995}
T.R. {Rimmele}, P.R. {Goode}, E.~{Harold}, R.T. {Stebbins}, \apjl \textbf{444},
  L119 (1995)

\bibitem{Skartlien2000}
R.~{Skartlien}, M.P. {Rast}, \apj \textbf{535}, 464 (2000)

\bibitem{Stein2001}
R.F. {Stein}, {\AA}.~{Nordlund}, \apj \textbf{546}, 585 (2001)

\bibitem{Germano1991}
M.~{Germano}, U.~{Piomelli}, P.~{Moin}, W.H. {Cabot}, Physics of Fluids
  \textbf{3}, 1760 (1991)

\bibitem{Moin1991}
P.~{Moin}, K.~{Squires}, W.~{Cabot}, S.~{Lee}, Physics of Fluids \textbf{3},
  2746 (1991)

\bibitem{Baudin2005}
F.~{Baudin}, R.~{Samadi}, M.~{Goupil}, T.~{Appourchaux}, C.~{Barban},
  P.~{Boumier}, W.J. {Chaplin}, P.~{Gouttebroze}, \aap \textbf{433}, 349 (2005)

\bibitem{Jacoutot2008a}
L.~{Jacoutot}, A.G. {Kosovichev}, A.~{Wray}, N.N. {Mansour}, \apjl
  \textbf{684}, L51 (2008)

\bibitem{Duvall1993a}
T.L. {Duvall}, Jr., S.M. {Jefferies}, J.W. {Harvey}, Y.~{Osaki}, M.A.
  {Pomerantz}, \apj \textbf{410}, 829 (1993)

\bibitem{Toutain1998}
T.~{Toutain}, T.~{Appourchaux}, C.~{Fr{\"o}hlich}, A.G. {Kosovichev},
  R.~{Nigam}, P.H. {Scherrer}, \apjl \textbf{506}, L147 (1998)

\bibitem{Gabriel1992}
M.~{Gabriel}, \aap \textbf{265}, 771 (1992)

\bibitem{Nigam1998}
R.~{Nigam}, A.G. {Kosovichev}, P.H. {Scherrer}, J.~{Schou}, \apjl \textbf{495},
  L115+ (1998)

\bibitem{Mitra-Kraev2008}
U.~{Mitra-Kraev}, A.G. {Kosovichev}, T.~{Sekii}, \aap \textbf{481}, L1 (2008)

\bibitem{Roxburgh1997}
I.W. {Roxburgh}, S.V. {Vorontsov}, \mnras \textbf{292}, L33 (1997)

\bibitem{Severino2001}
G.~{Severino}, M.~{Magr{\`i}}, M.~{Oliviero}, T.~{Straus}, S.M. {Jefferies},
  \apj \textbf{561}, 444 (2001)

\bibitem{Wachter2005}
R.~{Wachter}, A.G. {Kosovichev}, \apj \textbf{627}, 550 (2005)

\bibitem{Georgobiani2003}
D.~{Georgobiani}, R.F. {Stein}, {\AA}.~{Nordlund}, \apj \textbf{596}, 698
  (2003)

\bibitem{Kosugi2007}
T.~{Kosugi}, K.~{Matsuzaki}, T.~{Sakao}, T.~{Shimizu}, Y.~{Sone},
  S.~{Tachikawa}, T.~{Hashimoto}, K.~{Minesugi}, A.~{Ohnishi}, T.~{Yamada},
  S.~{Tsuneta}, H.~{Hara}, K.~{Ichimoto}, Y.~{Suematsu}, M.~{Shimojo},
  T.~{Watanabe}, S.~{Shimada}, J.M. {Davis}, L.D. {Hill}, J.K. {Owens}, A.M.
  {Title}, J.L. {Culhane}, L.K. {Harra}, G.A. {Doschek}, L.~{Golub}, \solphys
  \textbf{243}, 3 (2007)

\bibitem{Tsuneta2008}
S.~{Tsuneta}, K.~{Ichimoto}, Y.~{Katsukawa}, S.~{Nagata}, M.~{Otsubo},
  T.~{Shimizu}, Y.~{Suematsu}, M.~{Nakagiri}, M.~{Noguchi}, T.~{Tarbell},
  A.~{Title}, R.~{Shine}, W.~{Rosenberg}, C.~{Hoffmann}, B.~{Jurcevich},
  G.~{Kushner}, M.~{Levay}, B.~{Lites}, D.~{Elmore}, T.~{Matsushita},
  N.~{Kawaguchi}, H.~{Saito}, I.~{Mikami}, L.D. {Hill}, J.K. {Owens}, \solphys
  \textbf{249}, 167 (2008)

\bibitem{Nigam1998a}
R.~{Nigam}, A.G. {Kosovichev}, \apjl \textbf{505}, L51+ (1998)

\bibitem{Duvall1998}
T.L. {Duvall}, Jr., A.G. {Kosovichev}, K.~{Murawski}, \apjl \textbf{505}, L55+
  (1998)

\bibitem{Murawski1998}
K.~{Murawski}, T.L. {Duvall}, Jr., A.G. {Kosovichev}, in \emph{Structure and
  Dynamics of the Interior of the Sun and Sun-like Stars}, \emph{ESA Special
  Publication}, vol. 418, ed. by {S.~Korzennik} (1998), \emph{ESA Special
  Publication}, vol. 418, pp. 825

\bibitem{Braun1987}
D.C. {Braun}, T.L. {Duvall}, Jr., B.J. {Labonte}, \apjl \textbf{319}, L27
  (1987)

\bibitem{Cally2009}
P.S. {Cally}, \mnras \textbf{395}, 1309 (2009)

\bibitem{Parchevsky2007a}
K.V. {Parchevsky}, A.G. {Kosovichev}, \apjl \textbf{666}, L53 (2007)

\bibitem{Parchevsky2010}
K.~{Parchevsky}, A.~{Kosovichev}, E.~{Khomenko}, V.~{Olshevsky}, M.~{Collados},
  ArXiv e-prints  (2010)

\bibitem{Nagashima2007}
K.~{Nagashima}, T.~{Sekii}, A.G. {Kosovichev}, H.~{Shibahashi}, S.~{Tsuneta},
  K.~{Ichimoto}, Y.~{Katsukawa}, B.~{Lites}, S.~{Nagata}, T.~{Shimizu}, R.A.
  {Shine}, Y.~{Suematsu}, T.D. {Tarbell}, A.M. {Title}, \pasj \textbf{59}, 631
  (2007)

\bibitem{Schussler2006}
M.~{Sch{\"u}ssler}, A.~{V{\"o}gler}, \apjl \textbf{641}, L73 (2006)

\bibitem{Braun1992}
D.C. {Braun}, C.~{Lindsey}, Y.~{Fan}, S.M. {Jefferies}, \apj \textbf{392}, 739
  (1992)

\bibitem{Hanasoge2008}
S.M. {Hanasoge}, \apj \textbf{680}, 1457 (2008)

\bibitem{Khomenko2009}
E.~{Khomenko}, M.~{Collados}, \aap \textbf{506}, L5 (2009)

\bibitem{Kosovichev1998}
A.G. {Kosovichev}, V.V. {Zharkova}, \nat \textbf{393}, 317 (1998)

\bibitem{Kosovichev2006a}
A.G. {Kosovichev}, in \emph{Solar MHD Theory and Observations: A High Spatial
  Resolution Perspective}, \emph{Astronomical Society of the Pacific Conference
  Series}, vol. 354, ed. by {J.~Leibacher, R.~F.~Stein, \& H.~Uitenbroek}
  (2006), \emph{Astronomical Society of the Pacific Conference Series}, vol.
  354, pp. 154

\bibitem{Kosovichev2006b}
A.G. {Kosovichev}, in \emph{Proceedings of SOHO 18/GONG 2006/HELAS I, Beyond
  the spherical Sun}, \emph{ESA Special Publication}, vol. 624 (2006),
  \emph{ESA Special Publication}, vol. 624

\bibitem{Kosovichev2006c}
A.G. {Kosovichev}, \solphys \textbf{238}, 1 (2006)

\bibitem{Donea1999}
A.~{Donea}, D.C. {Braun}, C.~{Lindsey}, \apjl \textbf{513}, L143 (1999)

\bibitem{Donea2005}
A.~{Donea}, C.~{Lindsey}, \apj \textbf{630}, 1168 (2005)

\bibitem{Donea2006}
A.~{Donea}, D.~{Besliu-Ionescu}, P.S. {Cally}, C.~{Lindsey}, V.V. {Zharkova},
  \solphys \textbf{239}, 113 (2006)

\bibitem{Kosovichev2007}
A.G. {Kosovichev}, \apjl \textbf{670}, L65 (2007)

\bibitem{Tassoul1980}
M.~{Tassoul}, \apj Suppl. \textbf{43}, 469 (1980)

\bibitem{Gough1993}
D.O. {Gough}, in \emph{Astrophysical Fluid Dynamics - Les Houches 1987} (1993),
  pp. 399--560

\bibitem{Duvall1982}
T.L. {Duvall}, Jr., \nat \textbf{300}, 242 (1982)

\bibitem{Christensen-Dalsgaard1996}
J.~{Christensen-Dalsgaard}, W.~{Dappen}, S.V. {Ajukov}, E.R. {Anderson}, H.M.
  {Antia}, S.~{Basu}, V.A. {Baturin}, G.~{Berthomieu}, B.~{Chaboyer}, S.M.
  {Chitre}, A.N. {Cox}, P.~{Demarque}, J.~{Donatowicz}, W.A. {Dziembowski},
  M.~{Gabriel}, D.O. {Gough}, D.B. {Guenther}, J.A. {Guzik}, J.W. {Harvey},
  F.~{Hill}, G.~{Houdek}, C.A. {Iglesias}, A.G. {Kosovichev}, J.W. {Leibacher},
  P.~{Morel}, C.R. {Proffitt}, J.~{Provost}, J.~{Reiter}, E.J. {Rhodes}, Jr.,
  F.J. {Rogers}, I.W. {Roxburgh}, M.J. {Thompson}, R.K. {Ulrich}, Science
  \textbf{272}, 1286 (1996)

\bibitem{Christensen-Dalsgaard1988}
J.~{Christensen-Dalsgaard}, D.O. {Gough}, M.J. {Thompson}, in \emph{Seismology
  of the Sun and Sun-Like Stars}, \emph{ESA Special Publication}, vol. 286, ed.
  by {E.~J.~Rolfe} (1988), \emph{ESA Special Publication}, vol. 286, pp.
  493--497

\bibitem{Gough1986}
D.O. {Gough}, in \emph{NATO ASIC Proc. 169: Seismology of the Sun and the
  Distant Stars}, ed. by {D.~O.~Gough} (1986), pp. 125--140

\bibitem{Kosovichev1988e}
A.G. {Kosovichev}, K.V. {Parchevskii}, Soviet Astronomy Letters \textbf{14},
  201 (1988)

\bibitem{Duvall1988}
T.L. {Duvall}, Jr., J.W. {Harvey}, K.G. {Libbrecht}, B.D. {Popp}, M.A.
  {Pomerantz}, \apj \textbf{324}, 1158 (1988)

\bibitem{Christensen-Dalsgaard1982}
J.~{Christensen-Dalsgaard}, \mnras \textbf{199}, 735 (1982)

\bibitem{Schou1997}
J.~{Schou}, A.G. {Kosovichev}, P.R. {Goode}, W.A. {Dziembowski}, \apjl
  \textbf{489}, L197+ (1997)

\bibitem{Rozelot2004}
J.P. {Rozelot}, S.~{Lefebvre}, S.~{Pireaux}, A.~{Ajabshirizadeh}, \solphys
  \textbf{224}, 229 (2004)

\bibitem{Lefebvre2005}
S.~{Lefebvre}, A.G. {Kosovichev}, \apjl \textbf{633}, L149 (2005)

\bibitem{Lefebvre2007}
S.~{Lefebvre}, A.G. {Kosovichev}, J.P. {Rozelot}, \apjl \textbf{658}, L135
  (2007)

\bibitem{Dziembowski1990}
W.A. {Dziembowski}, A.A. {Pamyatnykh}, R.~{Sienkiewicz}, \mnras \textbf{244},
  542 (1990)

\bibitem{Gough1988}
D.O. {Gough}, A.G. {Kosovichev}, in \emph{Seismology of the Sun and Sun-Like
  Stars}, \emph{ESA Special Publication}, vol. 286, ed. by {E.~J.~Rolfe}
  (1988), \emph{ESA Special Publication}, vol. 286, pp. 195--201

\bibitem{Gough1990a}
D.O. {Gough}, A.G. {Kosovichev}, in \emph{IAU Colloq. 121: Inside the Sun},
  \emph{Astrophysics and Space Science Library}, vol. 159, ed. by
  {G.~Berthomieu \& M.~Cribier} (1990), \emph{Astrophysics and Space Science
  Library}, vol. 159, pp. 327

\bibitem{Kosovichev1999}
A.G. {Kosovichev}, Journal of Computational and Applied Mathematics
  \textbf{109}, 1 (1999)

\bibitem{Backus1968}
G.E. {Backus}, J.F. {Gilbert}, Geophysical Journal \textbf{16}, 169 (1968)

\bibitem{Tikhonov1977}
V.Y. Tikhonov, A. N.;~Arsenin, \emph{Solution of Ill-posed Problems}
  (Washington: Winston \& Sons, 1977)

\bibitem{Asplund2009}
M.~{Asplund}, N.~{Grevesse}, A.J. {Sauval}, P.~{Scott}, Ann. Rev. Astron.
  Astrophys. \textbf{47}, 481 (2009)

\bibitem{Bahcall2005}
J.N. {Bahcall}, A.M. {Serenelli}, S.~{Basu}, \apjl \textbf{621}, L85 (2005)

\bibitem{Elliott1998}
J.R. {Elliott}, A.G. {Kosovichev}, \apjl \textbf{500}, L199+ (1998)

\bibitem{Dappen2007}
W.~{D{\"a}ppen}, in \emph{Solar and Stellar Physics Through Eclipses},
  \emph{Astronomical Society of the Pacific Conference Series}, vol. 370, ed.
  by {O.~Demircan, S.~O.~Selam, \& B.~Albayrak} (2007), \emph{Astronomical
  Society of the Pacific Conference Series}, vol. 370, pp. 3

\bibitem{Lynden-Bell1967}
D.~{Lynden-Bell}, J.P. {Ostriker}, \mnras \textbf{136}, 293 (1967)

\bibitem{Schou1998}
J.~{Schou}, H.M. {Antia}, S.~{Basu}, R.S. {Bogart}, R.I. {Bush}, S.M. {Chitre},
  J.~{Christensen-Dalsgaard}, M.P. {di Mauro}, W.A. {Dziembowski},
  A.~{Eff-Darwich}, D.O. {Gough}, D.A. {Haber}, J.T. {Hoeksema}, R.~{Howe},
  S.G. {Korzennik}, A.G. {Kosovichev}, R.M. {Larsen}, F.P. {Pijpers}, P.H.
  {Scherrer}, T.~{Sekii}, T.D. {Tarbell}, A.M. {Title}, M.J. {Thompson},
  J.~{Toomre}, \apj \textbf{505}, 390 (1998)

\bibitem{Appourchaux2009}
T.~{Appourchaux}, P.~{Liewer}, M.~{Watt}, D.~{Alexander}, V.~{Andretta},
  F.~{Auch{\`e}re}, P.~{D'Arrigo}, J.~{Ayon}, T.~{Corbard}, S.~{Fineschi},
  W.~{Finsterle}, L.~{Floyd}, G.~{Garbe}, L.~{Gizon}, D.~{Hassler}, L.~{Harra},
  A.~{Kosovichev}, J.~{Leibacher}, M.~{Leipold}, N.~{Murphy}, M.~{Maksimovic},
  V.~{Martinez-Pillet}, B.S.A. {Matthews}, R.~{Mewaldt}, D.~{Moses},
  J.~{Newmark}, S.~{R{\'e}gnier}, W.~{Schmutz}, D.~{Socker}, D.~{Spadaro},
  M.~{Stuttard}, C.~{Trosseille}, R.~{Ulrich}, M.~{Velli}, A.~{Vourlidas}, C.R.
  {Wimmer-Schweingruber}, T.~{Zurbuchen}, Experimental Astronomy \textbf{23},
  1079 (2009)

\bibitem{DSilva1993}
S.~{D'Silva}, R.F. {Howard}, \solphys \textbf{148}, 1 (1993)

\bibitem{Brandenburg2005}
A.~{Brandenburg}, \apj \textbf{625}, 539 (2005)

\bibitem{Benevolenskaya1999}
E.E. {Benevolenskaya}, J.T. {Hoeksema}, A.G. {Kosovichev}, P.H. {Scherrer},
  \apjl \textbf{517}, L163 (1999)

\bibitem{Howe2008}
R.~{Howe}, Advances in Space Research \textbf{41}, 846 (2008)

\bibitem{Vorontsov2002}
S.V. {Vorontsov}, J.~{Christensen-Dalsgaard}, J.~{Schou}, V.N. {Strakhov}, M.J.
  {Thompson}, Science \textbf{296}, 101 (2002)

\bibitem{Howard1980}
R.~{Howard}, B.J. {Labonte}, \apjl \textbf{239}, L33 (1980)

\bibitem{Kosovichev1997b}
A.G. {Kosovichev}, J.~{Schou}, \apjl \textbf{482}, L207+ (1997)

\bibitem{Howe2000}
R.~{Howe}, J.~{Christensen-Dalsgaard}, F.~{Hill}, R.W. {Komm}, R.M. {Larsen},
  J.~{Schou}, M.J. {Thompson}, J.~{Toomre}, Science \textbf{287}, 2456 (2000)

\bibitem{Rempel2007}
M.~{Rempel}, \apj \textbf{655}, 651 (2007)

\bibitem{Spruit2003}
H.C. {Spruit}, \solphys \textbf{213}, 1 (2003)

\bibitem{Haber2002}
D.A. {Haber}, B.W. {Hindman}, J.~{Toomre}, R.S. {Bogart}, R.M. {Larsen},
  F.~{Hill}, \apj \textbf{570}, 855 (2002)

\bibitem{Basu2004}
S.~{Basu}, H.M. {Antia}, R.S. {Bogart}, \apj \textbf{610}, 1157 (2004)

\bibitem{Gough1976}
D.~Gough,   (1976)

\bibitem{Haber2000}
D.A. {Haber}, B.W. {Hindman}, J.~{Toomre}, R.S. {Bogart}, M.J. {Thompson},
  F.~{Hill}, \solphys \textbf{192}, 335 (2000)

\bibitem{Haber2004}
D.A. {Haber}, B.W. {Hindman}, J.~{Toomre}, M.J. {Thompson}, \solphys
  \textbf{220}, 371 (2004)

\bibitem{Komm2008}
R.~{Komm}, S.~{Morita}, R.~{Howe}, F.~{Hill}, \apj \textbf{672}, 1254 (2008)

\bibitem{Dikpati2006}
P.~Dikpati, M.;~Giman,   (2006)

\bibitem{Hindman2006}
B.W. {Hindman}, D.A. {Haber}, J.~{Toomre}, \apj \textbf{653}, 725 (2006)

\bibitem{Claerbout1968}
J.F. {Claerbout}, Geophysics \textbf{33}, 264 (1968)

\bibitem{Rickett2000}
J.E. {Rickett}, J.F. {Claerbout}, \solphys \textbf{192}, 203 (2000)

\bibitem{Lindsey1997}
C.~{Lindsey}, D.C. {Braun}, \apj \textbf{485}, 895 (1997)

\bibitem{Chou1999}
D.~{Chou}, H.~{Chang}, M.~{Sun}, B.~{Labonte}, H.~{Chen}, S.~{Yeh}, {The TON
  Team}, \apj \textbf{514}, 979 (1999)

\bibitem{Chen1998}
H.~{Chen}, D.~{Chou}, H.~{Chang}, M.~{Sun}, S.~{Yeh}, B.~{Labonte}, {The TON
  Team}, \apjl \textbf{501}, L139 (1998)

\bibitem{Braun2000}
D.C. {Braun}, C.~{Lindsey}, \solphys \textbf{192}, 307 (2000)

\bibitem{Chou2000a}
D.~{Chou}, T.L. {Duvall}, Jr., \apj \textbf{533}, 568 (2000)

\bibitem{Chou2007}
D.~{Chou}, in \emph{New Solar Physics with Solar-B Mission}, \emph{Astronomical
  Society of the Pacific Conference Series}, vol. 369, ed. by {K.~Shibata,
  S.~Nagata, \& T.~Sakurai} (2007), \emph{Astronomical Society of the Pacific
  Conference Series}, vol. 369, pp. 313

\bibitem{Nigam2010}
R.~{Nigam}, A.G. {Kosovichev}, \apj \textbf{708}, 1475 (2010)

\bibitem{Ben-Menahem1964}
A.~Ben-Menahem, Bulletin of the Seismological Society of America \textbf{54},
  1351 (1964)

\bibitem{Tindle1974}
C.~{Tindle}, K.~{Guthrie}, Journal of Sound Vibration \textbf{34}, 291 (1974)

\bibitem{Unno1989}
W.~{Unno}, Y.~{Osaki}, H.~{Ando}, H.~{Saio}, H.~{Shibahashi}, \emph{{Nonradial
  oscillations of stars}} (1989)

\bibitem{Bracewell1986}
R.N. {Bracewell}, \emph{{The Fourier Transform and its applications}} (1986)

\bibitem{Duvall1995}
T.L. {Duvall}, Jr., in \emph{GONG 1994. Helio- and Astro-Seismology from the
  Earth and Space}, \emph{Astronomical Society of the Pacific Conference
  Series}, vol.~76, ed. by {R.~K.~Ulrich, E.~J.~Rhodes Jr., \& W.~Dappen}
  (1995), \emph{Astronomical Society of the Pacific Conference Series},
  vol.~76, pp. 465--474

\bibitem{Birch2000}
A.C. {Birch}, A.G. {Kosovichev}, \solphys \textbf{192}, 193 (2000)

\bibitem{Birch2004}
A.C. {Birch}, A.G. {Kosovichev}, T.L. {Duvall}, Jr., \apj \textbf{608}, 580
  (2004)

\bibitem{Couvidat2006}
S.~{Couvidat}, A.C. {Birch}, A.G. {Kosovichev}, \apj \textbf{640}, 516 (2006)

\bibitem{Zhao2003a}
J.~{Zhao}, A.G. {Kosovichev}, in \emph{GONG+ 2002. Local and Global
  Helioseismology: the Present and Future}, \emph{ESA Special Publication},
  vol. 517, ed. by {H.~Sawaya-Lacoste} (2003), \emph{ESA Special Publication},
  vol. 517, pp. 417--420

\bibitem{Birch2001}
A.C. {Birch}, A.G. {Kosovichev}, G.H. {Price}, R.B. {Schlottmann}, \apjl
  \textbf{561}, L229 (2001)

\bibitem{Gizon2001}
L.~{Gizon}, A.C. {Birch}, R.I. {Bush}, T.L. {Duvall}, Jr., A.G. {Kosovichev},
  P.H. {Scherrer}, J.~{Zhao}, in \emph{Solar encounter. Proceedings of the
  First Solar Orbiter Workshop}, \emph{ESA Special Publication}, vol. 493, ed.
  by {B.~Battrick, H.~Sawaya-Lacoste, E.~Marsch, V.~Martinez Pillet, B.~Fleck,
  \& R.~Marsden} (2001), \emph{ESA Special Publication}, vol. 493, pp. 227--231

\bibitem{Kosovichev2003}
A.G. {Kosovichev}, T.L. {Duvall}, Jr., in \emph{Society of Photo-Optical
  Instrumentation Engineers (SPIE) Conference Series}, \emph{Presented at the
  Society of Photo-Optical Instrumentation Engineers (SPIE) Conference}, vol.
  4853, ed. by {S.~L.~Keil \& S.~V.~Avakyan} (2003), \emph{Presented at the
  Society of Photo-Optical Instrumentation Engineers (SPIE) Conference}, vol.
  4853, pp. 327--340

\bibitem{Kosovichev2000}
A.G. {Kosovichev}, T.L..J. {Duvall}, P.H. {Scherrer}, \solphys \textbf{192},
  159 (2000)

\bibitem{Zhao2010}
J.~{Zhao}, A.G. {Kosovichev}, T.~{Sekii}, \apj \textbf{708}, 304 (2010)

\bibitem{Parker1979}
E.N. {Parker}, \apj \textbf{230}, 905 (1979)

\bibitem{Zhao2001}
J.~{Zhao}, A.G. {Kosovichev}, T.L. {Duvall}, Jr., \apj \textbf{557}, 384 (2001)

\bibitem{Zhao2004}
J.~{Zhao}, A.G. {Kosovichev}, \apj \textbf{603}, 776 (2004)

\end{thebibliography}

\end{document}